\pdfoutput=1
\documentclass[12pt,a4paper]{article}
\usepackage[utf8]{inputenc} 
\usepackage{booktabs}
\usepackage{amsmath,amssymb,amsthm,mathtools,bm}
\usepackage[margin=0.82in]{geometry}
\usepackage{hyperref}
\usepackage{cite}
\usepackage{longtable}
\usepackage{xcolor}
\usepackage{tikz}

\counterwithin{equation}{section}

\newcommand{\cA}{\mathcal A}
\newcommand{\cB}{\mathcal B}

\newcommand{\cD}{\mathcal D}

\newcommand{\cH}{\mathcal H}

\newcommand{\cJ}{\mathcal J}
\newcommand{\cK}{\mathcal K}
\newcommand{\cL}{\mathcal L}
\newcommand{\cM}{\mathcal M}
\newcommand{\cN}{\mathcal N}

\newcommand{\cQ}{\mathcal Q}
\newcommand{\cR}{\mathcal R}
\newcommand{\cS}{\mathcal S}
\newcommand{\cT}{\mathcal T}
\newcommand{\cV}{\mathcal V}

\newcommand{\cZ}{\mathcal Z}
\newcommand{\Z}{\mathbb Z}

\newcommand{\dd}{\mathrm d}
\newcommand{\e}{\mathrm e}

\newcommand{\op}{\mathrm{op}}
\newcommand{\Pf}{\mathrm{Pf}}
\newcommand{\rank}{\mathrm{rank}}

\newcommand{\Tr}{\mathrm{Tr}}

\newcommand{\abs}[1]{\left|#1\right|}
\newcommand{\norm}[1]{\left\lVert#1\right\rVert}

\newcommand{\id}{\bm I}
\DeclareMathOperator{\diag}{diag}

\theoremstyle{plain}

\theoremstyle{definition}

\title{\Large \textbf{Dilaton-Axion Wormhole Partition Function \\ 
Away from the BPS Endpoint}}
\author{Soo-Jong Rey\\[0.2cm]
{\sl Kwangwoon University, Seoul 01897 Korea}}
\date{\tt sjrey@kw.ac.kr}
\begin{document}
\maketitle

\begin{abstract}
I determine the leading finite-radius correction to the two-boundary
response of a charged axion wormhole near its BPS endpoint. With
$\delta=a_*^2$, exact radial matching in a product
axion--dilaton/saxion system gives a boundary eigenvalue that vanishes
as $O(\delta)$ and an off-diagonal Hessian entry of order
$O(\delta^2)$. Their combination produces a cross-boundary covariance
linear in $\delta$, with a negative first sub-leading correction on the
branch studied here. I determine the semiclassical range in which the
quadratic approximation remains controlled and identify the first
finite-radius angular resonance,
$\delta^2\log(\delta/\delta_0)$. Under a uniform rescaling of all
active charges, the coefficient of the linear cross-boundary response
scales as the inverse square of the charge scale. I then average the
inverse response over the chiral Gaussian ensemble describing the BPS
endpoint. Boundary reduction preserves the hard-edge index.
Multiplication by the matched BPS sector weight converts the normalized
response into the leading finite-radius two-source coefficient.
\end{abstract}

\tableofcontents
\section{Introduction}

The finite-radius axion-wormhole branch departs from a fixed-charge BPS instanton.
A nonzero minimal radius couples the boundary data across the wormhole interior and generates a connected response between the two asymptotic ends~\footnote{I use the terms \emph{asymptotic boundary}, \emph{minimal three-sphere}, and \emph{wormhole interior} throughout. These correspond to ``mouth,'' ``neck,'' and ``throat,'' respectively, in the literature.}.
I determine the onset of this response, convert it into the corresponding fixed-charge source coefficient, and carry that coefficient into the axion-angle representation of the wormhole partition function.

\begin{figure}[htbp]
\centering
\includegraphics[width=0.65\textwidth]{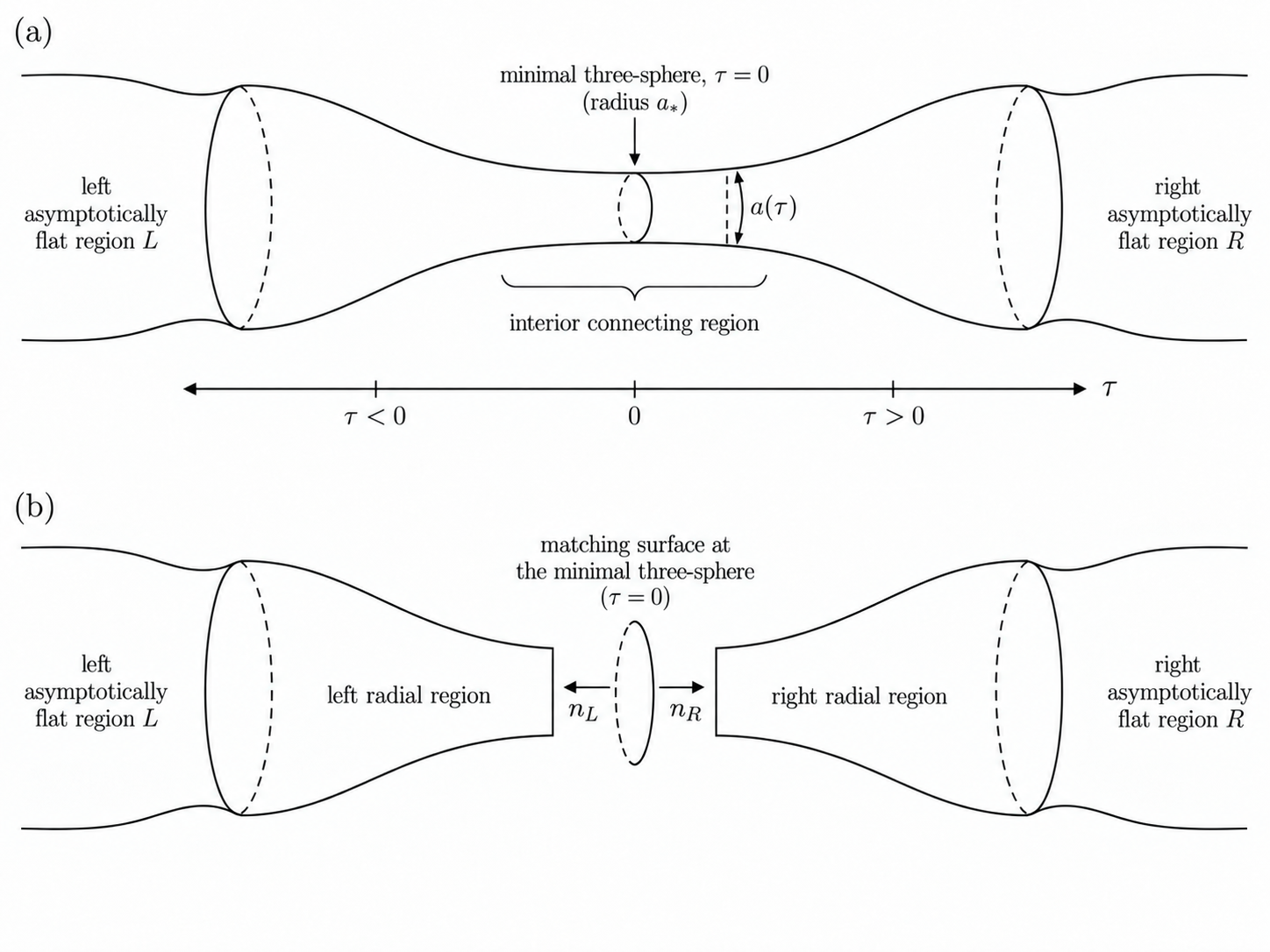}
\caption{\sl Geometry and radial split of the bipartite wormhole used in the calculation.
Panel (a) depicts a Euclidean wormhole with left and right asymptotically flat regions joined through the interior and a minimal three-sphere of radius $a_*$ at the reflection-symmetric point $\tau=0$.
Panel (b) depicts the split at $\tau=0$ into left and right radial regions; ${\bf n}_L$ and ${\bf n}_R$ are the outward normals at the matching surface.}
\label{fig:wormhole-geometry}
\end{figure}

I parameterize the saddle family by
\[
\delta=a_*^2,
\]
where $a_*$ is the radius of the minimal three-sphere.
I hold the axion charge $\bm\nu$ and the right asymptotic dilaton/saxion data fixed.
Smooth continuation through the interior then determines the left asymptotic data.
For one homogeneous fluctuation transverse to the background trajectory, let
\begin{equation}
q=(q_L,q_R)^T
\end{equation}
denote its two boundary values.
The reduced on-shell action and the connected boundary covariance take the form
\begin{equation}
I_{\mathrm{on}}^{(2)}
=
\frac{A}{2}q^T K(\delta)q,
\qquad
G(\delta)=A^{-1}K(\delta)^{-1},
\end{equation}
with $A>0$.
Exact radial matching in the product axion--dilaton/saxion system gives
\begin{equation}
K_{LL}=O(\delta),
\qquad
K_{RR}=O(1),
\qquad
K_{LR}=O(\delta^2).
\end{equation}
One eigenvalue of $K$ therefore vanishes linearly with $\delta$, while the second approaches a positive constant.
The inverse soft eigenvalue supplies one power of $\delta^{-1}$ and yields
\begin{equation}
G_{LR}(\delta)
=
g_{1,\bm\nu}\delta
+
g_{2,\bm\nu}\delta^2
+
O(\delta^3).
\end{equation}
For the subcritical product branch studied here,
\begin{equation}
g_{1,\bm\nu}>0,
\qquad
g_{2,\bm\nu}<0.
\end{equation}
The cross-boundary response consequently rises linearly from the BPS endpoint and bends downward at first sub-leading order.

Matched asymptotics fixes this power counting.
At fixed outer radius, the metric departs from the BPS profile at $O(\delta^2)$.
In the inner coordinate $a=\sqrt{\delta}\,\rho$, the reduced action instead carries an overall factor of $\delta$.
Exact inner--outer matching transports this $O(\delta)$ contribution to the asymptotic boundary form.
The boundary response therefore resolves the opening wormhole interior one order earlier than the fixed-radius metric expansion.

Papers I--III establish the three endpoint ingredients used in this calculation.
Paper-I constructs the physical fixed-charge BPS Hessian and factorizes it as
\begin{equation}
\mathcal H_{\bm\nu,0}
=
\mathcal Q_{\bm\nu}^{\dagger}\mathcal Q_{\bm\nu}.
\end{equation}
Paper-II organizes the coefficient sequence
\begin{equation}
C_{\bm\nu}^{ij}
\;\xrightarrow{\;R_b\;}\;
W_{\bm\nu}[b]
\;\xrightarrow{\;\sum_{\bm\nu}e^{i\bm\nu\cdot\bm\theta}\;}\;
Z_{\mathrm{wh}}(\bm\theta;b).
\end{equation}
Paper-III identifies the nonzero endpoint spectrum with squared singular values and supplies its Laguerre/chiral hard-edge measure.
These results fix the endpoint operator, the boundary reduction, and the endpoint spectral statistics.
The present paper computes the finite-radius deformation that joins them.

I first solve the homogeneous boundary problem through the first subleading order and determine the semiclassical range of the quadratic response.
I then derive the first angular resonance, the charge dependence of the linear coefficient, and the spectral average of the inverse response.
Matching the complete fixed-sector measure to its finite BPS weight converts the covariance slope into the leading unnormalized two-source coefficient.
The reduction data $b$ then produce $W_{\bm\nu}[b]$, and the charge sum produces the dependence on the asymptotic axion angle.

\subsection{Paper-I: the saddle and its physical fluctuations}

Paper-I formulates the saddle at fixed conserved axion charge~\cite{ReyBPS2026}.
At fixed charge, the BPS instanton is the endpoint of the non-BPS wormhole branch.
Physical fluctuations preserve the fixed charge.
Varying the charge moves to another saddle.
This distinction determines the fluctuation domain and the boundary terms in the variational problem.

Figure~\ref{fig:wormhole-geometry} shows the finite-radius geometry of the saddle and the radial split used below.
Away from the BPS endpoint, the minimal radius $a_*>0$ defines a smooth three-sphere at $\tau=0$.
Its extrinsic curvature vanishes, and the scale factor is reflection symmetric about this surface.
The axion and dilaton/saxion profiles obey the charged radial equations and need not share this reflection symmetry.
Splitting the smooth solution at $\tau=0$ produces left and right radial regions with opposite outward normals.
Their outward canonical momenta cancel when the two regions are joined.

Paper-I then constructs the physical Hessian at the BPS endpoint.
The Hamiltonian constraint removes the redundant metric variable.
Gauge fixing and the boundary conditions define the physical domain.
Collective zero modes are removed from the determinant and treated separately.
Linearization of the BPS equations on the remaining domain gives a first-order map $\mathcal Q_{\bm\nu}$.
The physical Hessian is its adjoint square:
\begin{equation}
 \mathcal H_{\bm\nu,0}
 =\mathcal Q_{\bm\nu}^{\dagger}\mathcal Q_{\bm\nu},
 \qquad
 \langle\eta,\mathcal H_{\bm\nu,0}\eta\rangle
 =\|\mathcal Q_{\bm\nu}\eta\|^2.
 \label{1-1}
\end{equation}
Here $\bm\nu$ denotes the fixed charge sector and $\eta$ denotes a physical perturbation.
The adjoint is taken on the same boundary domain.
Equation~\eqref{1-1} makes the BPS quadratic form positive semidefinite and identifies its nonzero eigenvalues with squared singular values of $\mathcal Q_{\bm\nu}$.
This factorization holds at the BPS endpoint. 

I use this BPS operator as the endpoint reference for the finite-radius problem.
The non-BPS solution changes the radial background, the radial interval, and the matching at the minimal three-sphere.
I compute these changes explicitly in a product axion--dilaton/saxion system.

Paper-I also identifies where the two-end response enters the long-distance source operator.
Each asymptotic boundary supports local source insertions.
Eliminating the wormhole interior produces a quadratic coefficient for these sources.
Expanding the total source over parent components gives the same-component and different-component terms of the same quadratic form.
I compute the response between the left and right asymptotic boundaries before assigning the two insertions to parent components.

\subsection{Paper-II: the coefficient and its charge transform}

Paper-II organizes the wormhole amplitude in the charge representation before transforming to the asymptotic axion angle~\cite{ReyCharge2026}.
For each fixed charge vector $\bm\nu$, the saddle calculation produces a two-end coefficient matrix $C_{\bm\nu}^{ij}$ in the space of boundary insertions.
A prescribed reduction $R_b$ then converts this matrix into the scalar sector coefficient
\[
C_{\bm\nu}^{ij}
\;\xrightarrow{\;R_b\;}\;
W_{\bm\nu}[b].
\]
The data $b$ specify the contractions, projections, insertions, source normalization, contour, and fields eliminated in this reduction.
Only after this operation, I form the axion-angle representation,
\begin{equation}
 Z_{\rm ax}(\bm\theta;b)
 =
 \sum_{\bm\nu}
 W_{\bm\nu}[b]\,
 \e^{i\bm\nu\cdot\bm\theta}.
 \label{1-2}
\end{equation}
The vector $\bm\theta$ is the compact Fourier variable conjugate to the quantized charge vector $\bm\nu$.
Its phase factor in Eq.~\eqref{1-2} is kinematic.
The charge dependence generated by the saddle, fluctuations, measure, contour, and reduction resides in $W_{\bm\nu}[b]$.

This ordering separates properties of the two-end coefficient from properties of the final charge transform.
Before reduction, a positive coefficient matrix constrains its mixed and diagonal source entries through quadratic-form inequalities.
After reduction, the scalar sequence $W_{\bm\nu}[b]$ may be positive, signed, or complex.
A positive sequence defines a charge probability law and obeys moment inequalities.
A complex sequence retains its phase sector by sector in Eq.~\eqref{1-2}.
The reduction therefore determines which analytic and positivity statements apply to the resulting wormhole partition function.

Paper-II also separates existence of the complexified coefficient from convergence of its charge sum.
Complex continuation of $\bm\theta$ first defines a complexified boundary-value problem.
The resulting coefficients must then decay sufficiently rapidly along the charge lattice for the Fourier series to converge.
For several axions, this condition is directional in charge space.
The dilute Bessel or Skellam form arises only after positivity, independent dilute events, charge symmetry, and unit-charge dominance have been imposed.

The present paper supplies the finite-radius input to this construction.
I first compute the deformation of the physical two-boundary response in a fixed charge sector.
For fixed boundary values, the quadratic action is governed by the boundary Hessian $K$.
After Gaussian integration over those boundary variables, the connected source response is governed by $K^{-1}$.
Multiplication by the complete fixed-sector weight gives the unnormalized two-end coefficient, and the Paper-II reduction $R_b$ then produces $W_{\bm\nu}[b]$.
Thus the finite-radius geometry is determined before the boundary reduction and before the charge transform in Eq.~\eqref{1-2}.

\subsection{Paper-III: the endpoint spectrum and its microscopic measure}

Paper-III resolves the BPS endpoint operator into its singular-value data~\cite{ReyChiral2026}.
For the fixed-charge physical Hessian
\[
\mathcal H_{\bm\nu,0}
=
\mathcal Q_{\bm\nu}^{\dagger}\mathcal Q_{\bm\nu},
\]
the associated chiral double is
\begin{equation}
 \mathcal D_{\bm\nu}
 =
 \begin{pmatrix}
 0 & \mathcal Q_{\bm\nu}^{\dagger}\\
 \mathcal Q_{\bm\nu} & 0
 \end{pmatrix},
 \qquad
 \operatorname{Spec}_{\ne0}\mathcal D_{\bm\nu}
 =
 \{\pm s_j\},
 \qquad
 \operatorname{Spec}_{\ne0}\mathcal H_{\bm\nu,0}
 =
 \{s_j^2\}.
 \label{1-3}
\end{equation}
The $s_j$ are the nonzero singular values of $\mathcal Q_{\bm\nu}$.
The adjoint-square structure pins the physical spectrum to the non-negative axis and makes zero a hard spectral edge.
A small $s_j^2$ represents a weak restoring direction of the BPS quadratic form and enhances any inverse response with nonzero overlap onto that mode.

Paper-III represents this endpoint by a finite chiral ensemble that preserves the index, antiunitary class, and endpoint boundary form of the physical BPS map.
For the central Gaussian representative, the squared singular values obey a Laguerre--Wishart law.
Its microscopic scaling limit at zero is the corresponding Bessel hard-edge ensemble.
The hard-edge index records the rectangular defect of the finite representation of the first-order map, while the Dyson class records its real, complex, or quaternionic structure.
These data determine the local endpoint statistics used below.

The D($-1$)/D3 ADHM system supplies the microscopic Type-IIB realization considered in Paper-III.
Its bifundamental sector contains a rectangular first-order block, and singular-value reduction produces the same Laguerre structure.
The matching to the gravitational endpoint is performed at the level of edge data.
The axion charge $\bm\nu$, the ADHM instanton number, the rectangular defect, and the gravitational hard-edge index are distinct labels.
A microscopic identification requires the charge dictionary, physical fluctuation domain, zero-mode quotient, hard-edge index, and antiunitary class to agree.
The resulting ADHM/chiral-Wishart integral represents the microscopic fixed-sector contribution to the coefficient discussed in Paper-II.

Paper-III supplies the statistical law of the BPS endpoint spectrum in each fixed charge sector.
The finite-radius deformation is a separate gravitational input.
In the present paper, radial matching determines the deformed boundary matrix and the effective source vectors before the endpoint ensemble is averaged.
The resulting inverse response depends on how these physical probes and the finite-radius deformation project onto the BPS endpoint eigenmodes.
Paper-III therefore supplies the endpoint spectral statistics, while the present calculation determines the observable that probes them.

This separation also fixes the roles of the two Bessel structures inherited from Papers II and III.
In Paper-II, modified Bessel functions arise in the dilute specialization of the charge-sector sum that constructs the axion-angle representation.
In Paper-III, the Bessel kernel describes microscopic singular-value correlations at the hard spectral edge within one fixed charge sector.
The first belongs to the sum over charge sectors; the second belongs to the spectrum inside a sector.
I keep these operations distinct throughout the finite-radius calculation.

\subsection{This Paper: the finite-radius response}

This paper computes the departure of the fixed-charge two-boundary response from its BPS endpoint.
I work charge sector by charge sector and parameterize the non-BPS saddle family by
\begin{equation}
 \delta=a_*^2 .
 \nonumber
\end{equation}
I hold the conserved charge vector and the right asymptotic dilaton/saxion data fixed.
Smooth continuation through the minimal three-sphere at $\tau=0$ then determines the left asymptotic data.
For each sector, I first compute the finite-radius boundary response, then apply the Paper-II reduction, and finally average the resulting response with the Paper-III endpoint measure.
This sequence keeps the geometric deformation, the boundary reduction, and the endpoint spectral average as distinct operations.

The leading response is fixed by matched radial asymptotics.
At fixed outer radius, the background metric first departs from its BPS profile at $O(\delta^2)$.
Near the shrinking minimal three-sphere, the rescaling $a=\sqrt{\delta}\,\rho$ instead produces a reduced inner action proportional to $\delta$.
Exact radial matching carries this inner contribution to the asymptotic boundary quadratic form.
For one homogeneous dilaton/saxion mode transverse to the background scaling trajectory, the boundary Hessian obeys
\begin{equation}
 K_{LL}=a_1\delta+O(\delta^2),
 \qquad
 K_{RR}=r_0+O(\delta^2),
 \qquad
 K_{LR}=b_2\delta^2+O(\delta^3),
 \qquad
 a_1,r_0>0,\quad b_2<0 .
 \nonumber
\end{equation}
One eigenvalue therefore softens linearly in $\delta$, while the second approaches a finite positive limit.
With $G=A^{-1}K^{-1}$, the connected cross-boundary covariance becomes
\begin{equation}
 G_{LR}(\delta)
 =
 g_{1,\bm\nu}\delta
 +
 g_{2,\bm\nu}\delta^2
 +
 O(\delta^3),
 \qquad
 g_{1,\bm\nu}>0,
 \qquad
 g_{2,\bm\nu}<0 .
 \nonumber
\end{equation}
The inverse soft eigenvalue supplies the factor $\delta^{-1}$ that converts the $O(\delta^2)$ mixed Hessian into an $O(\delta)$ cross-boundary response.
The exact radial solution fixes the first curvature of this response and makes it negative on the branch studied here.
I impose the finite endpoint matching
$Z_{\bm\nu}[0;\delta]\to Z_{\bm\nu,0}$
for the complete fixed-sector weight.
The corresponding unnormalized source coefficient then has the leading form
\begin{equation}
 C_{\bm\nu}^{LR}(\delta)
 =
 \delta\, Z_{\bm\nu,0} g_{1,\bm\nu}
 +
 o(\delta).
 \nonumber
\end{equation}

The same boundary formulation fixes the domain of this linear response.
A fixed positive quadratic term in the soft boundary direction lifts the small eigenvalue to $O(1)$ and moves the cross-boundary covariance from $O(\delta)$ to $O(\delta^2)$.
Cubic and quartic vertices control the opposite limit, where the growing soft-mode variance invalidates the Gaussian expansion.
Together with the outer small-radius expansion, they define the semiclassical interval
\begin{equation}
 \frac{C_{\rm ctl}}{A}\ll \delta \ll L^2 .
 \nonumber
\end{equation}
Within this interval, the homogeneous mode gives the linear response above.
The coupled angular problem develops its first finite-radius resonance in the $\ell=3$ sector,
\begin{equation}
 \delta^2\log(\delta/\delta_0),
 \nonumber
\end{equation}
whose first derivative vanishes at the BPS endpoint.\footnote{
Kehagias and Riotto~\cite{KehagiasRiotto2026} obtain logarithmic scalar running near the Euclidean AdS boundary from a Gauss--Bonnet tadpole and verify the Kontsevich--Segal--Witten allowability of their complex background metric.
Their logarithm belongs to the asymptotic background solution, whereas the logarithm here arises from a finite-radius fluctuation resonance about the near-BPS wormhole.
Both admit a radial-resonance organization and a finite-part prescription.
The exact radial transformation used here maps the near-BPS fluctuation equation to an asymptotically AdS form and provides a common analytic representation of the resonance.
The two gravitational saddles and their partition functions remain distinct.}

The leading geometric response has a simple charge scaling in the product system.
For a uniform rescaling of all active charges,
$\bm\nu\mapsto t\bm\nu$,
the stationary target-space trajectory keeps the same shape and
\begin{equation}
 g_{1,t\bm\nu}=t^{-2}g_{1,\bm\nu}.
 \nonumber
\end{equation}
Changing the relative charge magnitudes instead moves the stationary trajectory.
I determine that response by differentiating the background equations, the fluctuation operator, and the source map at fixed asymptotic data.
These charge-sensitivity equations propagate the change of charge composition into the finite-radius boundary coefficient.

I finally combine this deterministic deformation with the Paper-III endpoint ensemble.
The physical boundary reduction first fixes the endpoint matrix $M_{0,\bm\nu}$, its finite-radius perturbation $D_{1,\bm\nu}$, and the effective source vectors.
The inverse response then satisfies
\begin{equation}
 \left.\frac{d}{d\delta}M_{\bm\nu}(\delta)^{-1}\right|_{\delta=0}
 =
 -M_{0,\bm\nu}^{-1}
 D_{1,\bm\nu}
 M_{0,\bm\nu}^{-1}.
 \nonumber
\end{equation}
Small endpoint eigenvalues therefore enhance the finite-radius response according to their overlap with the deformation and the physical sources.
For the central chiral Gaussian representative, eliminating the unobserved columns produces another Wishart matrix with the same rectangularity.
The boundary projection therefore preserves the hard-edge index.
The resulting negative moments determine the spectral averages of the determinant and inverse response, together with their hard-edge integrability conditions.

The appendices derive the fixed-charge quadratic functional, the boundary Schur complements, the exact homogeneous and angular matching, the interaction bounds, the charge sensitivities, and the endpoint spectral averages.

\section{Geometry, boundary data, and the physical response}

I describe the finite-radius saddle in proper radial distance $\tau$,
with the minimal three-sphere at $\tau=0$.
The metric takes the form
\begin{equation}
 ds^2=\dd\tau^2+a(\tau)^2\dd\Omega_3^2,
 \qquad
 \delta=a_*^2,
 \qquad
 E=12\delta^2,
 \label{2-1}
\end{equation}
where $a_*=a(0)$ is the minimum of the scale factor.
The radial first integral gives
\[
 (a')^2=1-\frac{\delta^2}{a^4},
 \qquad
 a(-\tau)=a(\tau).
\]
Thus the metric is reflection symmetric and the minimal three-sphere is
totally geodesic.
The charged matter profile follows its own radial equations.
Its scalar momentum at $\tau=0$ is generally nonzero.
Cutting the smooth saddle at this surface produces left and right radial
problems with opposite outward normals.
Continuity of the fields together with cancellation of the two outward
canonical momenta reconstructs the complete two-boundary solution.

I formulate this family at fixed axion charge.
After Legendre transforming the axions, the fixed-charge Routhian
contains the dilaton/saxion potential $U_{\bm\nu}(\phi)$.
The dilaton/saxions evolve in this potential on the radial geometry
set by $a(\tau)$.
I vary $\delta$ while holding $\bm\nu$ and the right asymptotic
dilaton/saxion data fixed.
Smooth continuation through $\tau=0$ then determines the left
asymptotic values.
Their $\delta$ dependence contributes the canonical boundary work to
the derivative of the on-shell action.
All finite-radius derivatives below are taken along this fixed-charge
saddle family.

The metric symmetry and the matter boundary problem play different
roles.
At $\tau=0$, $a'=0$, while the scalar first integral fixes the allowed
scalar momentum through the charge potential.
A smooth charged profile can therefore cross the minimal sphere with
nonzero scalar derivative.
Parity decomposition of the fluctuations requires reflection symmetry
of the complete fluctuation operator and its boundary domain.
I instead impose continuity of the fluctuation fields and match their
canonical momenta directly across the minimal sphere.

The small-radius expansion contains an outer region and an inner
region with different power counting.
At fixed outer radius,
\[
 \frac{\delta^2}{a^4}\ll1,
\]
so the metric first departs from its BPS profile at $O(\delta^2)$.
Near the minimal three-sphere I set
\[
 a=\sqrt{\delta}\,\rho .
\]
Then $\delta^2/a^4=\rho^{-4}$ remains finite, while the reduced inner
action acquires an overall factor of $\delta$.
The inner problem therefore generates an $O(\delta)$ contribution to
the boundary quadratic form.
Exact radial matching carries this contribution through the overlap
region to the asymptotic boundaries.

This matched expansion defines the derivative at the BPS endpoint.
For fixed outer radius, the solution approaches the BPS profile as
$\delta\to0$.
For fixed $\rho=a/\sqrt{\delta}$, the inner solution retains a
nontrivial finite profile.
Their common overlap transfers the shrinking-interior dynamics to the
asymptotic response.
The resulting boundary variation contains both the deformation of the
bulk trajectory and the induced motion of the left asymptotic data.

I next reduce the quadratic problem to its physical boundary variables.
At fixed charge, I impose the Hamiltonian constraint and fix the gauge
before eliminating the remaining auxiliary and interior fields.
For a real bosonic quadratic form written in retained and eliminated
coordinates,
\begin{equation}
 \frac12
 \begin{pmatrix}x\\u\end{pmatrix}^{T}
 \begin{pmatrix}
 A & B\\
 B^{T} & D
 \end{pmatrix}
 \begin{pmatrix}x\\u\end{pmatrix},
 \qquad D^{-1}\ {\rm defined},
 \nonumber
\end{equation}
stationarity in $u$ gives the effective boundary Hessian
\begin{equation}
 A-BD^{-1}B^T .
 \nonumber
\end{equation}
I use this Schur-complement reduction throughout the finite-radius
calculation.

The radial split at $\tau=0$ provides an equivalent construction.
I evaluate the left and right on-shell actions as functions of their
common matching value and add them.
Extremizing this sum over the matching value imposes canonical-momentum
matching and reproduces the smooth full saddle.
Eliminating the matching variable before or after assembling the two
radial regions gives the same physical outer-boundary functional.
The matching surface is therefore an intermediate device; the final
boundary Hessian acts on the asymptotic source variables.

For one homogeneous dilaton/saxion mode transverse to the background
scaling trajectory, let $q_L$ and $q_R$ denote its asymptotic boundary
values.
I collect them into
\[
 q=(q_L,q_R)^T
\]
and define the physical boundary Hessian $K$ and connected covariance
$G$ by
\begin{equation}
 I_{\rm on}^{(2)}
 =
 \frac{\cA}{2}q^TKq,
 \qquad
 q=(q_L,q_R)^T,
 \qquad
 G=\cA^{-1}K^{-1}.
 \label{2-2}
\end{equation}
Here $\cA>0$ is the overall normalization inherited from the reduced
action.
The diagonal entries $K_{LL}$ and $K_{RR}$ measure the quadratic cost
of fixing each asymptotic value, while $K_{LR}$ measures their coupling
in the fixed-boundary action.

I first solve the Dirichlet problem with $q_L$ and $q_R$ fixed.
I then integrate these boundary variables with the Lebesgue measure
specified by the source functional and with linear sources.
Completing the Gaussian square gives
\[
 \frac{1}{2\cA}J^TK^{-1}J
\]
in the source exponent.
Two source derivatives therefore give $G$.
The distinction is essential:
$K_{LR}$ is the mixed coefficient of the fixed-boundary on-shell
action, whereas $G_{LR}$ is the connected left--right response after
the boundary variables have been integrated.

The boundary state forms part of this observable.
Adding a positive quadratic boundary term shifts $K$ by a positive
matrix and changes $G$, while leaving the interior Dirichlet solution
unchanged.
I therefore use one fixed boundary measure and one fixed boundary state
when comparing charge sectors and when forming the subsequent
axion-angle representation.

Appendix~\ref{sec:near-bps-problem} defines the fixed-charge source
functional, fluctuation domain, and Gaussian reduction.
Appendix~\ref{sec:finite-radius-boundary} constructs the radial
value-to-momentum maps, glues the two radial regions, and derives the
physical boundary Hessian.
Appendix~\ref{sec:homogeneous-radial} solves the product
axion--dilaton/saxion model and performs the exact radial and angular
matching.

\section{Finite-radius response and its range of validity}

\subsection{Linear response from exact radial matching}

I now evaluate the finite-radius response in the product
axion--dilaton/saxion system.
For $p$ axions with canonically normalized dilaton/saxions $\phi_i$, the
fixed-charge potential is
\begin{equation}
 U_{\bm\nu}(\phi)
 =
 \sum_i \nu_i^2 \exp(2\beta_i\phi_i),
 \qquad
 \beta_i>0 .
 \nonumber
\end{equation}
Introduce
\begin{equation}
 \beta^{-2}=\sum_i\beta_i^{-2},
 \qquad
 n_i=\frac{\beta}{\beta_i},
 \qquad
 \phi_i=n_i\sigma+E_{iu}s^u ,
 \nonumber
\end{equation}
where the columns of $E$ form an orthonormal basis orthogonal to $n$.
The field $\sigma$ follows the common scaling direction, while the
$p-1$ fields $s^u$ describe relative dilaton/saxion displacements.
In these variables,
\begin{equation}
 U_{\bm\nu}
 =
 \exp(2\beta\sigma)V_{\bm\nu}(s).
 \nonumber
\end{equation}

The stationary transverse trajectory satisfies
$\partial_uV_{\bm\nu}=0$.
At this point the normalized transverse Hessian is
\begin{equation}
 \frac{\partial_u\partial_vV_{\bm\nu}}{V_{\bm\nu}}
 =
 4\beta^2\delta_{uv}.
 \nonumber
\end{equation}
Thus every homogeneous transverse direction obeys the same radial
equation on the product background.
I work on the subcritical branch
\begin{equation}
 0<\sqrt{3}\,\beta<1 .
 \nonumber
\end{equation}

Let $q_L$ and $q_R$ denote the asymptotic values of one transverse mode.
These boundary values displace the dilaton/saxions away from their
stationary relative configuration.
Solving the radial equation with $(q_L,q_R)$ fixed and evaluating the
outward canonical momenta determines the boundary Hessian $K$ of
Eq.~\eqref{2-2}.
The exact solution gives
\begin{equation}
 K_{LL}
 =
 a_1\delta+O(\delta^2),
 \qquad
 K_{RR}
 =
 r_0+O(\delta^2),
 \qquad
 K_{LR}
 =
 b_2\delta^2+O(\delta^3),
 \label{3-1}
\end{equation}
with
\begin{equation}
 a_1>0,
 \qquad
 r_0>0,
 \qquad
 b_2<0 .
 \nonumber
\end{equation}
The couplings, charge vector, and asymptotic data determine these
coefficients; Appendix~\ref{sec:homogeneous-radial} gives their closed
forms.
The unequal powers in $K_{LL}$ and $K_{RR}$ follow from the boundary
prescription: the right asymptotic data remain fixed along the saddle
family, whereas the left data are determined by smooth continuation.
The metric itself remains reflection symmetric.

Equation~\eqref{3-1} contains the mechanism of the leading response.
Its determinant is
\begin{equation}
 \det K
 =
 a_1r_0\delta+O(\delta^2),
 \nonumber
\end{equation}
so one eigenvalue softens linearly in $\delta$ while the other tends to
$r_0$.
Since
\begin{equation}
 (K^{-1})_{LR}
 =
 -\frac{K_{LR}}{\det K},
 \nonumber
\end{equation}
the connected covariance becomes
\begin{equation}
 G_{LR}
 =
 g_{1,\bm\nu}\delta
 +
 g_{2,\bm\nu}\delta^2
 +
 O(\delta^3),
 \qquad
 g_{1,\bm\nu}
 =
 -\frac{b_2}{\cA a_1r_0}>0,
 \qquad
 g_{2,\bm\nu}<0 .
 \label{3-2}
\end{equation}
The $O(\delta)$ soft eigenvalue supplies the inverse factor
$\delta^{-1}$ that converts the $O(\delta^2)$ mixed Hessian into an
$O(\delta)$ cross-boundary covariance.
The exact radial solution also fixes the first subleading coefficient.
Its negative sign makes the response bend downward as the wormhole
opens away from the BPS endpoint.

The same solution fixes the sign and regularity of the quadratic form.
For sufficiently small positive $\delta$,
\begin{equation}
 K_{LL}K_{RR}-K_{LR}^2>0 ,
 \nonumber
\end{equation}
and hence $K$ is positive in this transverse sector.
Together with $b_2<0$, this gives $G_{LR}>0$.
Direct differentiation of the exact on-shell action reproduces the
same canonical momenta and the same boundary Hessian.

The linear law in Eq.~\eqref{3-2} also exposes the role of the boundary
state.
The soft contribution to the on-shell action begins as
\begin{equation}
 \frac{\cA a_1}{2}\,\delta q_L^2 .
 \nonumber
\end{equation}
Adding a fixed positive quadratic boundary term for $q_L$ lifts this
soft eigenvalue to a finite constant.
The mixed entry remains $O(\delta^2)$, while the determinant acquires a
nonzero endpoint limit.
The corresponding cross-boundary covariance then begins at
$O(\delta^2)$.
The $O(\delta)$ law therefore follows from the soft boundary direction
specified by the source functional used here.

The normalized covariance is only one factor in the unnormalized
two-source coefficient.
Let $Z_{\bm\nu}[0;\delta]$ denote the complete zero-source sector
weight, including the determinant or Pfaffian factors, exact-mode
insertions, collective-coordinate measure, boundary normalization, and
the remaining fluctuation sectors.
I impose the BPS matching condition
\begin{equation}
 Z_{\bm\nu}[0;\delta]
 \longrightarrow
 Z_{\bm\nu,0}
 \qquad
 (\delta\to0^+).
 \nonumber
\end{equation}
Then
\begin{equation}
 C_{\bm\nu}^{LR}(\delta)
 =
 \delta B_{\bm\nu}+o(\delta),
 \qquad
 B_{\bm\nu}
 =
 Z_{\bm\nu,0}g_{1,\bm\nu}.
 \label{3-3}
\end{equation}
The finite limit belongs to the complete fixed-sector measure.
The isolated two-variable bosonic Gaussian contains the soft factor
$\delta^{-1/2}$ and therefore does not furnish this matching by itself.

Since $G_{LR}(0)=0$, continuity of the complete sector weight fixes the
endpoint slope in Eq.~\eqref{3-3}.
When $Z_{\bm\nu}[0;\delta]$ is differentiable, its first derivative
enters only at $O(\delta^2)$, together with $g_{2,\bm\nu}$.
Thus the leading coefficient factorizes into the BPS sector weight and
the positive geometric response $g_{1,\bm\nu}$.
Its phase is inherited entirely from $Z_{\bm\nu,0}$.
Equation~\eqref{3-3} is the finite-radius input to the Paper-II
reduction.

\subsection{Angular matching and the semiclassical interval}

The higher harmonics introduce a second finite-radius structure.
For $\ell\geq1$, I expand the fluctuations in scalar harmonics of the
round three-sphere and retain the coupled axion--dilaton/saxion radial
system.
Angular axion current prevents the fixed-charge elimination used in the
homogeneous sector, so the axion and dilaton/saxion fluctuations remain
coupled.
The radial equations, constraints, and boundary conditions together
select the physical solutions.

I organize the small-radius expansion around the BPS radial basis.
Successive finite-radius corrections shift the radial power by fixed
steps.
A logarithm appears when one such correction reaches another indicial
root with a nonzero compatibility projection.
Appendix~\ref{sec:homogeneous-radial} evaluates these projections and
matches the local Frobenius expansion to the exact angular solutions.

The first two low harmonics illustrate the distinction.
For $\ell=1$, the BPS system contains a repeated index and hence a
logarithmic companion, but the first finite-radius correction to the
growing branch closes with an ordinary power.
For $\ell=3$, the compatibility projection is nonzero and produces the
first finite-radius resonance,
\begin{equation}
 \delta^2\log(\delta/\delta_0),
 \nonumber
\end{equation}
at fixed outer radius.
Its derivative is
\begin{equation}
 \frac{\dd}{\dd\delta}
 \left[
 \delta^2\log(\delta/\delta_0)
 \right]
 =
 2\delta\log(\delta/\delta_0)+\delta
 \longrightarrow 0
 \qquad
 (\delta\to0^+).
 \nonumber
\end{equation}
The resonance therefore enters beyond the linear endpoint response and
leaves the slope in Eq.~\eqref{3-2} unchanged.

Finite-cut Dirichlet data fix the angular boundary problem.
After the exact field transformation used in
Appendix~\ref{sec:homogeneous-radial}, the transverse scalar
Sturm--Liouville operators have positive potentials, and the
Poincar\'e bound excludes a zero-frequency Dirichlet mode on the finite
interval.
The translational collective mode lies along the background scaling
direction and has vanishing transverse projection.
Spatial rotations act trivially on the radial background.
These facts isolate the transverse angular sector from the collective
coordinates treated separately in the measure.

The homogeneous soft mode sets the complementary limitation on the
quadratic approximation.
Its variance grows as the soft eigenvalue decreases, so cubic and
quartic boundary vertices eventually compete with the Gaussian action.
I compute these vertices after allowing the regular scalar fields and
the wormhole radius to relax.
Their contractions define the dimensionless control coefficient
$C_{\rm ctl}$.
For an outer length scale $L$, the controlled quadratic regime is
\begin{equation}
 \frac{C_{\rm ctl}}{\cA}
 \ll
 \delta
 \ll
 L^2 .
 \label{3-4}
\end{equation}
The lower bound suppresses interaction corrections generated by the
soft variance.
The upper bound keeps the outer small-radius expansion under control.
Appendix~\ref{sec:homogeneous-radial} gives the dependence of
$C_{\rm ctl}$ on the number of transverse modes and on the product
couplings.

At fixed $\cA$, the interaction bound is reached before the geometric
endpoint $\delta=0$.
A controlled approach to the BPS endpoint therefore uses the
correlated semiclassical limit
\begin{equation}
 \frac{\delta}{L^2}\longrightarrow0,
 \qquad
 \frac{\cA\delta}{C_{\rm ctl}}\longrightarrow\infty .
 \nonumber
\end{equation}
Within this overlap regime, Eq.~\eqref{3-2} gives the endpoint slope and
its first curvature, while the $\ell=3$ resonance enters only at the
higher order $\delta^2\log\delta$.

\section{Charge dependence and endpoint spectral response}

\subsection{Charge dependence across sectors}

I now determine how the linear finite-radius response varies across the
charge lattice.
All comparisons use the same right asymptotic dilaton/saxion data, the
same boundary state, and the same source prescription.
The leading unnormalized coefficient found in Eq.~\eqref{3-3} is
\[
 B_{\bm\nu}=Z_{\bm\nu,0}g_{1,\bm\nu},
\]
so the charge dependence separates naturally into the BPS sector weight
$Z_{\bm\nu,0}$ and the geometric response $g_{1,\bm\nu}$.

The product model gives an exact scaling law for the geometric factor.
At fixed active support, a uniform rescaling
$\bm\nu\mapsto t\bm\nu$ multiplies the fixed-charge potential by $t^2$
without changing the shape of the stationary target-space trajectory.
The normalized transverse fluctuation problem is therefore unchanged,
while the overall charge scale enters the radial response through
$V_{\bm\nu}$.
Since
\[
 V_{t\bm\nu}=t^2V_{\bm\nu},
\]
the linear coefficient obeys
\begin{equation}
 g_{1,t\bm\nu}
 =
 t^{-2}g_{1,\bm\nu}.
 \nonumber
\end{equation}
Appendix~\ref{sec:charge-phase} derives this relation from the exact
product solution.
The factor $g_{1,\bm\nu}$ is real and positive.
Uniform charge rescaling therefore changes the magnitude of
$B_{\bm\nu}$ through the geometric multiplier and introduces no new
phase.

Charge reversal gives a second exact relation.
The product potential depends on each charge through $\nu_i^2$.
Hence
\[
 \nu_i\mapsto-\nu_i
\]
leaves the stationary trajectory, the homogeneous boundary Hessian, and
$g_{1,\bm\nu}$ unchanged.
Sign-related sectors with the same charge magnitudes therefore carry
the same finite-radius geometric response.
Any difference between their full coefficients $B_{\bm\nu}$ resides in
the corresponding BPS sector factors $Z_{\bm\nu,0}$.

Changing the relative charge magnitudes is a different deformation.
It changes the stationary transverse point and moves the complete
background trajectory in target space even when the right asymptotic
data are held fixed.
I determine this response by differentiating the background equations,
the physical fluctuation operator, and the effective source map with
respect to the charge data.
The resulting sensitivity equations contain both the explicit charge
dependence of the fixed-charge potential and the induced displacement
of the saddle.
Appendix~\ref{sec:charge-phase} formulates these equations using
independent symmetric charge bilinears and imposes the physical locus
$t_{ij}=\nu_i\nu_j$ after differentiation.

The charge dependence therefore separates into two structures.
Uniform rescaling is fixed algebraically by the homogeneity of the
product solution and gives the inverse-square law above.
Redistribution among charge components changes the target-space
trajectory and is determined by the charge-sensitivity problem at fixed
outer data.
This distinction supplies the sector-by-sector finite-radius input for
the spectral analysis below.

\subsection{Endpoint spectral response}

Having determined the charge dependence of the geometric coefficient, I now
fix one charge sector and average its finite-radius response over the BPS
endpoint spectrum.
Let the reduced physical boundary matrix have the expansion
\begin{equation}
 M_{\bm\nu}(\delta)
 =
 M_{0,\bm\nu}
 +
 \delta D_{1,\bm\nu}
 +
 O(\delta^2),
 \nonumber
\end{equation}
where $M_{0,\bm\nu}$ is the positive BPS endpoint matrix and
$D_{1,\bm\nu}$ is the first finite-radius deformation obtained from the
gravitational boundary problem.
Differentiating
$M_{\bm\nu}M_{\bm\nu}^{-1}=1$ gives
\begin{equation}
 \left.
 \frac{\dd}{\dd\delta}
 M_{\bm\nu}(\delta)^{-1}
 \right|_{\delta=0}
 =
 -
 M_{0,\bm\nu}^{-1}
 D_{1,\bm\nu}
 M_{0,\bm\nu}^{-1}.
 \nonumber
\end{equation}
A soft endpoint eigenvalue therefore enters the linear inverse response
through two inverse factors.
For a physical source vector, the contribution of each endpoint mode is
fixed by its eigenvalue together with the matrix elements of
$D_{1,\bm\nu}$ and the source overlaps in the same eigenbasis.

The Paper-III ensemble supplies the statistics of $M_{0,\bm\nu}$.
The observable requires more than its marginal eigenvalue density:
it depends on the joint orientation of the endpoint matrix, the
finite-radius deformation, and the effective source vectors.
I therefore fix the physical reduction and the deformation before taking
the endpoint average.
Appendix~\ref{sec:statistics} formulates the corresponding joint matrix
averages and retains the correlations between $M_{0,\bm\nu}$ and
$D_{1,\bm\nu}$.

For the explicit endpoint law, I represent the Paper-III square-root
operator by a central Gaussian matrix
\begin{equation}
 \mathcal Q=(X\;Y),
 \qquad
 X\in\mathbb F^{(N+\kappa)\times m},
 \qquad
 Y\in\mathbb F^{(N+\kappa)\times(N-m)} .
 \nonumber
\end{equation}
The $m$ columns of $X$ span the retained boundary block, while the
$N-m$ columns of $Y$ are eliminated.
Their Schur complement gives
\begin{equation}
 M_{0,\bm\nu}
 =
 X^\dagger X
 -
 X^\dagger Y
 (Y^\dagger Y)^{-1}
 Y^\dagger X
 =
 X^\dagger P_Y^\perp X ,
 \qquad
 P_Y^\perp
 =
 1-Y(Y^\dagger Y)^{-1}Y^\dagger .
 \nonumber
\end{equation}
For the central Gaussian ensemble, $X$ and $Y$ are independent and the
law of $X$ is invariant under row rotations.
Moreover,
\begin{equation}
 \operatorname{rank}P_Y^\perp
 =
 (N+\kappa)-(N-m)
 =
 m+\kappa .
 \label{4-1}
\end{equation}
A row rotation therefore identifies the reduced matrix with
\begin{equation}
 M_{0,\bm\nu}
 \stackrel{d}{=}
 Z^\dagger Z,
 \qquad
 Z\in\mathbb F^{(m+\kappa)\times m},
 \nonumber
\end{equation}
where $\stackrel{d}{=}$ denotes identical probability law and $Z$ is again central Gaussian.
The boundary elimination preserves the rectangularity $\kappa$ and
therefore preserves the hard-edge index of the Paper-III ensemble.
Appendix~\ref{sec:charge-phase} derives this reduced Wishart law and its
eigenvalue density.

The determinant weight and the source response probe different functions
of this spectrum.
The bosonic determinant contains inverse square-root powers of the
endpoint eigenvalues, while the source covariance contains inverse
powers selected by the source overlaps.
Their ensemble averages are therefore controlled by different hard-edge
negative moments.
Appendix~\ref{sec:charge-phase} gives the corresponding integrability
conditions.

Two soft mechanisms enter the calculation at different stages.
The homogeneous transverse mode of Section~3 develops a deterministic
eigenvalue proportional to $\delta$ as the finite-radius saddle approaches
the BPS endpoint.
The Paper-III ensemble instead contains random small positive eigenvalues
of $M_{0,\bm\nu}$ already at $\delta=0$.
Exact radial matching determines the first mechanism.
The BPS hard-edge measure determines the second.
After the physical matrix block, $D_{1,\bm\nu}$, and the source vectors
have been fixed, their combination gives the endpoint-averaged
finite-radius inverse response.

\section{Outlook}

This paper establishes the local finite-radius expansion of the
two-boundary response about the BPS endpoint.
Exact radial matching determines the leading geometric response and its
first correction, while the BPS endpoint ensemble determines how the
same response is weighted by the hard-edge spectrum.
The result places the finite-radius deformation, the fixed-charge
coefficient, and the endpoint spectral measure within one calculational
sequence.

The immediate extension is to carry this expansion beyond the product
target.
For a general target-space metric, the stationary trajectory, its
transverse Hessian, and the source map evolve together along the radial
flow.
The charge-sensitivity equations developed here provide the natural
starting point for this calculation.
Higher radial orders will also generate further resonant channels beyond
the first $\ell=3$ logarithm.
Together, these extensions determine how much of the near-BPS structure
is universal and how much records the detailed target-space geometry.

A more structural question emerges once the finite-radius coefficient is
known over many charge sectors.
Each sector then carries two kinds of data.
The reduced coefficient $W_{\bm\nu}[b]$ determines how that sector enters
the axion-angle representation, while the endpoint measure determines
the distribution of physical fluctuation eigenvalues inside the same
sector.
Paper-II and Paper-III treated these structures separately.
The present paper supplies the finite-radius response that places them
on the same family of saddles.
This makes their joint dependence on charge a well-defined observable.

The resulting problem is naturally expressed in two complementary
representations.
In the charge representation, one follows the sector dependence of the
response and its spectral data directly.
In the conjugate axion-angle representation, the same information is
reorganized by the Fourier transform over the charge lattice.
The relevant observables retain the correlation between the
sector-resolved spectral data and their Fourier transform to the
axion-angle representation.
Such observables probe more than the marginal charge distribution or the
marginal endpoint spectrum: they probe how spectral information is
organized across charge sectors.

This distinction becomes increasingly sharp for several axions.
Charge now ranges over a lattice, the axion angle lies on the dual
torus, and the endpoint spectral measure can vary with both the
magnitude and direction of the charge vector.
Directional large-charge behavior, hard-edge data, and finite-radius
response coefficients can therefore be assembled into joint
charge--spectral observables.
Their transformation to the dual angular variables provides a natural
way to test which correlations remain visible after reduction and which
require access to the full sector-resolved data.

The endpoint ensemble also provides a microscopic route to this
question.
In a specified Type-IIB compactification, the ADHM description fixes the
hard-edge index, symmetry class, microscopic scale, and charge
dictionary.
Combining these data with the finite-radius deformation derived here
produces a microscopic joint measure for the sector label, endpoint
spectrum, and physical response.
This construction would connect the macroscopic throat geometry to a
sector-resolved microscopic description without collapsing the charge
sum and the hard-edge statistics into a single distribution.

The sequence developed in Papers I--III and completed here therefore
opens a new level of the problem.
The BPS endpoint fixes the physical spectral edge.
Finite radius determines how the wormhole probes that edge.
The charge reduction determines how each sector contributes to the
partition function.
The next stage is to determine the joint information carried by these
structures and to characterize how that information changes when one
passes between the charge, axion-angle, and spectral representations.

\section*{Acknowledgment}
I thank Jonas Haferkamp, Chinmay Nirkhe and Mark Zhandry for informative clarification on forrelation. I acknowledge that this research was in part undertaken at and financially supported by the Institute for Pure \& Applied Mathematics (IPAM, USA), the Simons Institute for the Theory of Computing (SIfTC, USA), and the Fields Institute for Research in Mathematical Sciences (FIRMS, Canada). This work was also supported by the National Research Foundation of Korea (NRF) (RS-2021-NR060112). 

\section*{AI Statement}
I used ChatGPT 5.6 to draw Figure 1 and to check selected
high-precision numerical calculations in Appendix C.5. 
\appendix
\section{Boundary data and the physical quadratic problem}
\label{sec:near-bps-problem}

\subsection{Notation, domains, and regulators}

I fix the charge sector $\bm\nu$, the saddle branch $\mathfrak s$, the
integration contour, and the source, boundary-state, measure, and
reduction data collected in $b$.
Throughout the appendices,
\begin{equation}
\delta=a_*^2 .
\end{equation}
The labels $B$ and $F$ denote bosonic and fermionic sectors, respectively.
The integer $\ell$ labels the $S^3$ harmonic.
In the radial calculation, $m$ labels the harmonic degeneracy; in the
later random-matrix calculation, the same letter denotes a matrix
dimension.
These two uses are independent.
The indices $i,j$ label external sources.

I regulate the outer radius, the angular sum, and the radial problem
independently.
At fixed regulators, all Gaussian integrations and Schur complements are
finite dimensional.
I perform the algebraic eliminations at this regulated level and take
the continuum limits afterward.
Removing the radial regulator at fixed $\ell$ restores the infinite
radial spectrum, whereas the matching-surface block for each fixed
$(\ell,m)$ remains finite dimensional.

The quadratic objects below depend on
\begin{equation}
(\bm\nu,\mathfrak s,\delta,b,\phi_L,\phi_R).
\end{equation}
I display the charge label $\bm\nu$ and suppress the fixed branch
$\mathfrak s$, the fixed data $b$, the asymptotic values selected by the
saddle family, and repeated arguments in $\delta$ whenever no ambiguity
arises.
Superscripts $\Sigma$ and $\partial$ distinguish quantities defined on
the artificial matching surface from quantities defined on the physical
outer boundaries.

\begin{center}
\begin{longtable}{p{0.31\textwidth}p{0.62\textwidth}}
\toprule
Symbol & Definition \\
\midrule
$Z_{\bm\nu}[J;\delta]$
&
Full fixed-saddle source functional, including the classical weight,
fluctuation measure, and prescribed zero-mode insertions.
\\

$\cH_{\bm\nu,B}(\delta)$
&
Constraint-reduced bulk bosonic Hessian on the physical boundary domain.
\\

$\cN^{\rm out}_{\bm\nu,L/R,\ell}$
&
One-sided value-to-momentum map from data at the matching surface to the
outward canonical momentum on the left or right radial region.
\\

$\cK^\Sigma_{\bm\nu,B,\ell}$
&
Quadratic Hessian for the common matching-surface value on $\Sigma$ in
one harmonic sector.
\\

$\cK^\partial_{\bm\nu,B}$
&
Physical outer-boundary Hessian obtained after the radial interiors and
the artificial matching data have been eliminated.
\\

$M_{\bm\nu,B}$
&
Further Schur complement obtained after eliminating complementary
bosonic variables from the physical quadratic problem.
\\

$\mathbf b_{\bm\nu,i},\ \mathbf b^{\rm eff}_{\bm\nu,i}$
&
Bulk source vector and the corresponding effective source vector after
complementary bosonic variables have been eliminated.
\\

$\mathbf s_{\bm\nu,\ell m;i}$
&
Source vector induced on $\Sigma$ after the left and right radial
interiors have been eliminated.
\\

$\cA,\ \cA_{\bm\nu,Q},\ \cA_{\bm\nu,D}$
&
Overall reduced-action normalization and the source-independent
fluctuation factors generated by the two elimination schemes defined
below.
\\
\bottomrule
\end{longtable}
\end{center}
The distinction between the interface and the physical boundary is
essential.
In the homogeneous transverse example,
\begin{equation}
\cK^\partial_{\bm\nu,B}
=
\cA K,
\qquad
G=\cA^{-1}K^{-1},
\qquad
\cA>0 .
\end{equation}
The matrix $K$ therefore acts on the physical asymptotic values
$(q_L,q_R)$.
By contrast, $\cK^\Sigma_{\bm\nu,B,\ell}$ acts on the common field value
at the artificial radial cut.
Moving the cut changes $\cK^\Sigma$ while leaving the reconstructed
outer-boundary functional unchanged.

Bosonic Gaussian integrals use real fluctuation coordinates and real
harmonics.
A transpose denotes the real bilinear pairing used in these quadratic
forms.
A dagger denotes the Hermitian pairing used in the spectral analysis.
For analytic continuation, I continue the bilinear action along the
chosen integration contour.
On a real self-adjoint sector, the contour fixes the phase multiplying
$|\det H|^{-1/2}$.
On a complex sector, I choose and retain a definite branch of
$(\det H)^{-1/2}$ throughout the calculation.

\subsection{Boundary conditions and source coefficients}

I parameterize the non-BPS saddle family by
\begin{equation}
 \delta=a_*^2,
 \qquad
 E=12\delta^2 .
 \label{6-1}
\end{equation}
Let $\phi_L^A$ and $\phi_R^A$ denote the asymptotic dilaton/saxion
values.
Along the family I hold
$\bm\nu$, $\phi_R^A$, the reduction data $b$, the saddle branch
$\mathfrak s$, and the integration contour fixed.
For every $\delta>0$, smooth continuation through the minimal
three-sphere determines the left asymptotic value
$\phi_{L,\bm\nu,\mathfrak s}^A(\delta)$.
The finite-radius derivative therefore follows a definite curve in the
space of saddles.

For an on-shell quantity
$\mathcal O_{\bm\nu,\mathfrak s}$, I define this derivative by
\begin{equation}
 D_\delta\mathcal O_{\bm\nu,\mathfrak s}
 \equiv
 \frac{\dd}{\dd\delta}
 \mathcal O_{\bm\nu,\mathfrak s}
 \!\left[
 b,\phi_{L,\bm\nu,\mathfrak s}(\delta),\phi_R;\delta
 \right].
 \label{6-2}
\end{equation}
Thus $D_\delta$ includes the induced motion of the left asymptotic data.
Source derivatives have a different meaning: they act at fixed
$\delta$, fixed asymptotic values, and fixed saddle.
The boundary measure belongs to $b$ and is held fixed under both
operations.
Changing that measure defines a different observable.

For one selected saddle branch, write the reduced scalar sector
coefficient as
\begin{equation}
 W_{\bm\nu,\mathfrak s}[b;\delta]
 =
 \e^{-S^{\rm cl}_{\bm\nu,\mathfrak s}(\delta)}
 \Pi_{\bm\nu,\mathfrak s}[b;\delta] .
 \label{6-3}
\end{equation}
The classical exponential and the fluctuation/source prefactor are
therefore separated before the charge-sector sum is formed.
Assume that the classical action has a finite right derivative at the
BPS endpoint.
Its near-endpoint expansion is
\begin{equation}
 S^{\rm cl}_{\bm\nu,\mathfrak s}(\delta)
 =
 S^{\rm BPS}_{\bm\nu,\mathfrak s}
 +
 \widehat s_{1,\bm\nu,\mathfrak s}\,\delta
 +
 o(\delta),
 \label{6-4}
\end{equation}
with
\begin{equation}
 \widehat s_{1,\bm\nu,\mathfrak s}
 =
 \left.
 D_\delta S^{\rm cl}_{\bm\nu,\mathfrak s}
 \right|_{0^+}.
 \label{6-5}
\end{equation}
Because $\sqrt E=\sqrt{12}\,\delta$, the same expansion may be written
as
\begin{equation}
 s_{1,\bm\nu,\mathfrak s}
 =
 \frac{\widehat s_{1,\bm\nu,\mathfrak s}}{\sqrt{12}},
 \qquad
 S^{\rm cl}_{\bm\nu,\mathfrak s}(\delta)
 =
 S^{\rm BPS}_{\bm\nu,\mathfrak s}
 +
 s_{1,\bm\nu,\mathfrak s}\sqrt E
 +
 o(\sqrt E).
 \label{6-6}
\end{equation}
The coefficient in Eq.~\eqref{6-5} includes the boundary work generated
by the motion of $\phi_L^A(\delta)$.
I now fix $\mathfrak s$ and suppress the branch label.

The source functional distinguishes the unnormalized coefficient from
the connected response.
I define
\begin{equation}
 C_{\bm\nu}^{ij}(\delta)
 =
 \left.
 \partial_{J_i}\partial_{J_j}
 Z_{\bm\nu}[J;\delta]
 \right|_{J=0},
 \label{6-7}
\end{equation}
and
\begin{equation}
 G_{\bm\nu}^{ij}(\delta)
 =
 \left.
 \partial_{J_i}\partial_{J_j}
 \log Z_{\bm\nu}[J;\delta]
 \right|_{J=0}.
 \label{6-8}
\end{equation}
The corresponding one-source expectation is
\begin{equation}
 g_{\bm\nu}^{i}(\delta)
 =
 \left.
 \partial_{J_i}
 \log Z_{\bm\nu}[J;\delta]
 \right|_{J=0}.
 \label{6-9}
\end{equation}
Direct differentiation gives
\begin{equation}
 C_{\bm\nu}^{ij}(\delta)
 =
 Z_{\bm\nu}[0;\delta]
 \left[
 G_{\bm\nu}^{ij}(\delta)
 +
 g_{\bm\nu}^{i}(\delta)g_{\bm\nu}^{j}(\delta)
 \right].
 \label{6-10}
\end{equation}
For the centered insertions used in the quadratic calculation,
$g_{\bm\nu}^{i}=0$, and hence
\begin{equation}
 C_{\bm\nu}^{ij}(\delta)
 =
 Z_{\bm\nu}[0;\delta]\,
 G_{\bm\nu}^{ij}(\delta).
 \label{6-11}
\end{equation}
The connected response removes all source-independent factors.
The unnormalized coefficient retains the classical weight,
determinants or Pfaffians, exact-mode measures, collective-coordinate
factors, and boundary normalization.
Interactions can generate a nonzero one-source expectation and then
restore the second term in Eq.~\eqref{6-10}.

To isolate the fluctuation and source prefactor, define
\begin{equation}
 \widehat Z_{\bm\nu}[J;\delta]
 =
 \e^{S^{\rm cl}_{\bm\nu}(\delta)}
 Z_{\bm\nu}[J;\delta].
 \label{6-12}
\end{equation}
Its unnormalized two-source coefficient is
\begin{equation}
 \widehat C_{\bm\nu}^{ij}(\delta)
 =
 \left.
 \partial_{J_i}\partial_{J_j}
 \widehat Z_{\bm\nu}[J;\delta]
 \right|_{J=0}
 =
 \e^{S^{\rm cl}_{\bm\nu}(\delta)}
 C_{\bm\nu}^{ij}(\delta).
 \label{6-13}
\end{equation}
For the linear reduction considered here,
\begin{equation}
 \Pi_{\bm\nu}[b;\delta]
 =
 R_b\!\left[
 \widehat C_{\bm\nu}^{ij}(\delta)
 \right].
 \label{6-14}
\end{equation}
Equations~\eqref{6-3} and \eqref{6-14} reproduce the Paper-II order of
operations: the fixed-saddle source functional first supplies an
unnormalized two-end coefficient, the reduction $R_b$ converts that
coefficient into the scalar prefactor $\Pi_{\bm\nu}$, and the classical
exponential restores the full sector coefficient
$W_{\bm\nu}[b;\delta]$.
The reduction acts on $\widehat C_{\bm\nu}^{ij}$ itself, rather than on
a derivative of $\log\widehat Z_{\bm\nu}$.

The endpoint relation used in the main text now follows directly.
For a centered source component, take
\begin{equation}
 G_{\bm\nu}^{ij}(0)=0,
 \qquad
 G_{\bm\nu}^{ij}(\delta)
 =
 g_{1,\bm\nu}^{ij}\delta+o(\delta),
 \qquad
 Z_{\bm\nu}[0;\delta]
 \longrightarrow
 Z_{\bm\nu,0}.
 \label{6-15}
\end{equation}
Equation~\eqref{6-11} then gives
\begin{equation}
 C_{\bm\nu}^{ij}(\delta)
 =
 Z_{\bm\nu,0}g_{1,\bm\nu}^{ij}\delta+o(\delta),
 \qquad
 \left.
 \frac{\dd C_{\bm\nu}^{ij}}{\dd\delta}
 \right|_{0^+}
 =
 Z_{\bm\nu,0}g_{1,\bm\nu}^{ij}.
 \label{6-16}
\end{equation}
Thus the endpoint slope factorizes into the normalized geometric
response and the complete BPS sector weight.
The existence of the finite limit $Z_{\bm\nu,0}$ is a statement about
the full fixed-saddle measure, including exact modes, prescribed
insertions, and all remaining fluctuation factors.
The normalized response alone does not determine this weight.

\subsection{Constraint reduction and Gaussian integration}
\label{sec:quadratic-reduction}

The fluctuation domain is fixed by the constraints, gauge choice,
integration contour, outer boundary state, and regulator.
After constraint reduction, the centered fixed-saddle source functional
takes the form
\begin{align}
 Z_{\bm\nu}[J;\delta]
 ={}&
 \e^{-S^{\rm cl}_{\bm\nu}(\delta)}
 \cJ^{\rm exact}_{\bm\nu}(\delta)
 \int_{\Gamma_{\bm\nu,\delta}}
 \dd\eta\,\dd\Psi\,\dd\bar c_{\rm gh}\,\dd c_{\rm gh}
 \nonumber\\
 &\times
 \exp\!\left[
 -\frac12\eta^T\cH_{\bm\nu,B}\eta
 -\frac12\Psi^T\cD_{\bm\nu,F}\Psi
 -\bar c_{\rm gh}^T\cM_{\bm\nu,\rm gh}c_{\rm gh}
 +J_i\mathbf b_{\bm\nu,i}^T\eta
 +\frac12J_i\cV_{\bm\nu}^{ij}J_j
 -S_{\rm int}
 \right].
 \label{6-17}
\end{align}
All quadratic operators and source vectors in
Eq.~\eqref{6-17} depend on $\delta$.
The vectors $\mathbf b_{\bm\nu,i}$ describe linear source insertions,
while $\cV_{\bm\nu}^{ij}$ contains direct quadratic source terms.
The antisymmetric Grassmann form defines $\cD_{\bm\nu,F}$; an equivalent
complex Dirac representation gives the corresponding determinant.
First-order fermions carry their own boundary domain.

The factor $\cJ^{\rm exact}_{\bm\nu}$ contains the collective-coordinate
measure and the prescribed insertions that saturate exact fermion zero
modes.
I take it to be source independent in the formulas below.
A source-dependent saturation insertion must instead be included before
the source derivatives are taken.
An unsaturated fermion zero mode makes the corresponding saddle
coefficient vanish.

I now state explicitly how the physical bosonic Hessian entering
Eq.~\eqref{6-17} is obtained.
After eliminating the first-class constraints and linear multipliers,
let $u_{\rm aux}$ denote the remaining quadratic auxiliary variables.
For one angular sector, write
\begin{equation}
 I^{(2)}_{\ell}
 =
 \frac12
 \begin{pmatrix}
 \eta\\
 u_{\rm aux}
 \end{pmatrix}^{\!T}
 \begin{pmatrix}
 H_{dd}^{(\ell)} & H_{da}^{(\ell)}\\
 H_{ad}^{(\ell)} & H_{aa}^{(\ell)}
 \end{pmatrix}
 \begin{pmatrix}
 \eta\\
 u_{\rm aux}
 \end{pmatrix}
 +I^{(2)}_{\ell,\partial}.
 \label{6-18}
\end{equation}
When $H_{aa}^{(\ell)}$ is invertible, stationarity in
$u_{\rm aux}$ gives
\[
 u_{\rm aux}
 =
 -\bigl(H_{aa}^{(\ell)}\bigr)^{-1}
 H_{ad}^{(\ell)}\eta ,
\]
and hence
\begin{equation}
 \cH^{\rm phys}_{\bm\nu,B,\ell}
 =
 H_{dd}^{(\ell)}
 -
 H_{da}^{(\ell)}
 \bigl(H_{aa}^{(\ell)}\bigr)^{-1}
 H_{ad}^{(\ell)} .
 \label{6-19}
\end{equation}
Thus the physical quadratic operator is formed only after the
Hamiltonian and momentum constraints have been imposed.
The boundary quadratic form remains part of its domain.

At the BPS endpoint, this physical domain reduces to the endpoint domain
of Paper-I, and
\begin{equation}
 \cH_{\bm\nu,B}(0)
 =
 \cQ_{\bm\nu}^{\dagger}\cQ_{\bm\nu}.
 \label{6-20}
\end{equation}
Equation~\eqref{6-20} fixes the BPS endpoint operator.
The finite-radius Hessian is obtained by deforming the complete
constraint-reduced boundary-value problem.
The endpoint square also fixes no probability law for individual matrix
entries; the Paper-III ensemble is an additional microscopic input.

I next separate the finite bosonic block that carries the response from
its Gaussian complement.
Exact collective modes have already been extracted.
Choose a set of retained bosonic coordinates that is fixed for all
$\delta$, and let $P_B$ project onto this subspace.
Set $Q_B=1-P_B$ and define
\begin{equation}
 \cH_{\bm\nu,B,Q}(\delta)
 =
 Q_B\cH_{\bm\nu,B}(\delta)Q_B .
 \label{6-21}
\end{equation}
On a patch where $\cH_{\bm\nu,B,Q}$ is invertible, eliminating the
$Q_B$ variables gives
\begin{equation}
 M_{\bm\nu,B}(\delta)
 =
 P_B\cH_{\bm\nu,B}P_B
 -
 P_B\cH_{\bm\nu,B}Q_B
 \cH_{\bm\nu,B,Q}^{-1}
 Q_B\cH_{\bm\nu,B}P_B .
 \label{6-22}
\end{equation}
All blocks and source vectors are expressed in the same retained
coordinates as $\delta$ varies.
A change of basis within the retained subspace acts on
$M_{\bm\nu,B}$ by congruence and contributes the corresponding measure
Jacobian.

The endpoint scaling of a soft block follows directly from this Schur
complement.
Let $P_Z$ project onto an exact kernel of the endpoint operator that
remains in the retained problem, and set $Q_Z=1-P_Z$.
Self-adjointness gives
\[
 Q_Z\cH_{B,0}P_Z=0 .
\]
For
\[
 \cH_B(\delta)
 =
 \cH_{B,0}
 +
 \delta\cH_{B,1}
 +
 O(\delta^2)
\]
and a bounded inverse on the complementary subspace,
\begin{equation}
 M_Z(\delta)
 =
 \delta\,P_Z\cH_{B,1}P_Z
 +
 O(\delta^2).
 \label{6-23}
\end{equation}
The mixed blocks begin at $O(\delta)$, so their Schur contribution
starts at $O(\delta^2)$.
For differentiable input the remainder may be written as $o(\delta)$.
A contribution proportional to
$\delta^2\log(\delta/\delta_0)$ carries the same logarithm through the
finite-dimensional elimination.
At fixed regulator, the complementary endpoint spectrum has a finite
gap.
The continuum limit requires uniform control of the complementary
resolvent.

The fermionic reduction proceeds in parallel.
For a finite antisymmetric fermion matrix, choose a compatible
even-dimensional decomposition
$P_F+Q_F=1$ and a fixed Berezin orientation.
Eliminating the $Q_F$ variables gives
\begin{equation}
 M_{\bm\nu,F}
 =
 P_F\cD_{\bm\nu,F}P_F
 -
 P_F\cD_{\bm\nu,F}Q_F
 \bigl(Q_F\cD_{\bm\nu,F}Q_F\bigr)^{-1}
 Q_F\cD_{\bm\nu,F}P_F .
 \label{6-24}
\end{equation}
In a real Grassmann basis, $M_{\bm\nu,F}$ remains antisymmetric.
Appendix~\ref{app:block-identities} gives the corresponding determinant,
inverse, and Pfaffian identities.
On compatible reduced spaces,
\begin{align}
 \det{}'\cH_{\bm\nu,B}
 &=
 \det\cH_{\bm\nu,B,Q}\,
 \det M_{\bm\nu,B},
 \label{6-25}\\
 \Pf{}'\cD_{\bm\nu,F}
 &=
 \Pf\!\bigl(Q_F\cD_{\bm\nu,F}Q_F\bigr)\,
 \Pf M_{\bm\nu,F}.
 \label{6-26}
\end{align}
The primes refer only to exact modes already extracted together with
their measures.
Kernel removal in a continuum gluing formula must be performed
compatibly across the factors.

Gaussian elimination also transforms the sources.
Define
\begin{equation}
 \mathbf b_{\bm\nu,i,P}
 =
 P_B\mathbf b_{\bm\nu,i},
 \qquad
 \mathbf b_{\bm\nu,i,Q}
 =
 Q_B\mathbf b_{\bm\nu,i}.
 \label{6-27}
\end{equation}
The complementary response and the effective retained source are
\begin{align}
 G^{ij}_{\bm\nu,Q}
 &=
 \mathbf b_{\bm\nu,i,Q}^{\dagger}
 \cH_{\bm\nu,B,Q}^{-1}
 \mathbf b_{\bm\nu,j,Q},
 \label{6-28}\\
 \mathbf b^{\rm eff}_{\bm\nu,i}
 &=
 \mathbf b_{\bm\nu,i,P}
 -
 P_B\cH_{\bm\nu,B}Q_B
 \cH_{\bm\nu,B,Q}^{-1}
 \mathbf b_{\bm\nu,i,Q}.
 \label{6-29}
\end{align}
Completing the square produces both quantities in the same elimination
step.
The block inverse then gives the exact decomposition
\begin{equation}
 \mathbf b_{\bm\nu,i}^{\dagger}
 \cH_{\bm\nu,B}^{-1}
 \mathbf b_{\bm\nu,j}
 =
 G^{ij}_{\bm\nu,Q}
 +
 \bigl(\mathbf b^{\rm eff}_{\bm\nu,i}\bigr)^{\dagger}
 M_{\bm\nu,B}^{-1}
 \mathbf b^{\rm eff}_{\bm\nu,j}.
 \label{6-30}
\end{equation}
The complementary determinant, the complementary source response, and
the effective source vector are therefore parts of one Gaussian
reduction and enter the coefficient together.

Collect all remaining source-independent factors into
$\cZ_{\bm\nu,\rm rem}$.
These include the ghost sector, self-dual sectors, exact-mode measure,
prescribed saturation insertions, and any additional factors outside
the displayed bosonic and fermionic blocks.
At finite regulator define
\begin{equation}
 \cA_{\bm\nu,Q}
 =
 \cZ_{\bm\nu,\rm rem}
 \frac{
 \Pf\!\bigl(Q_F\cD_{\bm\nu,F}Q_F\bigr)
 }{
 \sqrt{|\det\cH_{\bm\nu,B,Q}|}
 } .
 \label{6-31}
\end{equation}
On a real self-adjoint patch, the contour phase is included in
$\cZ_{\bm\nu,\rm rem}$.
For a complex saddle, the same expression is understood with the
determinant branch fixed by the chosen contour.

The connected quadratic source response is
\begin{equation}
 \cR_{\bm\nu}^{ij}
 =
 \cV_{\bm\nu}^{ij}
 +
 G^{ij}_{\bm\nu,Q}
 +
 \bigl(\mathbf b^{\rm eff}_{\bm\nu,i}\bigr)^{\dagger}
 M_{\bm\nu,B}^{-1}
 \mathbf b^{\rm eff}_{\bm\nu,j}.
 \label{A-response}
\end{equation}
The three terms have distinct origins: a direct quadratic source term,
a response generated inside the eliminated complement, and propagation
through the retained physical block.

After removing the classical exponential as in Eq.~\eqref{6-12}, the
complete quadratic source functional becomes
\begin{equation}
 \widehat Z^{(2)}_{\bm\nu}[J;\delta]
 =
 \cA_{\bm\nu,Q}
 \frac{\Pf M_{\bm\nu,F}}
 {\sqrt{|\det M_{\bm\nu,B}|}}
 \exp\!\left(
 \frac12J_i\cR_{\bm\nu}^{ij}J_j
 \right).
 \label{A-full-gaussian}
\end{equation}
Two source derivatives give the corresponding unnormalized coefficient,
\begin{equation}
 \widehat C_{\bm\nu}^{ij}
 =
 \cA_{\bm\nu,Q}
 \frac{\Pf M_{\bm\nu,F}}
 {\sqrt{|\det M_{\bm\nu,B}|}}
 \cR_{\bm\nu}^{ij}
 +
 \widehat C_{\bm\nu,\rm higher}^{ij}.
 \label{6-32}
\end{equation}
Thus $\cR_{\bm\nu}^{ij}$ is the connected Gaussian response, while the
prefactor supplies its source-independent quadratic weight.
The term
$\widehat C_{\bm\nu,\rm higher}^{ij}$
contains interaction corrections and source-dependent contributions
beyond the quadratic functional.

For a fixed saddle, the finite-radius coefficient is therefore specified
by the data
\begin{equation}
 \left(
 M_{\bm\nu,B},
 M_{\bm\nu,F},
 \mathbf b^{\rm eff}_{\bm\nu,i},
 G^{ij}_{\bm\nu,Q},
 \cA_{\bm\nu,Q},
 \cV_{\bm\nu}^{ij},
 R_b
 \right).
 \label{6-33}
\end{equation}
The classical saddle fixes the exponential weight.
Constraint reduction and radial matching fix the physical quadratic
matrices and source response.
The complete field measure fixes the normalization.
The reduction $R_b$ then converts the resulting unnormalized source
coefficient into the scalar fixed-charge coefficient used in the main
text.

\section{Finite-radius expansion and boundary Hessians}
\label{sec:finite-radius-boundary}
\subsection{Local coefficients, matching, and finite elimination}

Fix the saddle branch, integration contour, and boundary family defined
in Appendix~\ref{sec:near-bps-problem}.
I express the fluctuations, source vectors, and boundary data in one
common coordinate system for all $\delta$.
At fixed outer radius, the local quadratic data admit the expansions
\begin{align}
 \cH^{\rm loc}_{\bm\nu,B}(\delta)
 &=
 \cH^{(0)}_{\bm\nu,B}
 +\delta\cH^{(1)}_{\bm\nu,B}
 +\delta^2\cH^{(2)}_{\bm\nu,B}
 +\cdots,
 \label{7-1}\\
 \cD^{\rm loc}_{\bm\nu,F}(\delta)
 &=
 \cD^{(0)}_{\bm\nu,F}
 +\delta\cD^{(1)}_{\bm\nu,F}
 +\delta^2\cD^{(2)}_{\bm\nu,F}
 +\cdots,
 \label{7-2}\\
 \mathbf b^{\rm loc}_{\bm\nu,i}(\delta)
 &=
 \mathbf b^{(0)}_{\bm\nu,i}
 +\delta\mathbf b^{(1)}_{\bm\nu,i}
 +\delta^2\mathbf b^{(2)}_{\bm\nu,i}
 +\cdots .
 \label{7-3}
\end{align}
Individual coefficients may vanish in a specified background.
These outer expansions cease to be uniform near the shrinking minimal
three-sphere.
The physical boundary matrices are therefore obtained by matching the
outer expansion to the corresponding inner problem.
The radial analysis of
Appendix~\ref{sec:homogeneous-calculation} supplies both the ordinary
power corrections and the resonant logarithms.

The finite-dimensional Gaussian eliminations preserve this asymptotic
structure.

\paragraph{Finite elimination and logarithmic expansions.}
Let
\begin{equation}
 \mathsf M(\delta)
 =
 \begin{pmatrix}
  \mathsf A(\delta) & \mathsf B(\delta)\\
  \mathsf C(\delta) & \mathsf D(\delta)
 \end{pmatrix}
\end{equation}
be a finite matrix whose blocks admit asymptotic expansions in integer
powers of $\delta$ and $\log(\delta/\delta_0)$ with finite limits as
$\delta\to0^+$.
Suppose
\begin{equation}
 \mathsf D(\delta)=\mathsf D_0+o(1),
 \qquad
 \det\mathsf D_0\neq0 .
\end{equation}
Then the Schur complement
\begin{equation}
 \mathsf A-\mathsf B\mathsf D^{-1}\mathsf C
\end{equation}
admits an asymptotic expansion in the same class.
If all four blocks are analytic in $\delta$, the Schur complement is
analytic in $\delta$.

To see this, write
\begin{equation}
 \mathsf D=\mathsf D_0+\mathsf E,
 \qquad
 \mathsf E=o(1).
\end{equation}
For sufficiently small $\delta$,
$\|\mathsf D_0^{-1}\mathsf E\|<1$, and hence
\begin{equation}
 \mathsf D^{-1}
 =
 \sum_{n\geq0}
 \bigl(-\mathsf D_0^{-1}\mathsf E\bigr)^n
 \mathsf D_0^{-1}.
\end{equation}
Expanding this series to any fixed asymptotic order and substituting it
into
$\mathsf A-\mathsf B\mathsf D^{-1}\mathsf C$
preserves the powers of $\delta$ and the logarithms carried by the
input blocks.
The finite-limit hypothesis excludes an unsuppressed
$\delta^0\log^m\delta$ divergence in the inverted block.
For analytic input, matrix inversion is analytic in a neighborhood of
the invertible matrix $\mathsf D_0$, and the Schur complement is
analytic there.

This finite-dimensional argument applies directly at fixed regulator.
Its continuum counterpart additionally requires a fixed operator domain
and uniform control of the inverse on the eliminated subspace.

For an analytic reduced bosonic block, I write
\begin{align}
 M_{\bm\nu,B}(\delta)
 &=
 M_{0,\bm\nu}
 +\delta D_{1,\bm\nu}
 +\delta^2D_{2,\bm\nu}
 +\cdots,
 \label{7-4}\\
 \mathbf b^{\rm eff}_{\bm\nu,i}(\delta)
 &=
 \mathbf b^{\rm eff}_{0,\bm\nu,i}
 +\delta\mathbf b^{\rm eff}_{1,\bm\nu,i}
 +\cdots,
 \label{7-5}\\
 \cV^{ij}_{\bm\nu}(\delta)
 &=
 \cV^{ij}_{0,\bm\nu}
 +\delta\cV^{ij}_{1,\bm\nu}
 +\cdots,
 \label{7-6}\\
 G^{ij}_{\bm\nu,Q}(\delta)
 &=
 G^{ij}_{0,\bm\nu,Q}
 +\delta G^{ij}_{1,\bm\nu,Q}
 +\cdots .
 \label{7-7}
\end{align}
The matrix $D_{1,\bm\nu}$ is the first finite-radius deformation that
enters the endpoint inverse response in the main text.
The angular problem also generates nonanalytic terms.
The first resonant contribution found below occurs in the $\ell=3$
sector and has strength
\begin{equation}
 \eta_{3,\log}(\delta)
 =
 \delta^2\log(\delta/\delta_0),
 \qquad
 \delta_0>0 .
 \label{7-8}
\end{equation}
Its first derivative vanishes at $\delta=0^+$.
Its contribution to an observable is fixed only after the corresponding
boundary map and source vectors have been applied.

Separate the matrix structure from its scalar strength:
\begin{equation}
 M(\delta)
 =
 M_0+\eta(\delta)D+o(\eta),
 \qquad
 \eta(\delta)\longrightarrow0 .
 \label{7-9}
\end{equation}
Let $M_0$ be Hermitian positive definite and let
\[
 g=\lambda_{\min}(M_0)>0 .
\]
Weyl's bound gives
\begin{equation}
 \|M(\delta)-M_0\|_{\op}<g
 \quad\Longrightarrow\quad
 M(\delta)>0 .
 \label{7-10}
\end{equation}
For the affine deformation
$M_0+\eta D$ with $\eta\geq0$, the first zero generated by a negative
direction of $D$ occurs at
\begin{equation}
 \eta_c
 =
 \inf_{\substack{v\neq0\\v^\dagger Dv<0}}
 \frac{v^\dagger M_0v}{-v^\dagger Dv}.
 \label{7-11}
\end{equation}
Each admissible Rayleigh quotient is the value of $\eta$ at which the
quadratic form first vanishes along that direction; their infimum gives
the first crossing.
Higher-order terms in $M(\delta)$ shift this affine estimate.
For the resonant term it is convenient, on $0<\delta<\delta_0$, to use
the positive strength
\[
 \widetilde\eta_{3,\log}(\delta)
 =
 \delta^2\log(\delta_0/\delta)
\]
and absorb the overall minus sign into $D$.

The source-independent nonzero-mode factor varies with the same
quadratic data.
At finite regulator,
\begin{align}
 \partial_\delta\log\cT^{\rm nz}_{\bm\nu}
 ={}&
 \frac12
 \Tr\!\left[
  (\cD_{F,\rm nz})^{-1}
  \partial_\delta\cD_{F,\rm nz}
 \right]
 +
 \Tr\!\left[
  (\cM_{\rm gh,nz})^{-1}
  \partial_\delta\cM_{\rm gh,nz}
 \right]
 \nonumber\\
 &-
 \frac12
 \Tr\!\left[
  (\cH_{B,\rm nz})^{-1}
  \partial_\delta\cH_{B,\rm nz}
 \right]
 +
 \partial_\delta\log\cT^{\rm extra}_{\bm\nu}.
 \label{7-12}
\end{align}
The factor $\cT^{\rm nz}_{\bm\nu}$ includes the associated measure
factors and contour phases.
In the continuum, its determinant traces are defined with the same
renormalization prescription used for the corresponding functional
determinants.

For the finite reduced matrices,
\begin{align}
 \partial_\delta
 \log\frac{\Pf M_F}{\sqrt{\det M_B}}
 &=
 \frac12\Tr\!\left(
 M_F^{-1}\partial_\delta M_F
 \right)
 -
 \frac12\Tr\!\left(
 M_B^{-1}\partial_\delta M_B
 \right),
 \label{7-13}\\
 \partial_\delta M_B^{-1}
 &=
 -M_B^{-1}
 (\partial_\delta M_B)
 M_B^{-1}.
 \label{7-14}
\end{align}
These identities hold on an invertible patch with the Pfaffian and
determinant branches fixed by the chosen contour.
A zero crossing marks the boundary of that patch and requires the
corresponding zero direction to be extracted and treated separately.

Equations~\eqref{7-1}--\eqref{7-14} propagate the local finite-radius
expansion through Gaussian elimination, determinants, and inverse
responses.
The radial boundary-value problem determines the matrices and source
vectors entering these identities. 

\subsection{One radial region and its value-to-momentum map}
I now construct the one-sided
value-to-momentum maps and glue the two radial regions to obtain the
physical outer-boundary Hessian.

Consider one scalar harmonic sector.
I use real harmonics $Y_{\ell m}$ on the unit $S^3$, normalized with
respect to $\dd\Omega_3/\mathcal V_3$, with
\begin{equation}
 -\nabla^2Y_{\ell m}
 =
 \ell(\ell+2)Y_{\ell m}.
 \label{7-15}
\end{equation}
Fix $(\ell,m)$, $\bm\nu$, and $\delta$.
After the angular decomposition, the coupled components in this sector
form a radial vector $y(\tau)$.
The same construction applies block by block to the remaining spin
sectors with their corresponding harmonics and multiplicities.

After constraint reduction, the quadratic action on one radial region is
\begin{align}
 I^{(2)}_{\bm\nu,B,\ell m}[y]
 ={}&
 \frac12\int\dd\tau\,
 \left[
 \dot y^\dagger A_\ell\dot y
 +\dot y^\dagger B_\ell y
 +y^\dagger B_\ell^\dagger\dot y
 +y^\dagger C_\ell y
 \right]
 \nonumber\\
 &+\frac12
 y_\Sigma^\dagger B_{\Sigma,\ell}y_\Sigma .
 \label{7-16}
\end{align}
The matrices $A_\ell$, $B_\ell$, and $C_\ell$ depend on
$(\tau,\bm\nu,\delta)$.
They are local radial coefficients and are unrelated to the block
matrices used in the Schur-complement algebra of the preceding
subsection.
On a real self-adjoint sector,
$A_\ell=A_\ell^\dagger$,
$C_\ell=C_\ell^\dagger$, and $A_\ell$ is invertible.

The canonical radial momentum is
\begin{equation}
 p
 =
 A_\ell\dot y+B_\ell y .
 \label{7-17}
\end{equation}
The Euler--Lagrange equation gives
\[
 \dot p
 =
 B_\ell^\dagger\dot y+C_\ell y .
\]
Eliminating $\dot y$ converts the second-order radial equation into the
first-order Hamiltonian system
\begin{equation}
 \frac{\dd}{\dd\tau}
 \binom{y}{p}
 =
 \begin{pmatrix}
 -A_\ell^{-1}B_\ell
 &
 A_\ell^{-1}
 \\
 C_\ell-B_\ell^\dagger A_\ell^{-1}B_\ell
 &
 B_\ell^\dagger A_\ell^{-1}
 \end{pmatrix}
 \binom{y}{p}.
 \label{7-18}
\end{equation}
Introducing a spectral parameter $z$ replaces
$C_\ell$ by $C_\ell-zW_\ell$, where $W_\ell>0$ is the radial
eigenvalue weight.
The Gaussian response studied in this paper is evaluated at $z=0$.

For either radial region $\alpha=L,R$, let the columns of
$Y_{\alpha,\ell}(\tau,z)$ solve the homogeneous equation and satisfy the
prescribed physical outer-boundary condition.
Define the corresponding momentum matrix by
\begin{equation}
 \Pi_{\alpha,\ell}
 =
 A_\ell\dot Y_{\alpha,\ell}
 +
 B_\ell Y_{\alpha,\ell}.
 \label{7-19}
\end{equation}
On every interval where $Y_{\alpha,\ell}$ is invertible, the
value-to-momentum map is
\begin{equation}
 \cN_{\bm\nu,\alpha,\ell}(\tau,z;\delta)
 =
 \Pi_{\alpha,\ell}(\tau,z)
 Y_{\alpha,\ell}(\tau,z)^{-1}.
 \label{7-20}
\end{equation}
Differentiating Eq.~\eqref{7-20} and using
Eq.~\eqref{7-18} gives the matrix Riccati equation
\begin{equation}
 \dot{\cN}
 =
 C_\ell-zW_\ell
 -
 \bigl(B_\ell^\dagger-\cN\bigr)
 A_\ell^{-1}
 \bigl(B_\ell-\cN\bigr).
 \label{7-21}
\end{equation}
For real $z$ and a self-adjoint boundary domain, $\cN$ is Hermitian.
For complex $z$, Eq.~\eqref{7-21} defines its analytic continuation with
the displayed operator ordering.
In that continuation,
$B_\ell^\dagger-\cN$ is generally distinct from
$(B_\ell-\cN)^\dagger$.
A zero of $\det Y_{\alpha,\ell}$ produces a pole of this Riccati chart.
The underlying first-order system remains regular and continues through
the pole in another boundary coordinate.

The two radial regions carry opposite outward orientations at the
artificial matching surface $\Sigma$.
I therefore convert the canonical momentum on each side to the outward
momentum before gluing and denote the resulting maps by
$\cN^{\rm out}_{\bm\nu,L,\ell}$ and
$\cN^{\rm out}_{\bm\nu,R,\ell}$.
For the common matching value $y_{\Sigma,\ell m}$, the two one-sided
on-shell actions add.
Including the explicit quadratic term localized at $\Sigma$ gives the
interface Hessian
\begin{equation}
 \cK^\Sigma_{\bm\nu,B,\ell}(z;\delta)
 =
 \cN^{\rm out}_{\bm\nu,L,\ell}(z;\delta)
 +
 \cN^{\rm out}_{\bm\nu,R,\ell}(z;\delta)
 +
 \cB_{\bm\nu,\Sigma,\ell}(z;\delta).
 \label{7-22}
\end{equation}
Variation with respect to the common matching value is therefore the
linearized gluing condition.
When both one-sided Dirichlet problems are invertible, a nontrivial
zero mode of the glued system exists precisely when
\begin{equation}
 \det
 \cK^\Sigma_{\bm\nu,B,\ell}(0;\delta)
 =
 0 .
 \label{7-23}
\end{equation}
A kernel of either one-sided Dirichlet problem must first be extracted
with its own measure; after that extraction,
Eq.~\eqref{7-23} applies to the remaining invertible blocks.

The same map follows directly from the one-sided on-shell action.
Holding the physical outer data fixed,
\begin{align}
 \partial_{y_\Sigma}
 I_L^{\rm on}(y_\Sigma)
 &=
 p^{\rm out}_{L,\Sigma},
 \label{7-24}\\
 \partial_{y_\Sigma}^2
 I_L^{\rm on}(y_\Sigma)
 &=
 \cN^{\rm out}_{\bm\nu,L,\ell}(0;\delta)
 \label{7-25}
\end{align}
at quadratic order.
The right region obeys the analogous relations.

A further simplification follows only when the complete left and right
boundary-value problems are related by reflection.
This requires the background fields, fluctuation operator, boundary
conditions, and boundary terms to transform into one another.
Then the two outward maps coincide and
\begin{equation}
 \cK^\Sigma_{\bm\nu,B,\ell}(0;\delta)
 =
 2\,\partial_{y_\Sigma}^2 I_{\rm half}^{\rm on}
 +
 \cB_{\bm\nu,\Sigma,\ell}(0;\delta).
 \label{7-26}
\end{equation}
Reflection symmetry of the metric alone does not imply this relation.
A generic charged scalar trajectory requires the two independent
one-sided maps in Eq.~\eqref{7-22}.

The matrix
$\cK^\Sigma_{\bm\nu,B,\ell}$
is the complete finite-dimensional quadratic form for the matching
value in this harmonic sector.
The next subsection integrates this matching value and combines its
determinant and inverse response with the two Dirichlet interiors.

\subsection{Gaussian gluing for one harmonic}

I now perform the gluing at fixed radial discretization and angular
cutoff.
For each side $\alpha=L,R$, decompose the radial field as
\[
 y_\alpha
 =
 y_\alpha^{\rm cl}[y_\Sigma]+\xi_\alpha ,
 \qquad
 \xi_\alpha|_\Sigma=0 ,
\]
where $y_\alpha^{\rm cl}[y_\Sigma]$ solves the homogeneous radial
equation with matching value $y_\Sigma$ and the prescribed physical
outer-boundary condition.
The fluctuations $\xi_\alpha$ therefore carry the Dirichlet Hessians
$\cH_{\bm\nu,L,D,\ell}$ and
$\cH_{\bm\nu,R,D,\ell}$.
Here $D$ denotes vanishing fluctuation at the artificial matching
surface $\Sigma$, with the physical outer-boundary condition unchanged.

Integrating the two Dirichlet interiors produces three objects.
The direct quadratic source term is
$\cV^{ij}_{\bm\nu,\ell m}$.
Propagation entirely inside the Dirichlet interiors gives
$G^{ij}_{\bm\nu,D,\ell m}$.
The coupling of the sources to the remaining matching value gives the
vector $\mathbf s_{\bm\nu,\ell m;i}$.
The dependence on $y_\Sigma$ is then governed by the interface Hessian
$\cK^\Sigma_{\bm\nu,B,\ell}$ derived in the preceding subsection.

Let $n_\ell$ denote the dimension of the matching vector in this
harmonic block.
The regulated bosonic source functional is
\begin{align}
 Z^{(2)}_{\bm\nu,B,\ell m}[J;\delta]
 ={}&
 \frac{\mathfrak n_{\bm\nu,\ell m}}
 {\sqrt{
 \det\cH_{\bm\nu,L,D,\ell}\,
 \det\cH_{\bm\nu,R,D,\ell}
 }}
 \exp\!\left[
 \frac12J_i
 \left(
 \cV^{ij}_{\bm\nu,\ell m}
 +
 G^{ij}_{\bm\nu,D,\ell m}
 \right)
 J_j
 \right]
 \nonumber\\
 &\times
 \int
 \frac{\dd^{n_\ell}y_\Sigma}{(2\pi)^{n_\ell/2}}
 \exp\!\left[
 -\frac12
 y_\Sigma^T
 \cK^\Sigma_{\bm\nu,B,\ell}(0;\delta)
 y_\Sigma
 +
 J_i\mathbf s_{\bm\nu,\ell m;i}^Ty_\Sigma
 \right]
 \nonumber\\
 ={}&
 \frac{
 \mathfrak n_{\bm\nu,\ell m}
 }{
 \sqrt{
 \det\cH_{\bm\nu,L,D,\ell}\,
 \det\cH_{\bm\nu,R,D,\ell}\,
 \det\cK^\Sigma_{\bm\nu,B,\ell}(0;\delta)
 }}
 \exp\!\left[
 \frac12J_i
 \cR^{ij}_{\bm\nu,\ell m}
 J_j
 \right].
 \label{7-27}
\end{align}
The normalization
$\mathfrak n_{\bm\nu,\ell m}$
contains the Jacobian of the decomposition and the regulated integration
measure.
It equals unity in normalized Cartesian block coordinates.
Equation~\eqref{7-27} is the bosonic harmonic block; the classical
saddle weight, fermions, ghosts, exact modes, and the remaining
source-independent factors enter in the full assembly below.

The connected quadratic response of this harmonic is
\begin{equation}
 \cR^{ij}_{\bm\nu,\ell m}
 =
 \cV^{ij}_{\bm\nu,\ell m}
 +
 G^{ij}_{\bm\nu,D,\ell m}
 +
 \mathbf s_{\bm\nu,\ell m;i}^T
 \bigl[
 \cK^\Sigma_{\bm\nu,B,\ell}(0;\delta)
 \bigr]^{-1}
 \mathbf s_{\bm\nu,\ell m;j}.
 \label{B-harmonic-response}
\end{equation}
The three terms correspond respectively to the direct source coupling,
propagation inside the two Dirichlet interiors, and propagation through
the common matching value.

Equation~\eqref{B-harmonic-response} follows from the shift
\begin{equation}
 y_\Sigma
 \longrightarrow
 y_\Sigma
 +
 \bigl(\cK^\Sigma_{\bm\nu,B,\ell}\bigr)^{-1}
 \sum_i J_i\mathbf s_{\bm\nu,\ell m;i}.
 \label{7-28}
\end{equation}
Completing the square generates simultaneously the interface
contribution to the source response and the factor
$\bigl(\det\cK^\Sigma_{\bm\nu,B,\ell}\bigr)^{-1/2}$.
Thus the determinant gluing and the matching-surface covariance arise
from the same Gaussian integration.
The next subsection assembles these harmonic blocks and converts the
finite determinant identity into the regulated functional determinant
of the complete quadratic problem.

\subsection{Assembly, functional determinants, and the full source coefficient}

I now assemble the harmonic blocks into the regulated quadratic
functional.
At scalar-harmonic cutoff $L_{\max}$, the matching space is the direct
sum over $(\ell,m)$, and its Hessian is
\begin{equation}
 \cK^{\Sigma,[L_{\max}]}_{\bm\nu,B}(\delta)
 =
 \bigoplus_{\ell=0}^{L_{\max}}
 \bigoplus_{m=1}^{(\ell+1)^2}
 \cK^\Sigma_{\bm\nu,B,\ell}(0;\delta).
 \label{B-direct-sum}
\end{equation}
Gauge directions and collective modes are removed on the compatible
reduced space.
The remaining spin sectors contribute analogous direct-sum blocks.
At fixed regulator, determinants multiply across these blocks, inverses
act blockwise, and the matching-surface source vector is the direct sum
of the vectors $\mathbf s_{\bm\nu,\ell m;i}$.

Suppressing the angular cutoff, the assembled connected response is
\begin{equation}
 \cR^{ij}_{\bm\nu}
 =
 \cV^{ij}_{\bm\nu}
 +
 G^{ij}_{\bm\nu,D}
 +
 \mathbf s_{\bm\nu,i}^{\dagger}
 \bigl(\cK^\Sigma_{\bm\nu,B}\bigr)^{-1}
 \mathbf s_{\bm\nu,j},
 \qquad
 G^{ij}_{\bm\nu,D}
 =
 \sum_{\ell,m}
 G^{ij}_{\bm\nu,D,\ell m}.
 \label{B-assembled-response}
\end{equation}
Here $\cV^{ij}_{\bm\nu}$ contains each direct quadratic source term
once.
A physical source may excite several harmonics, and multiplication of
their Gaussian functionals adds their connected quadratic responses.
The harmonic labels $(\ell,m)$ remain independent of the charge label
$\bm\nu$.

The finite-dimensional determinant products acquire a definite
regularized meaning in the continuum.
For a positive elliptic operator $T$ with nonzero discrete spectrum,
let $q_T$ denote its mass dimension and form the dimensionless operator
$T/\mu_{\rm ren}^{q_T}$.
I define
\begin{equation}
 \zeta_T(s;\mu_{\rm ren})
 =
 \Tr{}'
 \left(
 \frac{T}{\mu_{\rm ren}^{q_T}}
 \right)^{-s},
 \qquad
 \log\det{}'_{\zeta}
 \left(
 \frac{T}{\mu_{\rm ren}^{q_T}}
 \right)
 =
 -
 \left.
 \partial_s\zeta_T(s;\mu_{\rm ren})
 \right|_{s=0}.
 \label{B-zeta-definition}
\end{equation}
The trace is defined first in its convergence domain and then continued
to $s=0$.
A pole at $s=0$ is treated by the chosen finite-part prescription.
For a complex elliptic operator, the same construction uses a fixed
spectral cut.
A second-order bulk Hessian and the first-order
Dirichlet-to-Neumann operator on $\Sigma$ carry different mass
dimensions and therefore require their own powers of
$\mu_{\rm ren}$.
The symbol $\zeta$ here denotes the regularization scheme.

At finite outer radius, the bosonic gluing theorem relates the full
operator to the two Dirichlet interiors and the matching-surface
operator.
For an invertible elliptic boundary-value problem,
\begin{equation}
 \det_{\zeta}\cH_{\bm\nu,B}(\delta)
 =
 \e^{\cL_{\bm\nu,\rm loc}(\delta)}
 \det_{\zeta}\cH_{\bm\nu,L,D}(\delta)
 \det_{\zeta}\cH_{\bm\nu,R,D}(\delta)
 \det_{\zeta}\cK^\Sigma_{\bm\nu,B}(\delta).
 \label{7-28}
\end{equation}
Each determinant carries the dimensionless normalization appropriate
to its operator.
The local factor
$\cL_{\bm\nu,\rm loc}$
is fixed by the operator and its boundary conditions and survives the
continuum gluing relation~\cite{BFK,Forman,ReshetikhinVertman}.
In finite Cartesian block algebra this local factor vanishes and the
ordinary Schur-complement determinant identity is recovered.

For a noncompact outer limit, I define the determinants by the same
relative-determinant or subtraction-and-counterterm prescription on all
three factors.
When kernels occur, I first introduce a common spectral shift, apply
Eq.~\eqref{7-28} to the invertible problem, and then extract the
zero-mode powers together with their compatible kernel measures.
This procedure preserves the finite kernel factors generated by
gluing.

The fermionic sector requires its own first-order boundary split.
At the same regulator, elimination of the complementary Grassmann
coordinates produces
\[
 \Pf\cD_{\bm\nu,F,Q_D}\,
 \Pf M^\Sigma_{\bm\nu,F},
\]
with a fixed Berezin orientation.
I collect ghosts, self-dual sectors, exact-mode measures, prescribed
zero-mode saturation factors, counterterms, fermionic gluing
normalizations, and all remaining source-independent factors into
$\cZ_{\bm\nu,\rm rem}$.
The source-independent contribution from the two Dirichlet interiors is
then
\begin{equation}
 \cA_{\bm\nu,D}
 =
 \cZ_{\bm\nu,\rm rem}\,
 \mathfrak n_{\bm\nu}\,
 \e^{-\cL_{\bm\nu,\rm loc}/2}
 \frac{
 \Pf\cD_{\bm\nu,F,Q_D}
 }{
 \sqrt{
 \left|
 \det_{\zeta}\cH_{\bm\nu,L,D}\,
 \det_{\zeta}\cH_{\bm\nu,R,D}
 \right|
 }
 }.
 \label{B-dirichlet-factor}
\end{equation}
At finite regulator, the zeta determinants in
Eq.~\eqref{B-dirichlet-factor} are replaced by the corresponding
ordinary determinants.
The factor $\mathfrak n_{\bm\nu}$ collects the regulated
change-of-variable and measure Jacobians.
In normalized Cartesian coordinates,
$\mathfrak n_{\bm\nu}=1$ and
$\cL_{\bm\nu,\rm loc}=0$.
For a real self-adjoint problem, the absolute values display the
Gaussian magnitude while the contour phase resides in
$\cZ_{\bm\nu,\rm rem}$.
For a complex saddle, the chosen determinant branches replace this
notation.

After removing the classical exponential, the assembled quadratic
source functional is
\begin{equation}
 \widehat Z^{(2)}_{\bm\nu}[J;\delta]
 =
 \underbrace{
 \cA_{\bm\nu,D}
 \frac{
 \Pf M^\Sigma_{\bm\nu,F}
 }{
 \sqrt{
 \left|
 \det\cK^\Sigma_{\bm\nu,B}
 \right|
 }
 }
 }_{
 \widehat Z^{(2)}_{\bm\nu}[0;\delta]
 }
 \exp\!\left[
 \frac12
 J_i\cR^{ij}_{\bm\nu}J_j
 \right].
 \label{B-full-source-functional}
\end{equation}
Its Gaussian contribution to the unnormalized two-source coefficient is
therefore
\begin{align}
 \widehat C_{\bm\nu}^{ij}
 ={}&
 \cA_{\bm\nu,D}
 \frac{
 \Pf M^\Sigma_{\bm\nu,F}
 }{
 \sqrt{
 \left|
 \det\cK^\Sigma_{\bm\nu,B}
 \right|
 }
 }
 \left[
 \cV^{ij}_{\bm\nu}
 +
 G^{ij}_{\bm\nu,D}
 +
 \mathbf s_{\bm\nu,i}^{\dagger}
 \bigl(\cK^\Sigma_{\bm\nu,B}\bigr)^{-1}
 \mathbf s_{\bm\nu,j}
 \right]
 +
 \widehat C_{\bm\nu,\rm higher}^{ij}.
 \label{7-29}
\end{align}
The bracket is the connected Gaussian response
$\cR^{ij}_{\bm\nu}$.
Its prefactor is the complete quadratic weight
$\widehat Z^{(2)}_{\bm\nu}[0;\delta]$.
The first term in Eq.~\eqref{7-29} follows by differentiating
Eq.~\eqref{B-full-source-functional}; the final term collects
interaction corrections and source-dependent insertions from the full
functional.

This assembly preserves the distinction between determinant weight and
inverse response.
The same interface matrix
$\cK^\Sigma_{\bm\nu,B}$
enters the former through its determinant and the latter through its
inverse.
The next reduction acts on the matching trace itself and produces the
smaller physical boundary block used in the finite-radius and spectral
analysis.

\subsection{Further boundary elimination}

The matching-surface field generally contains more components than the
final observable requires.
Split the trace coordinates as
\begin{equation}
 y_\Sigma=(x,v),
\end{equation}
where $x=P_Ry_\Sigma$ denotes the retained trace variables and $v$
denotes the complementary trace variables.
The projector $P_R$ acts on the finite matching space.
It is distinct from the bulk projector $P_B$ of
Appendix~\ref{sec:quadratic-reduction}, which acts before radial
elimination.

In fixed $(x,v)$ coordinates, write
\begin{equation}
 \cK^\Sigma_{\bm\nu,B}
 =
 \begin{pmatrix}
  \mathsf A & \mathsf B\\
  \mathsf C & \mathsf D
 \end{pmatrix},
 \qquad
 \mathbf s_i
 =
 \binom{\mathbf s_{i,x}}{\mathbf s_{i,v}} .
 \label{7-30}
\end{equation}
For invertible $\mathsf D$, stationarity in $v$ gives
\begin{equation}
 v
 =
 -\mathsf D^{-1}\mathsf Cx
 +
 \mathsf D^{-1}\mathbf s_{i,v}J_i .
\end{equation}
Substitution produces the reduced trace Hessian
\begin{equation}
 M_B^\Sigma
 =
 \mathsf A
 -
 \mathsf B\mathsf D^{-1}\mathsf C .
 \label{7-31}
\end{equation}
The same elimination gives
\begin{equation}
 \mathbf s_i^{\rm eff}
 =
 \mathbf s_{i,x}
 -
 \mathsf B\mathsf D^{-1}\mathbf s_{i,v},
\end{equation}
the contact response
\begin{equation}
 \mathbf s_{i,v}^\dagger
 \mathsf D^{-1}
 \mathbf s_{j,v},
\end{equation}
and the Gaussian determinant factor
$(\det\mathsf D)^{-1/2}$.
The reduced Hessian, effective source, contact response, and determinant
therefore form one Schur-complement operation.

This reduction also fixes the first finite-radius deformation of the
retained matrix.
Keep the $(x,v)$ coordinates fixed as $\delta$ varies and let subscripts
$0,1$ denote the endpoint value and first derivative.
Differentiating Eq.~\eqref{7-31} gives
\begin{align}
 M_1
 ={}&
 \mathsf A_1
 -
 \mathsf B_1\mathsf D_0^{-1}\mathsf C_0
 -
 \mathsf B_0\mathsf D_0^{-1}\mathsf C_1
 \nonumber\\
 &+
 \mathsf B_0\mathsf D_0^{-1}
 \mathsf D_1
 \mathsf D_0^{-1}\mathsf C_0 .
 \label{7-32}
\end{align}
For a Hermitian quadratic form, define the endpoint stationary extension
\begin{equation}
 \mathfrak J_0x
 =
 \binom{x}{-\mathsf D_0^{-1}\mathsf C_0x}.
\end{equation}
Then
\begin{equation}
 M_1
 =
 \mathfrak J_0^\dagger
 (\cK^\Sigma)_1
 \mathfrak J_0 .
 \label{7-33}
\end{equation}
Equation~\eqref{7-33} is the stationary-value derivative of the complete
trace quadratic form.
When the stationary extension is differentiated, the terms containing
$\partial_\delta\mathfrak J_\delta$ cancel because the action is
stationary in the eliminated $v$ direction.
The surviving term is therefore the explicit derivative of the complete
quadratic form evaluated on the endpoint stationary extension.
Expanding Eq.~\eqref{7-33} gives Eq.~\eqref{7-32}.

If the retained subspace is an exact endpoint kernel, self-adjointness
makes the endpoint mixed blocks vanish,
$\mathsf B_0=\mathsf C_0=0$.
Equation~\eqref{7-33} then reduces to the direct projection of
$(\cK^\Sigma)_1$ onto that kernel.
This is the trace-space version of the endpoint projection used in
Appendix~\ref{sec:quadratic-reduction}.

The successive reductions used above are associative.

\paragraph{Order independence of elimination.}
Let $x$ denote the final retained coordinates, $v$ additional trace
coordinates, and $u$ interior coordinates.
Assume that every block inverted in the successive eliminations is
invertible.
With one fixed coordinate system, measure, and source functional,
eliminating $u$ and then $v$ gives the same effective Hessian, source
vector, contact response, and Gaussian normalization as eliminating
$(v,u)$ in one step.

To see this, in real coordinates, write
\begin{equation}
 H
 =
 \begin{pmatrix}
  H_{xx} & H_{xv} & H_{xu}\\
  H_{vx} & H_{vv} & H_{vu}\\
  H_{ux} & H_{uv} & H_{uu}
 \end{pmatrix},
 \qquad
 \mathbf b_i
 =
 (\mathbf b_{i,x},\mathbf b_{i,v},\mathbf b_{i,u}).
 \label{7-34}
\end{equation}
Set $y=(x,v)$.
Eliminating $u$ first gives
\begin{align}
 K
 &=
 H_{yy}
 -
 H_{yu}H_{uu}^{-1}H_{uy},
 \label{7-35}\\
 \mathbf s_i
 &=
 \mathbf b_{i,y}
 -
 H_{yu}H_{uu}^{-1}\mathbf b_{i,u},
 \label{7-36}\\
 G_D^{ij}
 &=
 \mathbf b_{i,u}^T
 H_{uu}^{-1}
 \mathbf b_{j,u}.
 \label{7-37}
\end{align}
Decompose $K$ on $(x,v)$ as in Eq.~\eqref{7-30}.
Eliminating $v$ then gives
\begin{align}
 M
 &=
 \mathsf A-\mathsf B\mathsf D^{-1}\mathsf C,
 \label{7-39}\\
 \mathbf b_i^{\rm eff}
 &=
 \mathbf s_{i,x}
 -
 \mathsf B\mathsf D^{-1}\mathbf s_{i,v},
 \label{7-40}\\
 G_Q^{ij}
 &=
 G_D^{ij}
 +
 \mathbf s_{i,v}^T
 \mathsf D^{-1}
 \mathbf s_{j,v}.
 \label{7-41}
\end{align}
Now combine the eliminated coordinates as $w=(v,u)$.
Their stationary equation is
\begin{equation}
 w
 =
 H_{ww}^{-1}
 \bigl(
 J_i\mathbf b_{i,w}-H_{wx}x
 \bigr).
\end{equation}
Direct elimination gives
\begin{align}
 M
 &=
 H_{xx}
 -
 H_{xw}H_{ww}^{-1}H_{wx},
 \label{7-42}\\
 \mathbf b_i^{\rm eff}
 &=
 \mathbf b_{i,x}
 -
 H_{xw}H_{ww}^{-1}\mathbf b_{i,w},
 \label{7-43}\\
 G_Q^{ij}
 &=
 \mathbf b_{i,w}^T
 H_{ww}^{-1}
 \mathbf b_{j,w}.
 \label{7-44}
\end{align}
Both procedures solve the same stationary equations with the same
retained coordinate $x$.
Uniqueness therefore gives identical stationary solutions and identical
quadratic forms in $x$ and $J$.
The determinant identity
\begin{equation}
 \det H
 =
 \det H_{uu}\,
 \det\mathsf D\,
 \det M
 =
 \det H_{ww}\,
 \det M
 \label{7-45}
\end{equation}
gives the same Gaussian normalization.
Thus the effective matrix, source vector, and contact response agree in
the same retained coordinates.

Under a real invertible reparameterization
$x=T\widetilde x$,
\begin{equation}
 \widetilde M
 =
 T^\dagger MT,
 \qquad
 \widetilde{\mathbf b}_i^{\rm eff}
 =
 T^\dagger\mathbf b_i^{\rm eff},
 \qquad
 \dd x
 =
 |\det T|\,\dd\widetilde x .
 \label{7-46}
\end{equation}
The Gaussian functional is invariant under this simultaneous
transformation.
Equality of individual matrix entries therefore refers to one fixed
coordinate system.

This order-independence statement connects the interface construction to the physical outer-boundary Hessian.
Retain the asymptotic values $(q_L,q_R)$ and eliminate both radial
interiors together with every artificial matching variable.
The resulting quadratic matrix is
$\cK^\partial_{\bm\nu,B}$.
One may perform these eliminations sequentially or in a single Schur
complement; the reconstructed outer functional is identical.

This statement also fixes the role of the artificial cut.
Moving $\Sigma$ changes the one-sided Dirichlet factors,
$\cK^\Sigma_{\bm\nu,B}$, and the induced matching-surface sources.
Their assembled outer-boundary functional remains fixed.
For example, a free scalar on a unit interval cut at $r$ has
\begin{equation}
 K^\Sigma(r)
 =
 \frac1r+\frac1{1-r}.
\end{equation}
The interface Hessian varies explicitly with $r$, while elimination of
the interface reproduces the same fixed endpoint functional for every
$0<r<1$.
The physical object is therefore
$\cK^\partial_{\bm\nu,B}$ and its associated source response.
The matrix $\cK^\Sigma_{\bm\nu,B}$ is the intermediate gluing operator
that constructs it.

\subsection{Charge sensitivities of the radial problem}

The preceding construction determines the response in one fixed charge
sector.
To compare neighboring sectors while preserving the same boundary
prescription, I differentiate the background and radial boundary-value
problem with respect to the charge data.

Introduce independent symmetric bilinears $t_{ij}$, $i\leq j$, through
\begin{equation}
 U(t,\phi)
 =
 \sum_{i\leq j}
 t_{ij}\,\mathcal U_{ij}(\phi),
 \qquad
 \mathcal U_{ij}
 =
 (2-\delta_{ij})\mathcal G_\chi^{ij}(\phi).
\end{equation}
The physical charge sectors lie on the rank-one locus
\[
 t_{ij}=\nu_i\nu_j.
\]
I first differentiate in the independent symmetric variables $t_{ij}$
and then restrict the result to this physical locus.
This procedure resolves changes in the relative charge components
without conflating them with changes of boundary data.

Let $t_A$ denote one such charge parameter.
Every derivative $\partial_A$ below includes both the explicit
$t_A$-dependence of the radial coefficients and the induced
$t_A$-dependence of the background solution at fixed prescribed outer
data.
For real spectral parameter $z$, define
\begin{equation}
 R_\ell=A_\ell^{-1},
 \qquad
 D_\ell=B_\ell-\cN .
\end{equation}
The Riccati equation~\eqref{7-21} becomes
\begin{equation}
 \dot\cN
 =
 C_\ell-zW_\ell
 -
 D_\ell^\dagger R_\ell D_\ell .
\end{equation}
Using
\begin{equation}
 \partial_A R_\ell
 =
 -R_\ell(\partial_A A_\ell)R_\ell,
\end{equation}
the first sensitivity
\begin{equation}
 \cN_A\equiv\partial_A\cN
\end{equation}
satisfies the linear inhomogeneous equation
\begin{equation}
 \dot\cN_A
 =
 D_\ell^\dagger R_\ell\cN_A
 +
 \cN_A R_\ell D_\ell
 +
 \cS_A ,
 \label{7-49}
\end{equation}
with
\begin{align}
 \cS_A
 ={}&
 \partial_A C_\ell
 -
 z\,\partial_A W_\ell
 -
 (\partial_A B_\ell)^\dagger R_\ell D_\ell
 -
 D_\ell^\dagger R_\ell(\partial_A B_\ell)
 \nonumber\\
 &+
 D_\ell^\dagger R_\ell
 (\partial_A A_\ell)
 R_\ell D_\ell .
 \label{7-50}
\end{align}
Thus the charge derivative of the nonlinear Riccati problem reduces to
a linear transport equation once the original radial solution is known.
Its boundary condition follows by differentiating the same prescribed
outer-boundary condition used to construct $\cN$.

For complex $z$, I differentiate the ordered Riccati equation directly.
The two factors flanking $R_\ell$ then retain their displayed ordering
rather than being related by Hermitian conjugation.
Successive charge derivatives obey linear inhomogeneous equations whose
sources are determined recursively by the lower derivatives.

The two one-sided sensitivities assemble at the artificial interface in
the same way as the original maps.
At fixed $\Sigma$ and in fixed trace coordinates,
\begin{equation}
 \partial_A\cK^\Sigma_{\bm\nu,B,\ell}
 =
 \partial_A\cN^{\rm out}_{\bm\nu,L,\ell}
 +
 \partial_A\cN^{\rm out}_{\bm\nu,R,\ell}
 +
 \partial_A\cB_{\bm\nu,\Sigma,\ell}.
 \label{7-51}
\end{equation}
If the matching surface moves under the chosen parametrization, its
endpoint motion contributes the corresponding boundary-evaluation
terms before Eq.~\eqref{7-51} is applied.
Any subsequent trace-space elimination is differentiated with the
stationary-extension identity~\eqref{7-33}.
After the artificial trace has been removed, this procedure gives the
charge sensitivity of the physical outer-boundary Hessian and of its
effective source vectors.

The sensitivity equations therefore separate two operations used later
in the charge analysis.
Uniform charge rescaling in the product system is fixed directly by its
homogeneity.
Redistribution among charge components moves the target-space
trajectory and is computed by Eqs.~\eqref{7-49}--\eqref{7-51}.
Both comparisons use the same outer data and the same boundary
functional.

\subsection{The deformation and the observable}

The BPS square fixes the endpoint operator, while the finite-radius
boundary-value problem fixes its physical deformation.
A continuation of the matrix factorization by itself carries no such
information.
This distinction can be seen directly at finite dimension.

Consider a reduced positive matrix written as
\begin{equation}
 M(\delta)
 =
 \cQ(\delta)^\dagger
 G_{\rm diff}(\delta)
 \cQ(\delta)
 +
 \Delta(\delta),
 \qquad
 G_{\rm diff}(0)=1,
 \qquad
 \Delta(0)=0 .
 \label{7-52}
\end{equation}
Let
\begin{equation}
 M_0=\cQ_0^\dagger\cQ_0>0.
\end{equation}
For any Hermitian matrix $D_1$, define
\begin{equation}
 R_Q
 =
 \frac12
 \cQ_0M_0^{-1}D_1 .
 \label{7-53}
\end{equation}
Then
\begin{equation}
 (\cQ_0+\delta R_Q)^\dagger
 (\cQ_0+\delta R_Q)
 =
 M_0+\delta D_1+O(\delta^2).
 \label{7-54}
\end{equation}
Thus an arbitrary Hermitian first variation can be absorbed into a
chosen deformation of the square root.
The physical $D_1$ is instead determined by the gravitational data:
the constraints, differential operator, target-space geometry,
boundary domain, radial matching, trace elimination, and source map.
Equation~\eqref{7-54} therefore provides a representation of a known
deformation, not a rule for deriving it.

Exact symmetries constrain this physical matrix after the boundary
problem has been fixed.
For a symmetry $g$ of the fixed-charge saddle and its boundary data,
choose an orthonormal basis on the retained finite space.
Then
\begin{equation}
 U(g)M_{\bm\nu}(\delta)U(g)^{-1}
 =
 M_{\bm\nu}(\delta).
 \label{7-55}
\end{equation}
Every coefficient in the finite-radius expansion commutes with
$U(g)$.
Rotational symmetry therefore separates distinct angular
representations at quadratic order.
The response within each representation remains fixed by its field
content, boundary state, and source normalization.

Schur elimination also fixes the sign of a common class of induced
deformations.
Consider a positive eliminated channel,
\begin{equation}
 \cH_{\rm ext}(\delta)
 =
 \begin{pmatrix}
  M_0 & \gamma(\delta)V\\
  \gamma(\delta)V^\dagger & H_{\rm ch}
 \end{pmatrix},
 \qquad
 H_{\rm ch}>0,
 \qquad
 \gamma(\delta)\in\mathbb R .
 \label{7-56}
\end{equation}
Eliminating that channel gives
\begin{equation}
 M_{\rm eff}
 =
 M_0
 -
 \gamma(\delta)^2
 VH_{\rm ch}^{-1}V^\dagger .
 \label{7-57}
\end{equation}
The induced correction is negative semidefinite.
For an indefinite or contour-rotated channel, the same Schur
construction uses the continued quadratic form and contour; its sign is
then fixed by that physical continuation.

The observable depends on more than the matrix deformation.
Let
\begin{equation}
 \mathbf b^{\rm eff}(\delta)
 =
 \mathbf b_0+\delta\mathbf b_1+O(\delta^2),
 \qquad
 M(\delta)
 =
 M_0+\delta D_1+O(\delta^2).
\end{equation}
Differentiating the inverse contraction gives
\begin{align}
 \left.
 \frac{\dd}{\dd\delta}
 \left[
 (\mathbf b^{\rm eff})^\dagger
 M^{-1}
 \mathbf b^{\rm eff}
 \right]
 \right|_0
 ={}&
 \mathbf b_1^\dagger M_0^{-1}\mathbf b_0
 +
 \mathbf b_0^\dagger M_0^{-1}\mathbf b_1
 \nonumber\\
 &-
 \mathbf b_0^\dagger
 M_0^{-1}D_1M_0^{-1}
 \mathbf b_0 .
 \label{7-58}
\end{align}
The first two terms measure the motion of the physical source vector.
The third measures the deformation of the quadratic matrix itself.
For the complete connected response,
\begin{equation}
 \cR^{ij}
 =
 \cV^{ij}
 +
 G_Q^{ij}
 +
 (\mathbf b_i^{\rm eff})^\dagger
 M^{-1}
 \mathbf b_j^{\rm eff},
\end{equation}
the first variation also contains
$\partial_\delta\cV^{ij}$ and
$\partial_\delta G_Q^{ij}$.
For the unnormalized coefficient, the derivative of the zero-source
Gaussian weight multiplies this connected response as prescribed by
Appendix~\ref{sec:near-bps-problem}.
The physical first variation therefore combines the matrix deformation,
source motion, complementary propagation, direct source term, and
fluctuation normalization.

The three small-$\delta$ structures occur at different orders.
For the homogeneous physical outer-boundary Hessian,
\begin{equation}
 K_{LR}=O(\delta^2),
\end{equation}
while one eigenvalue of $K$ is $O(\delta)$.
Inversion converts these two facts into a cross-boundary covariance
\begin{equation}
 G_{LR}=O(\delta).
\end{equation}
The coupled $\ell=3$ sector instead produces the resonant term
\begin{equation}
 \delta^2\log(\delta/\delta_0),
\end{equation}
whose first $\delta$-derivative vanishes at the BPS endpoint.
The finite-cut scalar--gravity Hessian encountered in the angular
matching problem begins at $O(\delta^3)$ in its own boundary variables.
These orders belong to different quadratic objects at different stages
of the reduction.
The exact radial calculation below identifies each object before any
comparison of powers is made.

\section{Exact radial solution and validity of the expansion}
\label{sec:homogeneous-radial}

Exact radial matching propagates the near-throat solution to the two
asymptotic boundaries and determines the physical boundary Hessian.
I first derive the fixed-charge homogeneous quadratic problem in a form
that remains regular at the minimal three-sphere.

\subsection{Fixed-charge action at the minimal three-sphere}
\label{sec:homogeneous-calculation}

For fixed axion charge in $(d+1)$ Euclidean dimensions, take round
$S^d$ slices,
\begin{equation}
 \dd s^2
 =
 N(\tau)^2\dd\tau^2
 +
 a(\tau)^2\dd\Omega_d^2 .
 \label{8-1}
\end{equation}
Let $\phi^A$ denote the dilaton/saxions with target metric $G_{AB}$.
The axion kinetic matrix is $\mathcal G^\chi_{ij}$ and its inverse is
$\mathcal G_\chi^{ij}$.
After fixing the conserved axion momenta $\nu_i$, the Legendre transform
produces the charge potential
\begin{equation}
 U_{\bm\nu}(\phi)
 =
 \nu_i\mathcal G_\chi^{ij}(\phi)\nu_j .
 \label{8-2}
\end{equation}
The homogeneous fixed-charge Euclidean action is
\begin{align}
 I_R
 =
 \frac{\mathcal V_d}{2\kappa_{d+1}^2}
 \int\dd\tau\,
 \Bigg[
 &-\frac{d(d-1)a^{d-2}}{N}(a')^2
 -d(d-1)Na^{d-2}
 +2\Lambda Na^d
 \nonumber\\
 &+\frac{a^d}{2N}
 G_{AB}(\phi)\phi^{A\prime}\phi^{B\prime}
 +\frac{N}{2a^d}U_{\bm\nu}(\phi)
 \Bigg]
 +I_{\rm outer}+I_\Sigma .
 \label{8-3}
\end{align}
Here $\mathcal V_d=\operatorname{Vol}(S^d)$.
I use the Einstein--Hilbert convention
$-\int\sqrt g\,(R-2\Lambda)$.
All explicit small-radius solutions below use $\Lambda=0$.

Before the charge Legendre transform, the axion contribution is
\begin{equation}
 -\frac{\mathcal V_d}{2\kappa_{d+1}^2}
 \frac{a^d}{2N}
 \mathcal G^\chi_{ij}
 \chi^{i\prime}\chi^{j\prime}.
\end{equation}
After removing the common prefactor, its canonical momentum is
\[
 p_{\chi,i}
 =
 -\frac{a^d}{N}
 \mathcal G^\chi_{ij}\chi^{j\prime}.
\]
I identify the fixed canonical momenta with $\nu_i$.
The Routhian $L-p_{\chi,i}\chi^{i\prime}$ then gives the positive term
$NU_{\bm\nu}/(2a^d)$ in Eq.~\eqref{8-3}.
This fixes the Euclidean sign used throughout the fixed-charge
calculation.

Remove the common factor
$\mathcal V_d/(2\kappa_{d+1}^2)$ and define
\begin{equation}
 p_a
 =
 -2d(d-1)a^{d-2}\frac{a'}{N},
 \qquad
 p_A
 =
 \frac{a^d}{N}G_{AB}\phi^{B\prime}.
 \label{8-5}
\end{equation}
Variation of the lapse gives the Hamiltonian constraint
\begin{equation}
 \mathcal H
 =
 -\frac{p_a^2}{4d(d-1)a^{d-2}}
 +\frac{1}{2a^d}G^{AB}p_Ap_B
 +d(d-1)a^{d-2}
 -2\Lambda a^d
 -\frac{U_{\bm\nu}}{2a^d}
 =
 0 .
 \label{8-6}
\end{equation}
In four dimensions, take one scalar $\sigma$, set $\Lambda=0$, and use
proper-radial gauge $N=1$.
The scalar equation has the first integral
\begin{equation}
 (a^3\sigma')^2-U_{\bm\nu}(\sigma)
 =
 -E,
 \qquad
 (a')^2
 =
 1-\frac{E}{12a^4}.
 \label{8-7}
\end{equation}
The second relation follows immediately by inserting the first into
Eq.~\eqref{8-6}.
For $E>0$, the turning point satisfies
$E=12a_*^4$.
These equations reproduce the radial system of
Ref.~\cite{Rey1991} and fix the normalization used below.

I now reduce the homogeneous scalar--gravity system directly at the
minimal sphere.
I solve the Hamiltonian constraint for the momentum along the
background scalar trajectory and retain the scale factor and transverse
dilaton/saxions as canonical variables.

Choose coordinates $(\sigma,s^u)$ in a tubular neighborhood of the
background trajectory, with $s^u=0$ on that trajectory.
At the trajectory take
\begin{equation}
 G_{\sigma\sigma}=1,
 \qquad
 G_{\sigma u}=0,
 \qquad
 p_u=0,
 \qquad
 \sigma'\neq0 .
 \label{8-8}
\end{equation}
These relations fix the metric only on the trajectory.
Derivatives of the target metric remain in the quadratic fluctuation
problem for a general target space.

Multiplying Eq.~\eqref{8-6} by $2a^d$ and isolating the scalar momentum
defines
\begin{equation}
 \mathcal R
 =
 \frac{a^2p_a^2}{2d(d-1)}
 -2d(d-1)a^{2d-2}
 +4\Lambda a^{2d}
 +U_{\bm\nu}.
\end{equation}
The constraint is quadratic in $p_\sigma$.
Its two branches are
\begin{equation}
 p_\sigma
 =
 \frac{
 -G^{\sigma u}p_u
 +
 \varsigma
 \sqrt{
 (G^{\sigma u}p_u)^2
 +
 G^{\sigma\sigma}
 \left(
 \mathcal R-G^{uv}p_up_v
 \right)
 }
 }{
 G^{\sigma\sigma}
 },
 \qquad
 \varsigma=\pm1 .
 \label{C-general-constraint}
\end{equation}
Gauge fixing
$\sigma(\tau)=\bar\sigma(\tau)$
turns the constrained action into
\[
 I_{\rm red}
 =
 \frac{\mathcal V_d}{2\kappa_{d+1}^2}
 \int\dd\tau\,
 \left[
 p_mx^{m\prime}
 +
 \bar\sigma' p_\sigma
 \right]
 +
 I_{\rm outer}+I_\Sigma ,
\]
where $x^m=(a,s^u)$.
For a general target metric, the second variation of this expression
contains derivatives of
$G^{\sigma\sigma}$,
$G^{\sigma u}$, and
$G^{uv}$.

The explicit calculation below uses a target metric that is locally a
product along the scaling direction,
\[
 \dd\ell^2
 =
 \dd\sigma^2
 +
 G_{uv}(s)\dd s^u\dd s^v .
\]
Equation~\eqref{C-general-constraint} then reduces to
\begin{equation}
 p_\sigma
 =
 \varsigma F,
 \qquad
 F=\sqrt{\Xi},
 \label{8-9}
\end{equation}
with
\begin{equation}
 \Xi
 =
 \frac{a^2p_a^2}{2d(d-1)}
 -
 G^{uv}p_up_v
 -
 2d(d-1)a^{2d-2}
 +
 4\Lambda a^{2d}
 +
 U_{\bm\nu}(\sigma,s).
 \label{8-10}
\end{equation}
The sign $\varsigma$ selects the branch of the constrained Hamiltonian.
On an interval where $\sigma$ is monotonic, define
\begin{equation}
 \omega
 \equiv
 \varsigma\bar\sigma'
 >0 .
 \label{8-11}
\end{equation}
On the background,
$\varsigma=\operatorname{sgn}p_\sigma
=\operatorname{sgn}\bar\sigma'$.
The reduced first-order action becomes
\begin{equation}
 I_{\rm red}
 =
 \frac{\mathcal V_d}{2\kappa_{d+1}^2}
 \int\dd\tau\,
 \left[
 p_mx^{m\prime}
 +
 \omega
 F(x,p;\bar\sigma,\bm\nu)
 \right]
 +
 I_{\rm outer}+I_\Sigma .
 \label{8-12}
\end{equation}
This constraint reduction leaves the scale-factor fluctuation explicit
and remains regular at the turning point $a'=0$ as long as
$p_\sigma\neq0$.

Let $Q^m$ and $P_m$ denote fluctuations of $x^m$ and $p_m$.
Let $F_{xx}$, $F_{xp}$, and $F_{pp}$ denote the corresponding Hessian
blocks of $F$, evaluated on the background.
When the transverse frame varies along the trajectory, its connection is
included in $D_\tau Q$.
The quadratic first-order action is
\begin{align}
 I_{\rm hom}^{(2)}
 =
 \frac{\mathcal V_d}{4\kappa_{d+1}^2}
 \int\dd\tau\,
 \Big[
 &2P^T D_\tau Q
 +
 \omega Q^TF_{xx}Q
 +
 2\omega Q^TF_{xp}P
 \nonumber\\
 &+
 \omega P^TF_{pp}P
 \Big]
 +
 I_{\rm hom,bdy}^{(2)} .
 \label{8-13}
\end{align}
This first-order form is the regular description of the constrained
homogeneous fluctuation problem.
On any patch where $F_{pp}$ is invertible, stationarity in $P$ gives
\begin{equation}
 P
 =
 -F_{pp}^{-1}
 \left(
 \omega^{-1}D_\tau Q
 +
 F_{px}Q
 \right).
 \label{8-14}
\end{equation}
Substituting Eq.~\eqref{8-14} yields
\begin{equation}
 I_{\rm hom}^{(2)}
 =
 \frac{\mathcal V_d}{4\kappa_{d+1}^2}
 \int\dd\tau\,
 \left[
 -\frac{1}{\omega}
 \left(
 D_\tau Q+\omega F_{px}Q
 \right)^T
 F_{pp}^{-1}
 \left(
 D_\tau Q+\omega F_{px}Q
 \right)
 +
 \omega Q^TF_{xx}Q
 \right]
 +
 I_{\rm hom,bdy}^{(2)} .
 \label{8-15}
\end{equation}
In a parallel transverse frame, comparison with the radial action
in Eq.~\eqref{7-16} gives
\begin{equation}
 A
 =
 -\frac{1}{\omega}F_{pp}^{-1},
 \qquad
 B
 =
 -F_{pp}^{-1}F_{px},
 \qquad
 C
 =
 \omega
 \left(
 F_{xx}
 -
 F_{xp}F_{pp}^{-1}F_{px}
 \right).
 \label{8-16}
\end{equation}
The matrices in Eq.~\eqref{8-16} acquire the common factor
$\mathcal V_d/(2\kappa_{d+1}^2)$ when inserted into
Eq.~\eqref{7-16}.
For a non-product target metric, the corresponding coefficients follow
from the Hessian of
$\bar\sigma'p_\sigma$
in Eq.~\eqref{C-general-constraint}.

On the background,
\begin{equation}
 F
 =
 |p_\sigma|
 =
 a^d\omega,
 \qquad
 p_a
 =
 -2d(d-1)a^{d-2}a' .
 \label{8-17}
\end{equation}
The momentum Hessian separates into the scale-factor and transverse
blocks,
\begin{align}
 F_{p_ap_a}
 &=
 \frac{
 a^2\omega^2
 -
 2d(d-1)(a')^2
 }{
 2d(d-1)a^d\omega^3
 },
 \label{8-18}\\
 F_{p_up_v}
 &=
 -\frac{G^{uv}}{a^d\omega},
 \qquad
 F_{p_ap_u}=0 .
 \label{8-19}
\end{align}
At the minimal three-sphere,
$a'=0$.
For the branch with $p_\sigma\neq0$, one has $\omega\neq0$, so both
blocks of $F_{pp}$ are invertible.
Equation~\eqref{8-16} then gives, before the common gravitational
normalization is restored,
\begin{equation}
 A_{aa}\big|_*
 =
 -2d(d-1)a_*^{d-2},
 \qquad
 A_{uv}\big|_*
 =
 a_*^dG_{uv},
 \qquad
 A_{au}\big|_*
 =
 0 .
 \label{8-20}
\end{equation}
The scale-factor kinetic term carries the gravitational conformal sign,
whereas the transverse dilaton/saxion block is positive for
$G_{uv}>0$.
I integrate the scale direction on the gravitational contour.
Zero crossings of the physical reduced quadratic form control changes
of the Gaussian contour phase.

Away from the minimal sphere, the second-order representation
Eq.~\eqref{8-15} becomes singular when
\begin{equation}
 a^2\omega^2
 -
 2d(d-1)(a')^2
 =
 0 .
 \label{8-21}
\end{equation}
This is a singularity of the elimination of $P_a$.
The first-order system~\eqref{8-13} remains the canonical continuation
through such a point.
Equivalently, one may solve the Hamiltonian constraint using another
monotonic field as the clock.
Both descriptions reconstruct the same outer boundary map.

The four-dimensional one-scalar truncation makes the scalar--gravity
block explicit.
For $d=3$,
\begin{equation}
 F(a,p_a;\sigma)
 =
 \sqrt{
 \frac{a^2p_a^2}{12}
 -
 12a^4
 +
 4\Lambda a^6
 +
 U_{\bm\nu}(\sigma)
 }.
 \label{8-22}
\end{equation}
With
$\omega=|\sigma'|$,
$F=a^3\omega$, and
$f\equiv Q^a$,
Eq.~\eqref{8-15} becomes
\begin{equation}
 I_{\rm hom}^{(2)}
 =
 \frac{\mathcal V_3}{4\kappa_4^2}
 \int\dd\tau\,
 \left[
 -\frac{1}{\omega F_{p_ap_a}}
 \left(
 f'
 +
 \omega F_{p_aa}f
 \right)^2
 +
 \omega F_{aa}f^2
 \right]
 +
 I_{\rm hom,bdy}^{(2)} .
 \label{8-23}
\end{equation}
All partial derivatives of $F$ are taken at fixed $\sigma$ and fixed
remaining canonical variables.
Define
\begin{equation}
 \mathcal R_4
 =
 \frac{a^2p_a^2}{12}
 -
 12a^4
 +
 4\Lambda a^6
 +
 U_{\bm\nu}(\sigma),
 \qquad
 F=\sqrt{\mathcal R_4}.
\end{equation}
Direct differentiation gives
\begin{align}
 F_{p_ap_a}
 &=
 \frac{a^2}{12F}
 -
 \frac{a^4p_a^2}{144F^3},
 \label{8-25}\\
 F_{p_aa}
 &=
 \frac{ap_a}{6F}
 -
 \frac{
 a^2p_a(\mathcal R_4)_a
 }{
 24F^3
 },
 \label{8-26}\\
 F_{aa}
 &=
 \frac{(\mathcal R_4)_{aa}}{2F}
 -
 \frac{(\mathcal R_4)_a^2}{4F^3},
 \label{8-27}\\
 (\mathcal R_4)_a
 &=
 \frac{ap_a^2}{6}
 -
 48a^3
 +
 24\Lambda a^5,
 \qquad
 (\mathcal R_4)_{aa}
 =
 \frac{p_a^2}{6}
 -
 144a^2
 +
 120\Lambda a^4 .
 \label{8-28}
\end{align}
At the minimal sphere $p_a=0$, hence $F_{p_aa}=0$.
Using Eq.~\eqref{8-20}, the local quadratic density becomes
\begin{equation}
 \left.
 \mathcal L_{\rm hom}^{(2)}
 \right|_*
 =
 \frac{\mathcal V_3}{4\kappa_4^2}
 \left[
 -12a_*(f')^2
 +
 \omega_*F_{aa}\big|_*\,f^2
 \right].
 \label{8-29}
\end{equation}
Equation~\eqref{8-29} gives the local form at the minimal sphere.
The radial propagation is governed by Eq.~\eqref{8-23}.
After the Hamiltonian constraint and scalar-clock gauge are imposed, the
one-axion--dilaton truncation leaves one homogeneous scalar--gravity
degree of freedom.
Additional dilaton/saxions, form fields, harmonics, and internal modes
enlarge this reduced quadratic system.

For several axions, all explicit charge dependence begins with
\begin{equation}
 U_{\bm\nu}(\phi)
 =
 \nu_i
 \mathcal G_\chi^{ij}(\phi)
 \nu_j ,
\end{equation}
and therefore
\begin{equation}
 \nabla_u\nabla_vU_{\bm\nu}
 =
 \nu_i
 \nabla_u\nabla_v
 \mathcal G_\chi^{ij}(\phi)
 \nu_j .
 \label{8-31}
\end{equation}
Introduce the independent symmetric bilinears $t_{ij}$ of
Appendix~\ref{sec:finite-radius-boundary}.
At a fixed phase-space point,
\begin{equation}
 \frac{\partial F}{\partial t_{ij}}
 =
 \frac{
 (2-\delta_{ij})
 \mathcal G_\chi^{ij}(\phi)
 }{
 2F
 },
 \qquad
 i\leq j .
 \label{8-32}
\end{equation}
The factor $2$ for $i<j$ restores the two off-diagonal entries of the
original double sum.
The physical charge sensitivities follow by differentiating the
background together with Eqs.~\eqref{8-16} and~\eqref{8-32} at fixed
outer data.

At $p_u=0$, the transverse Hessian of the square-root Hamiltonian is
\begin{equation}
 \omega F_{;uv}
 =
 \frac{1}{2a^d}
 \nabla_u\nabla_vU_{\bm\nu}
 -
 \frac{1}{4a^{3d}\omega^2}
 (\nabla_uU_{\bm\nu})
 (\nabla_vU_{\bm\nu}).
 \label{8-33}
\end{equation}
On the stationary transverse trajectory used below,
$\nabla_uU_{\bm\nu}=0$, so the second term vanishes.
For a general target metric,
Eq.~\eqref{C-general-constraint} supplies the additional derivatives of
the inverse target metric.

At fixed $(a,\omega,\phi)$, the two numerators in
Eq.~\eqref{8-33} carry quadratic and quartic charge dependence,
respectively.
The background itself moves with $\bm\nu$, and the subsequent Schur
reductions introduce further inverse matrices.
The final physical boundary matrix therefore acquires its charge
dependence only after the background sensitivity and all eliminated
fields have been included.
The next subsection performs the small-radius inner--outer expansion
that turns these local coefficients into the finite-radius boundary
response.

\subsection{Small-radius background expansion}

The near-BPS problem contains two radial scales.
At fixed outer radius the background approaches the BPS solution, while
the region $a=O(a_*)$ retains a nontrivial limiting geometry.
I separate these regions by inner--outer matching.

In four dimensions define
\begin{equation}
 \delta\equiv a_*^2=\sqrt{\frac{E}{12}},
 \qquad
 E=12\delta^2 .
 \label{8-34}
\end{equation}
The metric equation therefore departs from its BPS form at
$O(\delta^2)$ at fixed nonzero outer radius.

I use the local multi-axion truncation
\begin{equation}
 \dd\ell_{\rm dilaton/saxion}^2
 =
 \dd\sigma^2
 +
 G_{uv}(s)\dd s^u\dd s^v,
 \qquad
 \mathcal G_\chi^{ij}(\sigma,s)
 =
 \e^{2\beta\sigma}H^{ij}(s),
 \label{8-35}
\end{equation}
for which
\begin{equation}
 U_{\bm\nu}(\sigma,s)
 =
 \e^{2\beta\sigma}V_{\bm\nu}(s),
 \qquad
 V_{\bm\nu}(s)
 =
 \nu_iH^{ij}(s)\nu_j .
 \label{8-36}
\end{equation}
Choose a transverse stationary trajectory
$s^u=\bar s^u_{\bm\nu}$ satisfying
\begin{equation}
 \nabla_uV_{\bm\nu}(\bar s_{\bm\nu})=0 .
 \label{8-37}
\end{equation}
At the minimal three-sphere define
\begin{equation}
 u_*
 =
 \frac{
 \e^{2\beta\sigma_*}
 V_{\bm\nu}(\bar s_{\bm\nu})
 }{E}
 \geq1 .
 \label{8-38}
\end{equation}
Equation~\eqref{8-7} gives
$p_\sigma^2=E(u_*-1)$ at the minimal sphere.
Thus $u_*>1$ is precisely the branch on which $\sigma$ remains a valid
clock there.
At $u_*=1$, the scalar momentum vanishes and the first-order constrained
system of the preceding subsection supplies the regular description.

To resolve the inner region, set
\begin{equation}
 a=a_*\rho(t),
 \qquad
 \tau=a_*t,
 \qquad
 \psi(t)=\sigma(t)-\sigma_*,
 \label{8-39}
\end{equation}
and define the normalized charge potential
\begin{equation}
 \widehat U_{\bm\nu}(\psi,s)
 =
 \e^{2\beta\psi}
 \frac{
 V_{\bm\nu}(s)
 }{
 V_{\bm\nu}(\bar s_{\bm\nu})
 } .
 \label{8-40}
\end{equation}
Since
$U_{\bm\nu}=12\delta^2u_*\widehat U_{\bm\nu}$,
the two first integrals become
\begin{equation}
 \dot\rho^{\,2}
 =
 1-\rho^{-4},
 \qquad
 \rho^6\dot\psi^{\,2}
 =
 12\left(
 u_*\widehat U_{\bm\nu}-1
 \right),
 \label{8-41}
\end{equation}
where a dot now denotes $\dd/\dd t$.
At $\rho=1$,
\begin{equation}
 \dot\psi_*^{\,2}=12(u_*-1).
\end{equation}
The inner equations contain no explicit power of $a_*$.
All small-radius scaling enters through the overall action
normalization and through matching to the outer solution.

A signed coordinate makes the complete two-sided geometry elementary:
\begin{equation}
 \dd y
 =
 \frac{2\delta}{a^3}\dd\tau,
 \qquad
 a^2
 =
 \delta\sec y,
 \qquad
 -\frac{\pi}{2}<y<\frac{\pi}{2},
 \qquad
 k=\sqrt3\,\beta,
 \qquad
 \theta_\beta=\frac{\pi k}{2}.
 \label{8-42}
\end{equation}
The right exterior has $y>0$ and the left exterior has $y<0$.
The metric becomes
\begin{equation}
 \dd s^2
 =
 \delta
 \left[
 \frac{\dd y^2}{4\cos^3y}
 +
 \frac{\dd\Omega_3^2}{\cos y}
 \right].
\end{equation}
For $\delta>0$ and $|y|<\pi/2$, every metric eigenvalue is positive and
real, so
\begin{equation}
 \sum_{\mu=1}^{4}
 \left|
 \operatorname{Arg}\lambda_\mu
 \right|
 =
 0<\pi .
\end{equation}
The real finite-radius branch therefore lies strictly inside the
four-dimensional Kontsevich--Segal--Witten allowable domain.
The logarithm encountered later arises from fluctuation matching.

On the right half,
\begin{equation}
 y=\arccos(\rho^{-2}),
\end{equation}
and the sign reverses on the left half.
The exact stationary matter profile is
\begin{equation}
 H\equiv\e^{-\beta\sigma}
 =
 \frac{\mathfrak c_{\bm\nu}}{\delta}\sin w,
 \qquad
 w=\alpha_*-ky,
 \qquad
 \theta_\beta<\alpha_*<\pi-\theta_\beta,
 \qquad
 \mathfrak c_{\bm\nu}
 =
 \sqrt{
 \frac{
 V_{\bm\nu}(\bar s_{\bm\nu})
 }{12}
 } .
 \label{8-43}
\end{equation}
Direct differentiation gives
\begin{equation}
 a^3\sigma'
 =
 \sqrt{12}\,\delta\cot w,
 \qquad
 u_*=\csc^2\alpha_*,
 \qquad
 a^3\mathcal G^\chi_{ij}\chi^{j\prime}
 =
 -\nu_i .
\end{equation}
The two radial halves meet smoothly at $y=0$.
Their outward canonical momenta carry opposite orientations at the
matching surface.

I now impose the saddle family used throughout the paper.
Fix the right asymptotic value
\begin{equation}
 h
 \equiv
 \e^{-\beta\sigma_R}
 >0 .
\end{equation}
At the right end $y\to\pi/2$, Eq.~\eqref{8-43} gives
\begin{equation}
 \epsilon
 =
 \arcsin
 \frac{\delta h}{\mathfrak c_{\bm\nu}},
 \qquad
 \alpha_*
 =
 \theta_\beta+\epsilon,
 \qquad
 \vartheta
 =
 \pi k,
 \qquad
 w_R=\epsilon,
 \qquad
 w_L=\vartheta+\epsilon .
 \label{8-44}
\end{equation}
I take the small-radius branch
$0<\epsilon<\pi-\vartheta$.
The left asymptotic value is then
\begin{equation}
 H_L
 =
 \frac{\mathfrak c_{\bm\nu}}{\delta}
 \sin(\vartheta+\epsilon)
 =
 \frac{
 \mathfrak c_{\bm\nu}\sin\vartheta
 }{\delta}
 +
 h\cos\vartheta
 +
 O(\delta).
 \label{8-45}
\end{equation}
Thus fixing the right data and varying $\delta$ moves the left
asymptotic dilaton/saxion value, exactly as prescribed in
Appendix~\ref{sec:near-bps-problem}.

Conversely, fixing both finite asymptotic values determines the
minimal radius.
Using
$H_R=\mathfrak c_{\bm\nu}\sin\epsilon/\delta$
and the sum and difference of $H_L$ and $H_R$ gives
\begin{equation}
 \delta
 =
 \frac{
 2\mathfrak c_{\bm\nu}
 }{
 \sqrt{
 \left[
 (H_L+H_R)/\cos\theta_\beta
 \right]^2
 +
 \left[
 (H_L-H_R)/\sin\theta_\beta
 \right]^2
 }
 } .
 \label{8-46}
\end{equation}
The two-ended boundary-value problem therefore fixes $\delta$ once both
asymptotic scalar values and the charge sector are specified.

For
\begin{equation}
 0<\beta<\beta_c=\frac1{\sqrt3},
\end{equation}
one has
$0<\theta_\beta<\pi/2$.
Along the fixed-right-data family,
\begin{equation}
 \alpha_*-\theta_\beta
 =
 O(\delta),
 \qquad
 u_*(\delta)
 \longrightarrow
 u_*^{(0)}
 \equiv
 \csc^2\theta_\beta
 >1 .
 \label{8-47}
\end{equation}
Hence $p_\sigma$ remains nonzero at the minimal sphere throughout the
near-BPS branch used below.

The outer expansion follows directly from the same exact solution.
On the right exterior, define
\begin{equation}
 u
 =
 \arcsin\frac{\delta}{a^2}.
\end{equation}
Since
$y=\pi/2-u$,
Eq.~\eqref{8-43} becomes
\begin{equation}
 \e^{-\beta\sigma(a)}
 =
 \frac{
 \sqrt{V_{\bm\nu}(\bar s_{\bm\nu})}
 }{
 \sqrt{12}\,\delta
 }
 \sin\!\left(
 \alpha_*-\theta_\beta+\sqrt3\,\beta u
 \right).
 \label{8-48}
\end{equation}
Both
\begin{equation}
 \alpha_*-\theta_\beta
 =
 \arcsin
 \frac{\delta h}{\mathfrak c_{\bm\nu}}
\end{equation}
and
$u=\arcsin(\delta/a^2)$
are odd functions of $\delta$.
Their sine divided by $\delta$ therefore has an expansion in even
powers of $\delta$.
At fixed $a>0$,
\begin{equation}
 \e^{-\beta\sigma(a)}
 =
 \e^{-\beta\sigma_\infty}
 +
 \frac{
 \beta
 \sqrt{V_{\bm\nu}(\bar s_{\bm\nu})}
 }{
 2a^2
 }
 +
 O(\delta^2),
 \qquad
 \e^{-\beta\sigma_\infty}=h .
 \label{8-49}
\end{equation}
The first two terms are the BPS profile.
Hence
\begin{equation}
 \sigma(a;\delta)
 =
 \sigma_{\rm BPS}(a)
 +
 O(\delta^2)
 \label{8-50}
\end{equation}
at every fixed nonzero outer radius.
The $O(\delta)$ effect found below therefore originates in the inner
region and in the motion of the left asymptotic endpoint, rather than
from an $O(\delta)$ deformation of the fixed-radius outer background.

The same distinction appears in the classical action.
Set
\begin{equation}
 \cA
 \equiv
 \frac{\mathcal V_3}{2\kappa_4^2}.
\end{equation}
After using the field equations, the two-sided bulk density is
\begin{equation}
 \frac{\dd I_R}{\cA}
 =
 6\delta
 \left(
 \csc^2w-\sec^2y
 \right)\dd y .
 \label{8-51}
\end{equation}
The matching-surface contributions from the two halves cancel.
The flat outer reference terms cancel the integrated
$-\sec^2y$ contribution.
The renormalized action is therefore
\begin{align}
 \frac{S_{\rm full}^{\rm ren}}{\cA}
 &=
 \frac{6\delta}{k}
 \left[
 \cot\epsilon
 -
 \cot(\vartheta+\epsilon)
 \right]
 \nonumber\\
 &=
 \frac{
 6\mathfrak c_{\bm\nu}
 }{
 kh
 }
 -
 \frac6k
 \cot\vartheta\,\delta
 +
 \frac{
 6h
 }{
 k\mathfrak c_{\bm\nu}
 }
 \left(
 \csc^2\vartheta-\frac12
 \right)
 \delta^2
 +
 O(\delta^3).
 \label{8-52}
\end{align}
The first term is the finite endpoint action of this truncation.
The linear term realizes the right derivative introduced in
Appendix~\ref{sec:near-bps-problem}.

The boundary first variation reproduces its coefficient directly.
The outward left scalar momentum, with the common normalization
$\cA$ removed, is
\begin{equation}
 p_{\sigma,L}^{\rm out}
 =
 -\sqrt{12}\,\delta
 \cot(\vartheta+\epsilon),
\end{equation}
while Eq.~\eqref{8-45} gives
\begin{equation}
 \frac{\dd\sigma_L}{\dd\delta}
 =
 \frac1{\beta\delta}
 +
 O(1).
\end{equation}
The right asymptotic value is fixed, and the two matching-surface
variations cancel.
Consequently
\begin{equation}
 \frac1{\cA}
 \frac{\dd S_{\rm full}^{\rm ren}}{\dd\delta}
 =
 p_{\sigma,L}^{\rm out}
 \frac{\dd\sigma_L}{\dd\delta}
 \longrightarrow
 -\frac6k\cot\vartheta .
 \label{8-53}
\end{equation}
This agrees with the linear term in Eq.~\eqref{8-52}.
The $O(\delta)$ action shift is therefore boundary work along the
fixed-right-data saddle family, while the local outer background begins
at $O(\delta^2)$.

I next extract the scaling of the physical transverse fluctuations.
Let $q^u$ be covariant fluctuations about
$s^u=\bar s^u_{\bm\nu}$.
Because the background has
$s^{u\prime}=0$ and
$\nabla_uU_{\bm\nu}=0$,
the transverse sector decouples at quadratic order from the lapse,
scale factor, and longitudinal fluctuation of $\sigma$.
In the inner variables its quadratic action is
\begin{align}
 I_{\perp}^{(2)}
 ={}&
 \frac{\mathcal V_3\delta}{4\kappa_4^2}
 \int\dd t\,
 \left[
 \rho^3
 G_{uv}
 D_tq^uD_tq^v
 +
 q^u
 \widehat{\mathbb V}_{uv}
 (t;\widehat{\bm\nu})
 q^v
 \right]
 \nonumber\\
 &+
 \frac{\mathcal V_3\delta}{4\kappa_4^2}
 q_\Sigma^u
 \widehat{\mathcal B}_{uv,\Sigma}
 q_\Sigma^v
 +
 O(\delta^2),
 \label{8-54}
\end{align}
where
\begin{equation}
 \widehat{\mathbb V}_{uv}
 =
 6u_*\rho^{-3}
 \e^{2\beta\psi}
 \mathfrak M_{uv}(\widehat{\bm\nu}),
 \label{8-55}
\end{equation}
and
\begin{equation}
 \mathfrak M_{uv}(\widehat{\bm\nu})
 =
 \frac{
 \nu_i
 \nabla_u\nabla_vH^{ij}(\bar s_{\bm\nu})
 \nu_j
 }{
 \nu_i
 H^{ij}(\bar s_{\bm\nu})
 \nu_j
 } .
 \label{8-56}
\end{equation}
Here
$\widehat{\bm\nu}=\bm\nu/|\bm\nu|$
records the charge ratios on a fixed active support.
Uniform rescaling cancels from
$\mathfrak M_{uv}$.
The overall inner action is proportional to
\begin{equation}
 \delta=a_*^2 .
\end{equation}
This is the origin of the linear finite-radius scale that survives after
matching to the asymptotic boundary.

The same counting applies to every fixed scalar harmonic.
Let
\begin{equation}
 -\nabla_{S^3}^2Y_{\ell m}
 =
 \ell(\ell+2)Y_{\ell m},
\end{equation}
and suppose the radial potential has a finite inner limit after the
rescaling
\begin{equation}
 \widehat{\cV}_\ell(t)
 \equiv
 a_*^2\cV_\ell(a_*t).
\end{equation}
Then
\begin{align}
 I_{\ell,\Sigma}^{(2)}
 =
 \frac{\mathcal V_3\delta}{4\kappa_4^2}
 \int\dd t\,
 \Big[
 &\rho^3
 (D_tq_\ell)^T
 G
 D_tq_\ell
 +
 \rho\,\ell(\ell+2)
 q_\ell^TGq_\ell
 \nonumber\\
 &+
 \rho^3
 q_\ell^T
 \widehat{\cV}_\ell
 q_\ell
 \Big]
 +
 O(\delta^2).
 \label{8-57}
\end{align}
Thus every fixed-$\ell$ inner quadratic block carries the same overall
factor $\delta$.
The subsequent outer propagation determines whether this local scale
appears as an $O(\delta)$ diagonal Hessian, a higher-order mixed
Hessian, or an inverse response enhanced by an endpoint soft mode.
Harmonics with $\ell=O(a_*^{-1})$ define a separate double-scaling
regime.
The calculation below keeps $\ell$ fixed and proceeds to the exact
product-model boundary map.

\subsection{Product axion--dilaton/saxion model and boundary covariance}

I now solve the homogeneous transverse problem exactly.
The product target makes the normalized transverse Hessian universal,
while the exact finite-radius solution propagates this local structure
to the two physical asymptotic boundaries.
This calculation determines the boundary Hessian $K$, its soft
eigenvalue, and the cross-boundary covariance used in the main text.

The model follows from the fixed-charge Legendre transform of the
$[SL(2,\mathbb R)/SO(1,1)]^p$ seed sectors of symmetric-space axion
wormholes~\cite{HertogTrigianteVanRiet2017}.
Their effective coupling obeys
\begin{equation}
 b_{\rm eff}^{-2}
 =
 \sum_i b_i^{-2}.
\end{equation}
With $b_i=2\beta_i$, this gives $b_{\rm eff}=2\beta$, where $\beta$
is defined below.
The Euclidean axion kinetic terms carry the negative sign before the
charge Legendre transform.
Fixing the axion momenta converts them into the positive charge
potential of the Routhian.

Restrict to the active support of $\bm\nu$ and label its $p$ components
by $i=1,\ldots,p$.
Take canonically normalized dilaton/saxions $\phi_i$ with
\begin{equation}
 \dd\ell_{\rm dilaton/saxion}^2
 =
 \sum_{i=1}^p\dd\phi_i^2,
 \qquad
 \mathcal G^\chi_{ij}
 =
 \delta_{ij}\e^{-2\beta_i\phi_i},
 \qquad
 \beta_i>0.
 \label{8-58}
\end{equation}
The inverse axion metric and fixed-charge potential are
\begin{equation}
 \mathcal G_\chi^{ij}
 =
 \delta^{ij}\e^{2\beta_i\phi_i},
 \qquad
 U_{\bm\nu}
 =
 \sum_i\nu_i^2\e^{2\beta_i\phi_i}.
 \label{8-59}
\end{equation}
Define
\begin{equation}
 \frac1{\beta^2}
 =
 \sum_i\frac1{\beta_i^2},
 \qquad
 n_i
 =
 \frac{\beta}{\beta_i},
 \qquad
 \sum_i n_i^2=1.
 \label{8-60}
\end{equation}
Choose an orthonormal basis $E_{iu}$,
$u=1,\ldots,p-1$, for the subspace perpendicular to $n_i$:
\begin{equation}
 \sum_iE_{iu}E_{iv}
 =
 \delta_{uv},
 \qquad
 \sum_i n_iE_{iu}
 =
 0.
 \label{8-61}
\end{equation}
The orthogonal rotation
\begin{equation}
 \phi_i
 =
 n_i\sigma+E_{iu}s^u
 \label{8-62}
\end{equation}
separates the scaling direction from the relative dilaton/saxion
directions.
Equation~\eqref{8-59} becomes
\begin{equation}
 U_{\bm\nu}
 =
 \e^{2\beta\sigma}V_{\bm\nu}(s),
 \qquad
 V_{\bm\nu}(s)
 =
 \sum_iX_i(s),
 \qquad
 X_i(s)
 =
 \nu_i^2
 \exp\!\left(
 2\beta_iE_{iu}s^u
 \right).
 \label{8-63}
\end{equation}

\paragraph{Universal transverse Hessian.}
\label{prop:product-Hessian}
At an interior stationary point of Eq.~\eqref{8-63},
the normalized transverse Hessian defined in Eq.~\eqref{8-56} is
\begin{equation}
 \mathfrak M_{uv}
 =
 4\beta^2\delta_{uv}.
 \label{8-64}
\end{equation}
Thus every transverse direction in the active product sector has the
same positive charge-induced quadratic coefficient.

To see this, observe that stationarity gives
\begin{equation}
 0
 =
 \frac12\partial_uV_{\bm\nu}
 =
 \sum_i\beta_iE_{iu}X_i .
 \label{8-65}
\end{equation}
The $(p-1)\times p$ matrix with entries
$\beta_iE_{iu}$ has rank $p-1$.
Its kernel is therefore one dimensional.
The positive vector with components $\beta_i^{-2}$ belongs to this
kernel because
\begin{equation}
 \sum_i
 \beta_iE_{iu}\frac1{\beta_i^2}
 =
 \sum_i\frac{E_{iu}}{\beta_i}
 =
 \frac1\beta
 \sum_i n_iE_{iu}
 =
 0 .
\end{equation}
Since every $X_i$ is positive on the active support,
Eq.~\eqref{8-65} fixes
\begin{equation}
 X_i
 =
 \frac{c}{\beta_i^2},
 \qquad
 c>0.
 \label{8-67}
\end{equation}
It follows that
\begin{align}
 \partial_u\partial_vV_{\bm\nu}
 &=
 4\sum_i
 \beta_i^2E_{iu}E_{iv}X_i
 =
 4c\,\delta_{uv},
 \\
 V_{\bm\nu}
 &=
 c\sum_i\frac1{\beta_i^2}
 =
 \frac{c}{\beta^2}.
\end{align}
Their ratio gives Eq.~\eqref{8-64}.

Henceforth
$V_{\bm\nu}$ denotes
$V_{\bm\nu}(\bar s_{\bm\nu})$.
The BPS profile obtained from Eq.~\eqref{8-49} is
\begin{equation}
 \e^{-\beta\sigma_{\rm BPS}(a)}
 =
 \e^{-\beta\sigma_\infty}
 +
 \frac{\beta\sqrt{V_{\bm\nu}}}{2a^2}.
 \label{8-70}
\end{equation}
Introduce the classical inverse-area scale
\begin{equation}
 \varpi_{\bm\nu}
 =
 \frac{
 2\e^{-\beta\sigma_\infty}
 }{
 \beta\sqrt{V_{\bm\nu}}
 }.
 \label{8-71}
\end{equation}
Then
\begin{equation}
 U_{\bm\nu}^{\rm BPS}(a)
 =
 \frac{4a^4}{\beta^2}
 \left(
 1+\varpi_{\bm\nu}a^2
 \right)^{-2}.
 \label{8-72}
\end{equation}
For an orthonormal homogeneous transverse mode, the quadratic field
equation on the outward BPS branch $a'=1$ is
\begin{equation}
 a^2q''
 +
 3aq'
 -
 \frac{2\mathfrak m_\perp}{\beta^2}
 \left(
 1+\varpi_{\bm\nu}a^2
 \right)^{-2}q
 =
 0,
 \qquad
 \mathfrak m_\perp=4\beta^2.
 \label{8-73}
\end{equation}
Thus
\begin{equation}
 a^2q''
 +
 3aq'
 -
 8
 \left(
 1+\varpi_{\bm\nu}a^2
 \right)^{-2}q
 =
 0 .
\end{equation}
A fundamental pair is
\begin{equation}
 q_+(a)
 =
 \frac{a^2}{1+\varpi_{\bm\nu}a^2},
 \qquad
 q_-(a)
 =
 a^{-4}
 \left(
 1+\varpi_{\bm\nu}a^2
 \right)^2.
 \label{8-75}
\end{equation}
Near the BPS endpoint these branches behave as $a^2$ and $a^{-4}$.
The homogeneous charged outer equation therefore contains no
Frobenius logarithm.

I now solve the finite-radius transverse problem exactly.
Using Eq.~\eqref{8-64} on the matter profile of Eq.~\eqref{8-43},
the quadratic action becomes
\begin{equation}
 I_q^{(2)}
 =
 \cA\delta
 \int_{-\pi/2}^{\pi/2}
 \left[
 q_y^2
 +
 2k^2\csc^2w\,q^2
 \right]\dd y,
 \qquad
 \frac{\dd^2q}{\dd w^2}
 =
 2\csc^2w\,q .
 \label{8-76}
\end{equation}
The coefficient $\cA$ is the positive normalization fixed in the
preceding subsection.
A convenient fundamental pair is
\begin{equation}
 u_1(w)
 =
 \cot w,
 \qquad
 u_2(w)
 =
 1-w\cot w,
 \qquad
 u_1u_2'-u_1'u_2
 =
 1.
 \label{8-77}
\end{equation}
Near the right BPS end,
\begin{equation}
 u_1(w)
 =
 w^{-1}+O(w),
 \qquad
 u_2(w)
 =
 \frac{w^2}{3}+O(w^4).
 \label{8-78}
\end{equation}
In the limiting right inner profile, set $\epsilon=0$ before taking
the overlap limit.
Then
\begin{equation}
 w
 =
 k\arcsin\frac{\delta}{a^2}
 =
 \frac{k\delta}{a^2}
 +
 O\!\left(
 \frac{\delta^3}{a^6}
 \right).
 \label{8-79}
\end{equation}
For
$q=C_1u_1+C_2u_2$,
comparison with the two BPS branches gives
\begin{equation}
 \frac{B_{\rm out}}{A_{\rm out}}
 =
 \frac{k^3}{3}
 \frac{C_2}{C_1}
 \delta^3
 +
 O(\delta^4).
 \label{8-80}
\end{equation}
This cubic ratio is a local overlap statement.
The physical two-boundary Hessian also depends on the global Dirichlet
data and on the displacement
$\epsilon(\delta)$ of the background.

Let $q_L$ and $q_R$ be the two asymptotic boundary values.
The on-shell Dirichlet action defines
\begin{equation}
 I_{q,\rm on}^{(2)}
 =
 \frac{\cA}{2}
 (q_L,q_R)
 K
 \binom{q_L}{q_R},
 \qquad
 \Delta
 =
 u_{1R}u_{2L}
 -
 u_{2R}u_{1L},
 \label{8-81}
\end{equation}
where
\begin{equation}
 w_R=\epsilon,
 \qquad
 w_L=\vartheta+\epsilon .
\end{equation}
The physical outer-boundary Hessian is
\begin{equation}
 \cK^\partial_{\bm\nu,B}
 =
 \cA K .
\end{equation}
It is distinct from the matching-surface Hessian
$\cK^\Sigma_{\bm\nu,B}$ of Appendix~\ref{sec:finite-radius-boundary}.

Primes below denote derivatives with respect to $w$.
With the common factor $\cA$ removed, the outward canonical momenta are
$2k\delta q_w(w_L)$ and
$-2k\delta q_w(w_R)$.
The exact Dirichlet-to-Neumann matrix is therefore
\begin{align}
 K_{LL}
 &=
 \frac{2k\delta}{\Delta}
 \left[
 -u_{2R}u_{1L}'
 +
 u_{1R}u_{2L}'
 \right],
 \nonumber\\
 K_{RR}
 &=
 -\frac{2k\delta}{\Delta}
 \left[
 u_{1R}'u_{2L}
 -
 u_{1L}u_{2R}'
 \right],
 \nonumber\\
 K_{LR}
 &=
 K_{RL}
 =
 -\frac{2k\delta}{\Delta}.
 \label{8-82}
\end{align}
For every permitted $\delta>0$,
$0<w_R<w_L<\pi$ and the integrand in
Eq.~\eqref{8-76} is strictly positive.
Hence $K$ is positive definite.
Equivalently, a zero eigenvector of $K$ would give a nonzero solution
with vanishing outward momenta at both ends; integration of the field
equation against that solution forces it to vanish.

The exact matrix now gives the near-BPS boundary scaling.
Define
\begin{equation}
 v_0
 =
 1-\vartheta\cot\vartheta,
 \qquad
 v_1
 =
 -\cot\vartheta
 +
 \vartheta\csc^2\vartheta .
\end{equation}
For the subcritical branch,
$0<\vartheta<\pi$, and
$v_0,v_1>0$.
Expanding Eq.~\eqref{8-82} at fixed right asymptotic data gives
\begin{align}
 K_{LL}
 &=
 a_1\delta
 +
 O(\delta^2),
 &
 a_1
 &=
 2k\frac{v_1}{v_0}
 >0,
 \nonumber\\
 K_{RR}
 &=
 r_0
 +
 \frac{kh}{\mathfrak c_{\bm\nu}}
 \delta^2
 +
 O(\delta^3),
 &
 r_0
 &=
 \frac{
 2k\mathfrak c_{\bm\nu}
 }{h}
 >0,
 \nonumber\\
 K_{LR}
 &=
 b_2\delta^2
 +
 O(\delta^3),
 &
 b_2
 &=
 -\frac{
 2kh
 }{
 \mathfrak c_{\bm\nu}v_0
 }
 <0 .
 \label{8-83}
\end{align}
Consequently
\begin{equation}
 \det K
 =
 a_1r_0\,\delta
 +
 O(\delta^2).
\end{equation}
The finite-radius variation of the on-shell quadratic action begins as
\begin{equation}
 I_{q,\rm on}^{(2)}(\delta)
 -
 I_{q,\rm on}^{(2)}(0)
 =
 \frac{\cA a_1}{2}
 \delta q_L^2
 +
 O(\delta^2|q|^2).
 \label{8-84}
\end{equation}
The two eigenvalues of the physical Hessian are
\begin{equation}
 \cA a_1\delta+O(\delta^2),
 \qquad
 \cA r_0+O(\delta^2).
\end{equation}
Thus one boundary combination becomes soft linearly in $\delta$, while
the orthogonal combination retains a finite restoring term.

The covariance follows by integrating the two boundary values with the
fixed Lebesgue measure and linear sources.
Since
\begin{equation}
 G
 =
 \cA^{-1}K^{-1},
\end{equation}
the exact cross-boundary response is
\begin{align}
 G_{LR}(\delta)
 &=
 \frac1{\cA}
 (K^{-1})_{LR}
 \nonumber\\
 &=
 \frac{
 \delta h^2
 }{
 2\cA k\mathfrak c_{\bm\nu}^{\,2}
 }
 \frac{
 \sin^2(\vartheta+\epsilon)
 }{
 \vartheta
 -
 \sin\vartheta
 \cos(\vartheta+2\epsilon)
 }
 \nonumber\\
 &=
 g_{1,\bm\nu}\delta
 +
 O(\delta^2),
 \label{8-85}\\
 g_{1,\bm\nu}
 &=
 \frac{
 h^2\sin^2\vartheta
 }{
 2\cA k\mathfrak c_{\bm\nu}^{\,2}
 \left[
 \vartheta-\sin\vartheta\cos\vartheta
 \right]
 }
 >0 .
 \label{8-86}
\end{align}
The same inverse may be written as
\begin{equation}
 (K^{-1})_{LR}
 =
 -\frac{1}{
 2k\delta
 \left(
 u_{1R}'u_{2L}'
 -
 u_{1L}'u_{2R}'
 \right)
 }.
\end{equation}
The denominator in Eq.~\eqref{8-86} is positive because
\begin{equation}
 \vartheta-\sin\vartheta\cos\vartheta
 =
 2\int_0^\vartheta\sin^2t\,\dd t
 >0 .
\end{equation}
The scaling is now transparent:
$K_{LR}=O(\delta^2)$,
whereas
$\det K=O(\delta)$.
The inverse therefore converts the quadratic off-diagonal Hessian entry
into the linear covariance
$G_{LR}=O(\delta)$.

The exact inverse also fixes the first subleading term.
Define
\begin{equation}
 v_2
 =
 2v_0\csc^2\vartheta,
 \qquad
 G_{LR,\bm\nu}
 =
 g_{1,\bm\nu}\delta
 +
 g_{2,\bm\nu}\delta^2
 +
 O(\delta^3),
 \qquad
 \frac{g_{2,\bm\nu}}{g_{1,\bm\nu}}
 =
 -\frac{h}{\mathfrak c_{\bm\nu}}
 \frac{v_2}{v_1}.
 \label{8-87}
\end{equation}
Indeed, the logarithmic derivative of the $\epsilon$-dependent factor
in Eq.~\eqref{8-85} at $\epsilon=0$ is
\begin{equation}
 2\cot\vartheta
 -
 \frac{
 2\sin^2\vartheta
 }{
 \vartheta-\sin\vartheta\cos\vartheta
 }
 =
 -\frac{v_2}{v_1}.
 \label{8-88}
\end{equation}
Since
\begin{equation}
 \epsilon'(0)
 =
 \frac{h}{\mathfrak c_{\bm\nu}},
\end{equation}
Eq.~\eqref{8-87} follows.
For $0<\vartheta<\pi$,
$v_0,v_1,v_2$ are positive, so
\begin{equation}
 g_{2,\bm\nu}<0.
\end{equation}
The normalized cross-boundary response therefore bends downward at its
first sub-leading order.

The unnormalized source coefficient requires the complete fixed-saddle
weight.
Let
\begin{equation}
 Z_{\bm\nu}[0;\delta]
 \longrightarrow
 Z_{\bm\nu,0}
\end{equation}
as $\delta\to0^+$ after exact modes, prescribed insertions, fermions,
ghosts, boundary normalization, and the remaining fluctuation factors
have been included.
Then Eq.~\eqref{8-85} and the centered-source identity of
Appendix~\ref{sec:near-bps-problem} give
\begin{equation}
 C_{\bm\nu}^{LR}(0)
 =
 0,
 \qquad
 \left.
 \partial_\delta
 C_{\bm\nu}^{LR}
 \right|_{0^+}
 =
 Z_{\bm\nu,0}
 g_{1,\bm\nu}.
 \label{8-89}
\end{equation}
The isolated two-boundary bosonic Gaussian does not possess this finite
endpoint weight: its soft eigenvalue contributes a factor
$\delta^{-1/2}$.
The finite quantity $Z_{\bm\nu,0}$ is therefore a property of the
complete fixed-charge measure.
If the prescribed insertions leave an exact fermion zero mode
unsaturated, the corresponding BPS sector weight vanishes and so does
the slope in Eq.~\eqref{8-89}.

All $p-1$ orthonormal transverse directions obey the same radial
equation and carry the same coefficient with the common boundary
normalization used here.
In the original $\phi_i$ coordinates their projector is
\begin{equation}
 EE^T
 =
 \id-nn^T .
\end{equation}

Finally, the result depends on the boundary state exactly as described
in the main text.
Adding
\begin{equation}
 \frac{\cA}{2}
 \left(
 \mu_Lq_L^2+\mu_Rq_R^2
 \right),
 \qquad
 \mu_L,\mu_R\geq0,
\end{equation}
replaces
\begin{equation}
 K
 \longrightarrow
 K+\diag(\mu_L,\mu_R).
\end{equation}
The cross covariance becomes
\begin{equation}
 G_{LR}^{(\mu)}
 =
 \begin{cases}
 -\dfrac{
 b_2\delta
 }{
 \cA a_1(r_0+\mu_R)
 }
 +O(\delta^2),
 &
 \mu_L=0,
 \\[6pt]
 -\dfrac{
 b_2\delta^2
 }{
 \cA\mu_L(r_0+\mu_R)
 }
 +O(\delta^3),
 &
 \mu_L>0 .
 \end{cases}
 \label{8-90}
\end{equation}
A positive quadratic term in the soft left-boundary direction lifts its
$O(\delta)$ eigenvalue to $O(1)$ and changes the cross response from
$O(\delta)$ to $O(\delta^2)$.
I use $\mu_L=\mu_R=0$ in the remainder of the calculation.
For this boundary state the bare two-variable Gaussian carries the
factor $\delta^{-1/2}$, while the physical finite-BPS matching concerns
the complete fixed-charge weight.
The finite-radius transverse quadratic form itself remains strictly
positive for every $\delta>0$ on the branch considered here.

\subsection{Cubic and quartic corrections}

The linearly soft boundary eigenvalue amplifies interaction corrections.
I therefore compute the cubic and quartic vertices of the effective soft
boundary action and use them below to determine the range of validity of
the Gaussian approximation.

At the stationary product point define
\begin{equation}
 \omega_i
 =
 \frac{\beta^2}{\beta_i^2},
 \qquad
 t_i
 =
 2\beta_iE_{iu}q^u .
\end{equation}
Equations~\eqref{8-60}, \eqref{8-63}, and \eqref{8-67} give
\begin{equation}
 \frac{X_i}{V}=\omega_i,
 \qquad
 \sum_i\omega_i=1,
 \qquad
 \omega_i=\frac{\beta^2}{\beta_i^2}.
\end{equation}
The exact transverse charge potential on the fixed background is
therefore
\begin{equation}
 \frac{V(\bar s+q)}{V(\bar s)}
 =
 \sum_i\omega_i\e^{t_i},
 \qquad
 \sum_i\omega_it_i=0,
 \qquad
 \sum_i\omega_it_i^2
 =
 4\beta^2|q|^2 .
 \label{8-91}
\end{equation}
The transverse kinetic metric is flat in these coordinates.
After subtracting the background action,
\begin{equation}
 \frac{\Delta I_{\perp}}{\cA\delta}
 =
 \int\dd y
 \left[
 |q_y|^2
 +
 3\csc^2w
 \left(
 \sum_i\omega_i\e^{t_i}-1
 \right)
 \right].
 \label{8-92}
\end{equation}
The quadratic term reproduces Eq.~\eqref{8-76}.
Expanding the exponential gives the direct cubic and quartic terms
\begin{equation}
 \frac{
 I_{\perp}^{(3)}
 +
 I_{\perp,\rm dir}^{(4)}
 }{
 \cA\delta
 }
 =
 \int\dd y\,\csc^2w
 \left[
 4\beta^2
 \sum_i
 \beta_i(E_i\cdot q)^3
 +
 2\beta^2
 \sum_i
 \beta_i^2(E_i\cdot q)^4
 \right].
 \label{8-93}
\end{equation}
Here
$E_i\cdot q\equiv E_{iu}q^u$.
Relaxation of the remaining physical fields first contributes at
quartic order.

Choose a unit transverse direction $e$ and denote the left soft
boundary amplitude by $z$.
At the BPS endpoint the stationary value of the regular right-boundary
combination is zero.
The leading soft profile is therefore
\begin{equation}
 q(w)
 =
 z e\,f(w),
 \qquad
 f(w)
 =
 \frac{1-w\cot w}{v_0},
 \qquad
 0\leq w\leq\vartheta,
 \qquad
 0\leq f(w)\leq1 .
 \label{8-94}
\end{equation}
The last inequality follows from monotonicity of
$1-w\cot w$ on $0<w<\pi$.
Define
\begin{equation}
 J_n
 =
 \int_0^\vartheta
 \csc^2w\,f(w)^n\dd w,
 \qquad
 T_n(e)
 =
 \sum_i
 \beta_i^{\,n-2}
 (E_i\cdot e)^n .
 \label{8-95}
\end{equation}
Since
$f(w)=O(w^2)$
at the right endpoint, the integrand defining $J_n$ is
$O(w^{2n-2})$.
In particular, $J_3$ and $J_4$ are finite.

Evaluating Eq.~\eqref{8-93} on the stationary linear profile gives
\begin{equation}
 \frac{I_{\rm soft}}{\cA\delta}
 =
 \frac{a_1z^2}{2}
 +
 \frac{\lambda_3z^3}{6}
 +
 \frac{\lambda_{4,\rm dir}z^4}{24}
 +
 \cdots,
 \qquad
 \lambda_3
 =
 8kT_3J_3,
 \qquad
 \lambda_{4,\rm dir}
 =
 16kT_4J_4 .
 \label{8-96}
\end{equation}
The $O(z^2)$ correction to the stationary profile contributes first at
$O(z^4)$, so the linear profile determines $\lambda_3$ exactly.
The quartic coefficient also receives exchange contributions from the
regular complementary fields.

Collect those regular coordinates in $u$ and write their quadratic
coupling to the soft mode as
\begin{equation}
 \frac12u^T\mathcal H_cu
 +
 \frac{z^2}{2}\mathcal J^Tu .
 \label{8-97}
\end{equation}
Stationarity gives
\[
 u
 =
 -\frac{z^2}{2}
 \mathcal H_c^{-1}\mathcal J ,
\]
and hence
\begin{equation}
 \lambda_4
 =
 \lambda_{4,\rm dir}
 -
 3\mathcal J^T
 \mathcal H_c^{-1}
 \mathcal J,
 \qquad
 |\lambda_4|
 \leq
 |\lambda_{4,\rm dir}|
 +
 3\|\mathcal H_c^{-1}\|
 \|\mathcal J\|^2 .
 \label{8-98}
\end{equation}
Equation~\eqref{8-98} is the Schur-complement correction to the quartic
vertex.
For a positive complementary quadratic form it lowers the direct
quartic coefficient.
For a contour-rotated gravitational coordinate, the coordinate and its
source are continued together.

For a one-dimensional contour-rotated gravitational mode,
\begin{equation}
-\frac12 H u^2+\frac12\mathcal J z^2u
\end{equation}
gives
\begin{equation}
\Delta I_{\rm eff}
=
+\frac{\mathcal J^2z^4}{8H}
\end{equation}
after stationary elimination.
I use this common contour continuation for the gravitational exchange
channels below.

I now evaluate the complementary relaxation explicitly in the
homogeneous product model at $d=3$ and $\Lambda=0$.
A transverse boundary displacement changes the individual scalar
first-integral constants.
The Hamiltonian constraint then shifts the minimal radius.
I fix the charges, the induced outer metrics, and the longitudinal
dilaton/saxion boundary data.
The left transverse displacement is $ze$.
The effective soft action is obtained after the regular right-boundary
combination and the interior fields have relaxed.
At the order required for the limiting cubic and quartic vertices, its
stationary right transverse value may be set to zero.
The two radial halves are joined smoothly, so their oriented
matching-surface terms cancel.

Let $\widehat\delta$ denote the displaced squared minimal radius and
introduce harmonic radial time $\xi$, increasing from the right end to
the left end, through
\[
 |\dd\xi|
 =
 \frac{\dd\tau}{a^3}.
\]
With $X=\log a$, the scalar and gravitational first integrals separate:
\begin{equation}
 \dot\phi_i^{\,2}
 -
 \nu_i^2\e^{2\beta_i\phi_i}
 =
 -r_i^2,
 \qquad
 \dot X^2
 =
 \e^{4X}
 -
 \widehat\delta^2,
 \qquad
 \sum_i r_i^2
 =
 12\widehat\delta^2 .
 \label{8-99}
\end{equation}
Here a dot denotes $\dd/\dd\xi$.
The branch connected continuously to the reference saddle has the exact
solution
\begin{equation}
 \begin{aligned}
 a^2
 &=
 \widehat\delta
 \csc(2\widehat\delta\xi),
 &
 0<\xi<
 \frac{\pi}{2\widehat\delta},
 \\
 \e^{-\beta_i\phi_i}
 &=
 \frac{|\nu_i|}{r_i}
 \sin(\beta_ir_i\xi+\epsilon_i),
 &
 r_i
 &=
 \frac{2k\widehat\delta}{\beta_i}\sqrt{b_i},
 \qquad
 \sum_i\omega_i b_i=1 .
 \end{aligned}
 \label{8-100}
\end{equation}
The final relation is equivalent to the Hamiltonian constraint because
$k^2=3\beta^2$ and
$\omega_i=\beta^2/\beta_i^2$.
The metric reflects about the middle of the harmonic interval, while
the phases $\epsilon_i$ impose the scalar boundary data.

Set
\[
 x_i
 =
 \beta_i(E_i\cdot e),
 \qquad
 t
 =
 \log\frac{\widehat\delta}{\delta},
 \qquad
 \eta
 =
 \frac{\delta h}{\mathfrak c_{\bm\nu}} .
\]
The orthogonality relations imply
\[
 \sum_i\omega_i x_i=0,
 \qquad
 \sum_i\omega_i x_i^2=\beta^2 .
\]
The exact right phases and the left Dirichlet conditions are
\begin{equation}
 \epsilon_i
 =
 \arcsin\!\left(
 \eta\e^t\sqrt{b_i}
 \right),
 \qquad
 \frac{
 \sin(\vartheta\sqrt{b_i}+\epsilon_i)
 }{
 \e^t\sqrt{b_i}
 }
 =
 \sin(\vartheta+\epsilon)\e^{-x_i z},
 \qquad
 \epsilon=\arcsin\eta .
 \label{8-101}
\end{equation}
These equations impose the physical boundary data before the
small-radius limit is taken.
The variation of the throat radius begins at $O(z^2)$.
Since the complete inner action carries the prefactor $\cA\delta$,
eliminating this radius response first changes the effective action at
$O(\cA\delta z^4)$.

The exact renormalized action at fixed outer data is
\begin{equation}
 \frac{S^{\rm ren}}{\cA}
 =
 \sum_i
 \frac{r_i}{\beta_i}
 \left[
 \cot\epsilon_i
 -
 \cot(\vartheta\sqrt{b_i}+\epsilon_i)
 \right].
 \label{8-102}
\end{equation}
Flat reference subtraction removes the outer gravitational divergence.
The Hamiltonian constraint cancels the remaining constant bulk terms.
At
$b_i=1$ and $t=0$,
Eq.~\eqref{8-102} reduces to the background result
Eq.~\eqref{8-52}.

The $\delta\to0$ boundary map takes a particularly simple form.
Define
\begin{equation}
 \begin{aligned}
 \mathcal F(b)
 &=
 \log
 \frac{
 \sin(\vartheta\sqrt b)
 }{
 \sqrt b\,\sin\vartheta
 },
 &
 \mathcal F(b_i)
 &=
 t-x_i z,
 \qquad
 \sum_i\omega_i b_i=1,
 \\
 F_1
 &=
 \mathcal F'(1)
 =
 -\frac{v_0}{2},
 &
 F_2
 &=
 \mathcal F''(1)
 =
 \frac{
 2-\vartheta\cot\vartheta
 -\vartheta^2\csc^2\vartheta
 }{4}.
 \end{aligned}
 \label{8-103}
\end{equation}
Since $0<\vartheta<\pi$, one has $v_0>0$ and therefore $F_1<0$.
The Jacobian of the boundary equations is invertible.
The implicit function theorem then gives a unique analytic solution for
$b_i(t,z)$ near the reference saddle.

Expanding
$\mathcal F(b_i)=t-x_i z$
together with
$\sum_i\omega_i b_i=1$
gives
\begin{equation}
 \begin{aligned}
 t
 &=
 t_2z^2+O(z^3),
 &
 t_2
 &=
 \frac{\beta^2F_2}{2F_1^2},
 \\
 b_i
 &=
 1
 -
 \frac{x_i}{F_1}z
 +
 \frac{
 F_2(\beta^2-x_i^2)
 }{
 2F_1^3
 }
 z^2
 +
 O(z^3).
 \end{aligned}
 \label{8-104}
\end{equation}
Thus the minimal radius responds only at second order in a transverse
boundary displacement.
This response contributes to the quartic vertex and leaves the
quadratic Hessian unchanged.

I next compute the scalar exchange at fixed harmonic interval.
The regular scalar complement is governed by
\[
 L
 =
 -\partial_w^2
 +
 2\csc^2w
\]
on $0<w<\vartheta$, with the regular $w^2$ branch at the right endpoint
and Dirichlet data at the left endpoint.
Set
\[
 P(w)
 =
 1-w\cot w,
 \qquad
 v(w)
 =
 v_0\cot w
 -
 \cot\vartheta\,P(w).
\]
The Dirichlet Green function is
\begin{equation}
 \begin{aligned}
 G_D(w,s)
 &=
 \frac{
 P(w_<)v(w_>)
 }{v_0},
 \qquad
 w_<
 =
 \min(w,s),
 \qquad
 w_>
 =
 \max(w,s),
 \\
 \|G_D\|_{L^2\to L^2}
 &\leq
 \frac1{
 2+\pi^2/\vartheta^2
 } .
 \end{aligned}
 \label{8-105}
\end{equation}
The Wronskian
\[
 P v'-P'v=-v_0
\]
fixes the derivative jump.
Since
$L\geq-\partial_w^2+2$
on the Dirichlet interval, the Poincar\'e inequality gives the displayed
resolvent bound.

Define
\[
 j_f(w)
 =
 \csc^2w\,f(w)^2 .
\]
The second-order source solution is
\begin{equation}
 \begin{aligned}
 \mathfrak r(w)
 &=
 -2
 \int_0^\vartheta
 G_D(w,s)j_f(s)\dd s
 \\
 &=
 \frac{
 -1+w^2\csc^2w-P(w)
 +4F_2P(w)/v_0
 }{
 2v_0^2
 },
 \\
 \mathcal I_f
 &=
 \int_0^\vartheta
 \int_0^\vartheta
 j_f(w)
 G_D(w,s)
 j_f(s)
 \dd w\,\dd s
 \\
 &=
 -\frac12
 \int_0^\vartheta
 j_f(w)\mathfrak r(w)\dd w
 >0 .
 \end{aligned}
 \label{8-106}
\end{equation}
Direct differentiation gives
\[
 L\mathfrak r=-2j_f,
 \qquad
 \mathfrak r(0)=\mathfrak r(\vartheta)=0.
\]
The apparent singular terms in the explicit expression cancel at
$w=0$.
In the product variables, each regular scalar channel has this same
fixed-interval Green function.
Its second-order displacement is
\[
 \delta^{(2)}\phi_i
 =
 \beta_i(E_i\cdot e)^2
 \mathfrak r(w)z^2 .
\]
Substitution into the Schur-complement formula
Eq.~\eqref{8-98} gives
\[
 \Delta\lambda_{4,\rm scalar}
 =
 -96kT_4\mathcal I_f,
 \qquad
 \mathcal J_{\rm scalar}^T
 \mathcal H_{\rm scalar}^{-1}
 \mathcal J_{\rm scalar}
 =
 32kT_4\mathcal I_f .
\]

The remaining relaxation is the change of the harmonic interval,
equivalently the throat radius.
For fixed $t$ solve the scalar boundary equations first, so that
$b_i=b_i(t,z)$ is determined by
$\mathcal F(b_i)=t-x_i z$.
Define
\[
 g(b)
 =
 \sqrt b\,
 \cot(\vartheta\sqrt b).
\]
Up to an additive constant, the leading constrained action is
\begin{equation}
 \mathcal S(t,z)
 =
 -\frac6k
 \e^t
 \sum_i\omega_i g(b_i)
 +
 3\pi\e^t
 \left(
 1-\sum_i\omega_i b_i
 \right).
 \label{8-107}
\end{equation}
Differentiating along the implicitly determined functions $b_i(t,z)$
gives
\[
 \partial_t\mathcal S
 =
 3\pi\e^t
 \left(
 1-\sum_i\omega_i b_i
 \right).
\]
Thus stationarity in $t$ is exactly the Hamiltonian constraint.

At the reference saddle,
\begin{equation}
 H_t
 \equiv
 \left.
 \partial_t^2\mathcal S
 \right|_0
 =
 -\frac{3\pi}{F_1}
 =
 \frac{6\pi}{v_0}
 >0,
 \qquad
 J_t
 \equiv
 \left.
 \partial_t\partial_z^2\mathcal S
 \right|_0
 =
 \frac{
 3\pi\beta^2F_2
 }{
 F_1^3
 } .
 \label{8-108}
\end{equation}
The scalar variables have already relaxed in these derivatives.
Eliminating $t$ therefore gives the quartic correction
\[
 -\frac{3J_t^2}{H_t}
 =
 -\frac{
 288\pi\beta^4F_2^2
 }{
 v_0^5
 },
\]
and reproduces the radius shift $t_2$ in Eq.~\eqref{8-104}.

Combining the direct term, scalar exchange, and radius exchange gives
the complete quartic coefficient of the homogeneous soft direction:
\begin{equation}
 \lambda_4
 =
 16kT_4J_4
 -
 96kT_4\mathcal I_f
 -
 \frac{
 288\pi\beta^4F_2^2
 }{
 v_0^5
 },
 \qquad
 \mathcal J^T\mathcal H_c^{-1}\mathcal J
 =
 32kT_4\mathcal I_f
 +
 \frac{
 96\pi\beta^4F_2^2
 }{
 v_0^5
 }.
 \label{8-109}
\end{equation}
The first exchange term comes from the regular scalar complement.
The second comes from the Hamiltonian constraint through the radius
response.
Both enter at
$O(\cA\delta z^4)$.
Equation~\eqref{8-109} is the quartic vertex used in the validity
analysis below.

The corresponding fields through second order are
\begin{equation}
 \begin{aligned}
 \phi_i(w;z)-\bar\phi_i(w)
 ={}&
 z(E_i\cdot e)f(w)
 \\
 &+
 z^2
 \left[
 \beta_i(E_i\cdot e)^2
 \mathfrak r(w)
 +
 \frac{t_2}{\beta_i}
 \bigl(1-f(w)\bigr)
 \right]
 +
 O(z^3),
 \\
 \widehat\delta
 &=
 \delta
 \left[
 1+t_2z^2+O(z^3)
 \right].
 \end{aligned}
 \label{8-110}
\end{equation}
The first term is the soft transverse profile.
The $\mathfrak r$ term is the regular scalar relaxation.
The term proportional to $1-f$ is the longitudinal response induced by
the change of radial scale.
At the left boundary,
$f(\vartheta)=1$ and
$\mathfrak r(\vartheta)=0$, so the prescribed displacement is purely
transverse and has no longitudinal component.
Equation~\eqref{8-101} fixes the physical right boundary data before
the inner limit is taken; Eq.~\eqref{8-110} is its expansion at fixed
rescaled coordinate.
The metric remains reflection symmetric while the scalar profile
carries the asymmetric boundary data.

\subsection{Range of the quadratic approximation}

The interaction vertices of the preceding subsection determine the
lower end of the Gaussian regime.
I first check them directly against the constrained tree action and
then derive uniform bounds for all transverse soft directions.

After solving the boundary equations for $b_i$ and imposing
$\sum_i\omega_i b_i=1$, the renormalized action relative to the
undeformed solution is
\begin{equation}
 \frac{\Delta S^{\rm ren}}{\cA\delta}
 =
 -\frac6k
 \left[
 \e^t\sum_i\omega_i g(b_i)
 -
 \cot\vartheta
 \right]
 +
 O(\delta).
 \label{8-111}
\end{equation}
Here $t=t(z)$ and $b_i=b_i(z)$ are the constrained solutions of
Eq.~\eqref{8-103}.
The second, third, and fourth derivatives of Eq.~\eqref{8-111} at
$z=0$ reproduce $a_1$, $\lambda_3$, and the complete quartic
coefficient in Eq.~\eqref{8-109}.
A direct power-series solution of the boundary equations gives the same
coefficients for the equal two-factor model and for the unequal
couplings $(0.6,0.7,0.9)$.\footnote{I checked the resulting algebraic
residuals below $10^{-45}$ at 65-digit precision.}
Independent inversion of the exact finite-radius boundary matrix
reproduces Eq.~\eqref{8-85}.

As a calibration case, take two equal factors,
\begin{equation}
 \beta_i=\sqrt2\,\beta,
 \qquad
 E=\frac1{\sqrt2}(1,-1)^T .
\end{equation}
Interchange of the two factors sends $z\to-z$ and therefore sets
$\lambda_3=0$.
At $k=1/2$,
\begin{align}
a_1&=\frac{\pi}{2},
\qquad
\lambda_3=0,
\nonumber\\
\lambda_4
&=\frac{\pi(\pi^2-10)}{48}
=-0.00853437199\ldots,
\label{8-112}\\
\lambda_{4,\rm dir}
&=0.10409776147\ldots,
\nonumber\\
\Delta\lambda_{4,\rm scalar}
&=-0.02684160432\ldots,
\nonumber\\
\Delta\lambda_{4,\rm length}
&=-0.08579052914\ldots,
\nonumber\\
t_2
&=-\frac{\pi^2-8}{96}.
\nonumber
\end{align}
The scalar and interval-length exchange terms outweigh the positive
direct quartic term and make the net quartic coefficient negative.
The quadratic coefficient remains positive.
The local soft potential therefore becomes less confining at quartic
order in this example.

Set
\begin{equation}
 \Lambda_*
 =
 \cA\delta .
\end{equation}
For one soft coordinate with fixed Lebesgue measure, use the effective
tree action
\begin{equation}
 \Lambda_*
 \left(
 \frac{a_1z^2}{2}
 +
 \frac{\lambda_3z^3}{6}
 +
 \frac{\lambda_4z^4}{24}
 \right).
\end{equation}
Expanding around the Gaussian measure gives
\begin{align}
 \frac{Z_{\rm soft}}{Z_{\rm G}}
 &=
 1
 +
 \frac1{\Lambda_*}
 \left[
 -\frac{\lambda_4}{8a_1^2}
 +
 \frac{5\lambda_3^2}{24a_1^3}
 \right]
 +
 O(\Lambda_*^{-2}),
 \nonumber\\
 \langle z^2\rangle_c
 &=
 \frac1{\Lambda_*a_1}
 \left[
 1
 +
 \frac1{\Lambda_*}
 \left(
 -\frac{\lambda_4}{2a_1^2}
 +
 \frac{\lambda_3^2}{a_1^3}
 \right)
 +
 O(\Lambda_*^{-2})
 \right].
 \label{8-113}
\end{align}
The cubic vertex shifts the mean already at first subleading order,
\begin{equation}
 \langle z\rangle
 =
 -\frac{\lambda_3}{2\Lambda_*a_1^2}
 +
 O(\Lambda_*^{-2}),
\end{equation}
and Eq.~\eqref{8-113} uses the connected variance after subtracting
this shift.
The quartic and cubic corrections remain perturbative when
\begin{equation}
 \frac{|\lambda_4|}
 {\Lambda_*a_1^2}
 \ll1,
 \qquad
 \frac{\lambda_3^2}
 {\Lambda_*a_1^3}
 \ll1.
 \label{8-114}
\end{equation}

I now keep all $d_\perp=p-1$ soft transverse directions.
The cubic radial coefficient is
\begin{equation}
 c_3=8kJ_3.
\end{equation}
The quartic coefficient receives the direct contribution and the
regular scalar exchange,
\begin{equation}
 c_4=16kJ_4-96k\mathcal I_f,
\end{equation}
while the radius response imposed by the Hamiltonian constraint gives
\begin{equation}
 c_g=-\frac{288\pi\beta^4F_2^2}{v_0^5}.
\end{equation}
These coefficients define the cubic and quartic vertex tensors,
\begin{equation}
 V_{abc}
 =
 c_3\sum_i
 \beta_i E_{ia}E_{ib}E_{ic},
\end{equation}
and
\begin{align}
 W_{abcd}
 ={}&
 c_4\sum_i
 \beta_i^2E_{ia}E_{ib}E_{ic}E_{id}
 \nonumber\\
 &+
 \frac{c_g}{3}
 \left(
 \delta_{ab}\delta_{cd}
 +\delta_{ac}\delta_{bd}
 +\delta_{ad}\delta_{bc}
 \right).
 \label{8-115}
\end{align}
The first term in $W_{abcd}$ contains the direct quartic vertex and the
regular scalar exchange. 
The second term is the isotropic radius response.

The resulting soft boundary action is
\begin{equation}
 \frac{I_{\rm soft}}{\cA\delta}
 =
 \frac{a_1}{2}z_az_a
 +\frac{1}{6}V_{abc}z_az_bz_c
 +\frac{1}{24}W_{abcd}z_az_bz_cz_d
 +\cdots .
\end{equation}
For a unit transverse vector $e_a$,
\begin{equation}
 \lambda_3(e)=V_{abc}e_ae_be_c,
 \qquad
 \lambda_4(e)=W_{abcd}e_ae_be_ce_d,
\end{equation}
which reproduces the directional cubic and quartic coefficients derived
in the preceding subsection.

Set
\begin{equation}
 S_\beta
 =
 \sum_i\beta_i^2,
 \qquad
 N_3
 =
 S_\beta-(3p-2)\beta^2,
 \qquad
 R_3
 =
 S_\beta-p^2\beta^2,
 \qquad
 R_4
 =
 S_\beta-(2p-1)\beta^2.
 \label{8-116}
\end{equation}
Since
\begin{equation}
 \left(\sum_i\beta_i^2\right)
 \left(\sum_i\beta_i^{-2}\right)
 \geq p^2,
\end{equation}
one has
$S_\beta\geq p^2\beta^2$.
Thus $N_3$, $R_3$, and $R_4$ are nonnegative for $p\geq2$.
Using
\begin{equation}
 EE^T=\id-nn^T
\end{equation}
gives the exact contractions
\begin{equation}
 V_{abc}V_{abc}
 =
 c_3^2N_3,
 \qquad
 V_{abb}V_{acc}
 =
 c_3^2R_3,
 \qquad
 W_{aabb}
 =
 c_4R_4
 +
 \frac{c_g}{3}
 d_\perp(d_\perp+2).
 \label{8-117}
\end{equation}
Repeated transverse indices are summed.

Gaussian Wick contraction then gives the first correction to the
multidimensional soft partition function,
\begin{equation}
 \frac{Z_{\rm soft}}{Z_{\rm G}}
 =
 1
 +
 \frac1{\Lambda_*}
 \left[
 -\frac{
 c_4R_4
 +
 c_gd_\perp(d_\perp+2)/3
 }{
 8a_1^2
 }
 +
 \frac{c_3^2N_3}{12a_1^3}
 +
 \frac{c_3^2R_3}{8a_1^3}
 \right]
 +
 O(\Lambda_*^{-2}).
 \label{8-118}
\end{equation}
The corresponding connected covariance is
\begin{equation}
 \langle z_a z_b\rangle_c
 =
 \frac1{\Lambda_*a_1}
 \left[
 \delta_{ab}
 +
 \frac1{\Lambda_*}
 \left(
 -\frac{W_{abcc}}{2a_1^2}
 +
 \frac{
 V_{acd}V_{bcd}
 +
 V_{abc}V_{cdd}
 }{
 2a_1^3
 }
 \right)
 +
 O(\Lambda_*^{-2})
 \right].
 \label{8-119}
\end{equation}
For $d_\perp=1$, Eqs.~\eqref{8-118} and \eqref{8-119} reduce to
Eq.~\eqref{8-113}.
The connected contraction in Eq.~\eqref{8-119} includes the subtraction
of the cubic mean shift.

Triangle and Cauchy--Schwarz inequalities give sufficient uniform
bounds on these corrections.
The partition-function correction is bounded by
$\mathcal C_Z/\Lambda_*$ and the operator norm of the relative
covariance correction by $\mathcal C_V/\Lambda_*$, with
\begin{equation}
 \begin{aligned}
 \mathcal C_Z
 &=
 \frac{
 |c_4|R_4
 +
 |c_g|d_\perp(d_\perp+2)/3
 }{
 8a_1^2
 }
 +
 \frac{c_3^2N_3}{12a_1^3}
 +
 \frac{c_3^2R_3}{8a_1^3},
 \\
 \mathcal C_V
 &=
 \frac{
 |c_4|R_4
 +
 |c_g|(d_\perp+2)/3
 }{
 2a_1^2
 }
 +
 \frac{
 c_3^2
 \left(
 N_3+\sqrt{N_3R_3}
 \right)
 }{
 2a_1^3
 }.
 \end{aligned}
 \label{8-120}
\end{equation}

A separate bound controls the field range over which the exponential
potential may be truncated.
Define
\begin{equation}
 b
 =
 \max_i
 \beta_i\|E_i\|,
 \qquad
 t_{\max}
 =
 2b|q|.
\end{equation}
Then $|t_i|\leq t_{\max}$, and Taylor's integral remainder gives
\begin{equation}
 \left|
 \sum_i\omega_i
 \left(
 \e^{t_i}
 -
 1-t_i-\frac12t_i^2
 \right)
 \right|
 \leq
 \frac{
 t_{\max}\e^{t_{\max}}
 }6
 \sum_i\omega_it_i^2 .
 \label{8-121}
\end{equation}
For the isotropic Gaussian soft block,
\begin{equation}
 \langle|z|^2\rangle_{\rm G}
 =
 \frac{
 d_\perp
 }{
 \cA a_1\delta
 }.
\end{equation}
Thus
\begin{equation}
 t_{\max,\rm rms}^2
 \lesssim
 \frac{
 4d_\perp b^2
 }{
 \cA a_1\delta
 }.
\end{equation}
The local field expansion is controlled by
$\cA a_1\delta\gg d_\perp b^2$, with fixed numerical factors absorbed
in the asymptotic symbol $\gg$.

The exchange coefficients remain bounded throughout every closed
coupling interval contained in $0<k<1$.
Indeed, monotonicity of $P(w)/w^2$ gives
\begin{equation}
 f(w)
 \leq
 \left(
 \frac{w}{\vartheta}
 \right)^2,
\end{equation}
and monotonicity of $\sin w/w$ on $0<w<\vartheta<\pi$ then gives
\begin{equation}
 J_n
 \leq
 \frac{
 \vartheta
 }{
 (2n-1)\sin^2\vartheta
 },
 \qquad
 n=3,4,
 \qquad
 \mathcal I_f
 \leq
 \frac{
 \vartheta
 }{
 5\sin^4\vartheta
 \left(
 2+\pi^2/\vartheta^2
 \right)
 }.
 \label{8-122}
\end{equation}
The exact finite-radius boundary map gives continuous corrections to
these endpoint vertices.

The upper end of the expansion is fixed by the outer geometric scale.
For an outer length $L$,
\begin{equation}
 \frac{\delta}{L^2}\ll1 .
\end{equation}
Combining this condition with the field-range and Gaussian bounds, I
define
\begin{equation}
 \mathcal C_{\rm ctl}
 =
 \max
 \left\{
 \frac{d_\perp b^2}{a_1},
 \mathcal C_Z,
 \mathcal C_V
 \right\},
 \qquad
 \frac{\mathcal C_{\rm ctl}}{\cA}
 \ll
 \delta
 \ll
 L^2 .
 \label{8-123}
\end{equation}
Fixed order-one constants are absorbed into the definition of the
asymptotic window in Eq.~\eqref{8-123}.
With this convention, the equal two-factor example at $k=1/2$ has
\begin{equation}
 \mathcal C_{\rm ctl}
 =
 \frac1{6\pi}.
\end{equation}
Its connected variance receives the relative correction
\begin{equation}
 \frac{
 10-\pi^2
 }{
 24\pi\Lambda_*
 }
 =
 \frac{
 0.00172942534\ldots
 }{
 \Lambda_*
 },
\end{equation}
in agreement with the negative net quartic coefficient in
Eq.~\eqref{8-112}.

The two inequalities in Eq.~\eqref{8-123} admit, for example,
\begin{equation}
 \delta
 =
 L
 \sqrt{
 \frac{
 \mathcal C_{\rm ctl}
 }{
 \cA
 }
 }
\end{equation}
whenever
$\cA L^2\gg\mathcal C_{\rm ctl}$.
For equal product couplings at fixed effective $\beta$,
$\mathcal C_{\rm ctl}$ grows at most as $p^2$.
A simultaneous large-$p$ limit therefore requires
\begin{equation}
 \cA\delta\gg p^2 .
\end{equation}

The product truncation includes the homogeneous dilaton/saxion, axion,
and radius responses through this order.
The controlled near-BPS limit is
\begin{equation}
 \frac{\delta}{L^2}\longrightarrow0,
 \qquad
 \frac{\cA\delta}{\mathcal C_{\rm ctl}}
 \longrightarrow\infty .
\end{equation}
At fixed $\cA$, the Gaussian regime terminates before the literal
endpoint $\delta=0$ is reached.
A complete endpoint treatment must then use the higher vertices and
the full fixed-charge measure.
The finite complete BPS weight entering Eq.~\eqref{8-89} remains a
separate matching datum.

\subsection{Coupled angular modes and radial resonance}

For $\ell>0$, the axion fluctuations remain dynamical.
The homogeneous fixed-charge Routhian eliminates only the conserved
radial axion momenta.
Angular dependence permits radial charge density to flow into the
angular current, so the axion and dilaton/saxion fluctuations must be
solved as a coupled system.

Let
\begin{equation}
 -\nabla_{S^3}^2Y_{\ell m}
 =
 \lambda_\ell Y_{\ell m},
 \qquad
 \lambda_\ell
 =
 \ell(\ell+2).
 \label{8-124}
\end{equation}
For one transverse dilaton/saxion direction write
\begin{equation}
 \delta\phi_i
 =
 E_{iu}q^uY_{\ell m},
 \qquad
 \delta\chi_i
 =
 \xi_iY_{\ell m}.
 \label{8-125}
\end{equation}
Before the fixed-charge Legendre transform, expansion of the Euclidean
axion kinetic term gives the mixed quadratic contribution
\begin{equation}
 I_{q\chi}^{(2)}
 \supset
 -2\cA
 \sum_i
 \beta_iE_{iu}\nu_i
 \int\dd\tau\,
 q^u\partial_\tau\xi_i .
 \label{8-126}
\end{equation}
For $\ell=0$, the perturbed radial charges remain conserved and the
axion derivatives may be eliminated at fixed charge.
For $\ell>0$, current conservation instead gives
\begin{equation}
 \partial_\tau(\delta\nu_i)
 =
 -\lambda_\ell a\,
 \mathcal G^\chi_{ii}\xi_i,
 \label{8-127}
\end{equation}
where
\[
 \nu_i
 =
 -a^3\mathcal G^\chi_{ii}\partial_\tau\chi_i
\]
in proper-radial gauge.
Thus $\delta\nu_i$ is no longer radially constant, and $\xi_i$ remains
part of the fluctuation system.

I first determine the BPS radial operator.
On the outward BPS branch I use $a$ as the radial coordinate, so
$a'=1$ and primes below denote $\dd/\dd a$.
Combining the stationary product relation derived in
Appendix~C.3 with the BPS background gives, on every active factor,
\begin{equation}
 \nu_i^2\e^{2\beta_i\bar\phi_i}
 =
 \frac{4a^4}{\beta_i^2}.
 \label{8-128}
\end{equation}
For the transverse axion block define
\begin{equation}
 c_i
 =
 \frac{\beta_i\nu_i}{2a^2}\xi_i,
 \qquad
 c_i
 =
 E_{iu}c^u .
 \label{8-129}
\end{equation}
For one transverse pair $(q,c)$, suppressing the common normalization
and harmonic degeneracy, the BPS quadratic Lagrangian is
\begin{equation}
 L_\ell^{(2)}
 =
 \frac12a^3q'^2
 -
 \frac12a^3c'^2
 +
 \frac{a\lambda_\ell}{2}(q^2-c^2)
 -
 4aq^2
 +
 4a^2q'c .
 \label{8-130}
\end{equation}
The negative axion kinetic term is the Euclidean contour convention
before the fixed-charge transform.
The Euler equations are
\begin{align}
 a^2q''
 +
 3aq'
 +
 (8-\lambda_\ell)q
 +
 4ac'
 +
 8c
 &=
 0,
 \nonumber\\
 a^2c''
 +
 3ac'
 -
 \lambda_\ell c
 +
 4aq'
 &=
 0.
 \label{8-132}
\end{align}

For a local power
\begin{equation}
 (q,c)=a^s(Q,C),
\end{equation}
set
\begin{equation}
 x=s(s+2).
\end{equation}
The indicial matrix is
\begin{equation}
 \mathbb M_\ell(s)
 =
 \begin{pmatrix}
 x+8-\lambda_\ell & 4(s+2)\\
 4s & x-\lambda_\ell
 \end{pmatrix}.
 \label{8-133}
\end{equation}
Its determinant factorizes exactly:
\begin{equation}
 \det\mathbb M_\ell(s)
 =
 [s-(\ell-2)]
 (s+\ell)
 [s-(\ell+2)]
 (s+\ell+4).
 \label{8-134}
\end{equation}
The coupled transverse axion--dilaton/saxion block therefore has the
four BPS indices
\begin{equation}
 s
 =
 \ell-2,
 \qquad
 -\ell,
 \qquad
 \ell+2,
 \qquad
 -\ell-4,
 \label{8-135}
\end{equation}
counted with multiplicity.
These are local radial indices.
The physical boundary spectrum follows only after the boundary domain
and the two-end matching are imposed.

The $\ell=1$ sector already contains a repeated BPS index.
Its roots are
\begin{equation}
 3,\quad -1,\quad -1,\quad -5 .
\end{equation}
At $s=-1$, $\mathbb M_1(-1)$ has rank one.
With
\begin{equation}
 v=\binom{1}{-1},
 \qquad
 v_1=\binom{1}{0},
\end{equation}
one finds
\begin{equation}
 \mathbb M_1(-1)v=0,
 \qquad
 \mathbb M_1(-1)v_1
 +
 \mathbb M_1'(-1)v
 =
 0.
\end{equation}
A complete local BPS basis is
\begin{equation}
 a^3\binom{1}{-1},
 \qquad
 a^{-1}\binom{1}{-1},
 \qquad
 a^{-1}
 \left[
 \binom{1}{-1}
 \log\frac{a}{a_{\rm ref}}
 +
 \binom{1}{0}
 \right],
 \qquad
 a^{-5}\binom{3}{5}.
\end{equation}
The fixed radius $a_{\rm ref}>0$ sets the logarithm's origin.
The Wronskian of these four solutions, with rows
$(q,c,q',c')$, is
\begin{equation}
 -128a^{-6}.
\end{equation}
Changing $a_{\rm ref}$ shifts the logarithmic solution by a multiple of
the ordinary $a^{-1}v$ solution.
This logarithm is already part of the BPS Frobenius basis.
Whether a finite-radius deformation excites it is a separate matching
question.

I now expand the finite-radius equations in the BPS overlap region.
Set
\begin{equation}
 a=a_*\rho,
 \qquad
 k^2=3\beta^2,
 \qquad
 f_\rho=\sqrt{1-\rho^{-4}}.
\end{equation}
For the limiting right inner background
$\alpha_*\to\theta_\beta$, the large-$\rho$ expansion is
\begin{align}
 f_\rho
 &=
 1-\frac12\rho^{-4}
 +
 O(\rho^{-8}),
 \nonumber\\
 \frac{\beta\sqrt U}{2a_*^2}
 &=
 \rho^2
 \left[
 1-\frac{1-k^2}{6}\rho^{-4}
 +
 O(\rho^{-8})
 \right],
 \nonumber\\
 \beta\dot\psi
 &=
 \frac2\rho
 \left[
 1-\frac{1+2k^2}{6}\rho^{-4}
 +
 O(\rho^{-8})
 \right].
 \label{8-138}
\end{align}
Here the dot denotes the inner coordinate $t$.
The finite phase $\epsilon(\delta)$ is restored when the local overlap
solution is matched to the physical outer data.

For the local resonance calculation define
\begin{equation}
 D_\rho
 =
 \rho\partial_\rho,
 \qquad
 w
 =
 k\arcsin(\rho^{-2}).
\end{equation}
The exact transverse pair equations on the limiting product background
are
\begin{align}
 0={}&
 \left[
 (1-\rho^{-4})D_\rho^2
 +
 2D_\rho
 -
 \lambda_\ell
 +
 8k^2\rho^{-4}\csc^2w
 \right]q
 \nonumber\\
 &+
 \left[
 4kf_\rho\rho^{-2}\csc w\,D_\rho
 +
 8k^2\rho^{-4}\csc w\cot w
 \right]c,
 \nonumber\\
 0={}&
 \left[
 (1-\rho^{-4})D_\rho^2
 +
 2D_\rho
 -
 \lambda_\ell
 +
 4k^2\rho^{-4}
 \right]c
 \nonumber\\
 &+
 4kf_\rho\rho^{-2}\csc w\,D_\rho q .
 \label{C-full-rho-pair}
\end{align}
In this overlap it is convenient to use the covariant axion variable
\begin{equation}
 c_i
 =
 \operatorname{sgn}(\nu_i)
 \e^{-\beta_i\bar\phi_i}\xi_i .
\end{equation}
Equation~\eqref{8-128} shows that this definition agrees with
Eq.~\eqref{8-129} at the BPS endpoint.

Expanding
$f_\rho$, $\csc w$, and $\cot w$ at large $\rho$ gives
\begin{equation}
 \mathbb M_\ell(D_\rho)
 +
 \rho^{-4}\mathbb N_\beta(D_\rho)
 +
 O(\rho^{-8}).
\end{equation}
Acting on $\rho^s(Q,C)$, the first correction is
\begin{equation}
 \mathbb N_\beta(s)
 =
 \begin{pmatrix}
 -s^2-\dfrac83(1-k^2)
 &
 -\dfrac23(4-k^2)s-\dfrac43(2+k^2)
 \\[6pt]
 -\dfrac23(4-k^2)s
 &
 -s^2+4k^2
 \end{pmatrix}.
 \label{8-139}
\end{equation}
The deformation lowers the radial power by four.
A logarithm is forced only when the shifted power coincides with another
BPS root and the corresponding forcing has nonzero projection onto the
left null space of the indicial matrix.

For $\ell=1$, the $s=3$ branch reaches the repeated root $s=-1$ at
first order.
Equation~\eqref{8-139} gives
\begin{equation}
 \mathbb N_\beta(3)v
 =
 (2k^2-1)v .
\end{equation}
The ordinary correction
\begin{equation}
 v_{\rm corr}
 =
 \binom{(1-2k^2)/4}{0}
\end{equation}
satisfies
\begin{equation}
 \mathbb M_1(-1)v_{\rm corr}
 +
 \mathbb N_\beta(3)v
 =
 0 .
\end{equation}
Thus the first finite-radius recurrence closes with an ordinary
$\rho^{-1}$ term.
The repeated BPS root still supplies its logarithmic companion, but no
new resonant logarithm is forced at this order.
Its coefficient is determined by the global boundary matching.

For $\ell=2$, the candidate resonance from $s=4$ to $s=0$ also
cancels: the forcing has zero projection onto the left null vector at
$s=0$.
The first nonvanishing finite-radius resonance occurs at $\ell=3$.
The BPS roots are
\begin{equation}
 s
 =
 5,\qquad
 1,\qquad
 -3,\qquad
 -7.
 \label{8-140}
\end{equation}
The $5\to1$ and $-3\to-7$ compatibility projections vanish.
The $1\to-3$ pair has right null vectors
\begin{equation}
 v_+
 =
 \binom{3}{1},
 \qquad
 v_-
 =
 \binom{-1}{1},
\end{equation}
while a left null vector of
$\mathbb M_3(-3)$ is
\begin{equation}
 w_-^T=(-3,1).
\end{equation}
Direct evaluation gives
\begin{align}
 w_-^T
 \mathbb N_\beta(1)v_+
 &=
 40-16k^2
 =
 40-48\beta^2,
 \nonumber\\
 w_-^T
 \mathbb M_3'(-3)v_-
 &=
 -32 .
 \label{8-143}
\end{align}
On the branch $0<k<1$, the first quantity is strictly positive.
The compatibility condition therefore requires a logarithm.

At order $\rho^{-3}$ write
\begin{equation}
 \rho^{-3}
 \left[
 \kappa_3\log\rho\,v_-
 +
 u
 \right].
\end{equation}
The recurrence becomes
\begin{equation}
 \mathbb M_3(-3)u
 +
 \kappa_3
 \mathbb M_3'(-3)v_-
 +
 \mathbb N_\beta(1)v_+
 =
 0 .
\end{equation}
Multiplication by $w_-^T$ removes the ordinary vector $u$ and fixes
\begin{equation}
 \kappa_3
 =
 -
 \frac{
 w_-^T\mathbb N_\beta(1)v_+
 }{
 w_-^T\mathbb M_3'(-3)v_-
 }
 =
 \frac{5-2k^2}{4}
 =
 \frac{5-6\beta^2}{4}.
\end{equation}
One convenient choice of ordinary completion gives
\begin{equation}
 \Psi_+(\rho)
 =
 \rho v_+
 +
 \kappa_3
 \rho^{-3}\log\rho\,v_-
 +
 \rho^{-3}
 \left[
 \binom{(10k^2-19)/12}{0}
 +
 b\,v_-
 \right]
 +
 \cdots,
 \qquad
 \kappa_3
 =
 \frac{5-6\beta^2}{4}.
 \label{8-144}
\end{equation}
Substitution into both coupled equations verifies the logarithmic and
ordinary terms.
The coefficient $b$ multiplies the homogeneous $\rho^{-3}v_-$
solution and is fixed by matching.
A regular change of fluctuation variables can shift this ordinary term,
whereas the resonance residue $\kappa_3$ is invariant.

The inner logarithm now determines the nonanalytic order seen at fixed
outer radius.
Since
\begin{equation}
 \rho=\frac{a}{a_*},
 \qquad
 a_*^2=\delta,
\end{equation}
normalize the leading branch as $a\,v_+=a_*\rho v_+$.
Equation~\eqref{8-144} then contains
\begin{equation}
 a_*\Psi_+(a/a_*)
 \supset
 \frac{
 \kappa_3\delta^2
 }{
 a^3
 }
 \log\frac{a}{\sqrt\delta}\,
 v_- .
 \label{8-145}
\end{equation}
Thus the first nonvanishing finite-radius angular resonance enters the
outer solution at
\begin{equation}
 O\!\left(
 \delta^2\log\delta
 \right).
\end{equation}
Its first derivative with respect to $\delta$ vanishes as
$\delta\to0^+$.
It therefore does not modify the linear endpoint slope obtained from
the homogeneous transverse mode.

The logarithm also carries the standard matching-scale freedom of a
resonant Frobenius expansion.
The BPS overlap operator is covariant under radial dilatations, while
the first finite-radius correction lowers the radial weight by four.
For two independent solutions labelled by their boundary data,
write the resonant part as
\begin{equation}
 \rho^{-3}
 \left[
 \kappa_3
 \log\frac{\rho}{\mu_\rho}
 +
 b_j(\mu_\rho)
 \right]
 v_-,
 \qquad
 j=1,2 .
 \label{8-146}
\end{equation}
Independence of the physical solution from the arbitrary overlap scale
$\mu_\rho$ gives
\begin{equation}
 \frac{\dd b_j}{\dd\log\mu_\rho}
 =
 \kappa_3 .
 \label{8-147}
\end{equation}
Hence
$b_1-b_2$ is matching-scale invariant.
The common running is the classical radial resonance; the physical
boundary map converts the invariant matched data and the residue
$\kappa_3$ into the finite-radius response.
The exact transformation in the next subsection supplies that boundary
map and reproduces the angular resonance directly.

\subsection{Exact angular matching and the gravity mode}

The coupled product system admits an exact field transformation that
separates the bulk angular equations.
I use this transformation to construct the physical two-boundary map at
finite radius while retaining the generic matter phase.

With
\[
 Q=(q,c)^T,
\]
the finite-radius quadratic density divided by $\cA\delta$ is
\begin{align}
 L_Q={}&
 q_y^2
 -
 (c_y+k\cot w\,c)^2
 +
 \frac{\lambda_\ell}{4\cos^2y}(q^2-c^2)
 -
 2k^2\csc^2w\,q^2
 \nonumber\\
 &-
 4k\csc w\,q(c_y+k\cot w\,c).
 \label{8-148}
\end{align}
Introduce
\begin{equation}
 Q
 =
 S(w)
 \binom{z_k}{z_0},
 \qquad
 S(w)
 =
 \begin{pmatrix}
 \csc w & -\cot w\\
 -\cot w & \csc w
 \end{pmatrix},
 \qquad
 \det S=1 .
 \label{8-149}
\end{equation}
Direct substitution gives the exact identity
\begin{align}
 L_Q={}&
 z_{k,y}^2-z_{0,y}^2
 +
 \frac{\lambda_\ell}{4\cos^2y}
 (z_k^2-z_0^2)
 -
 k^2z_k^2
 +
 \partial_y(Q^TTQ),
 \nonumber\\
 T(w)
 &=
 \begin{pmatrix}
 0 & -k\csc w\\
 -k\csc w & -k\cot w
 \end{pmatrix}.
 \label{8-150}
\end{align}
The two bulk fields therefore satisfy the same
P\"oschl--Teller equation with two fixed values of the parameter,
\begin{equation}
 \left[
 -\partial_y^2
 +
 \frac{\lambda_\ell}{4\cos^2y}
 \right]z_\eta
 =
 \eta^2z_\eta,
 \qquad
 \eta=k,0 .
 \label{8-151}
\end{equation}
The transformation removes the asymmetric matter phase from the bulk
operator.
That phase survives entirely in the boundary matrices $S(w)$ and
$T(w)$.

This separation is useful because reflection symmetry is obscured in
the original variables.
For a scalar equation $q''=Vq$, decomposition into even and odd parts
gives
\begin{equation}
 q_e''
 =
 V_eq_e+V_oq_o,
 \qquad
 q_o''
 =
 V_eq_o+V_oq_e .
 \label{8-152}
\end{equation}
For
\[
 V(y)=2k^2\csc^2(\alpha_*-ky),
\]
the odd part $V_o$ is nonzero at generic $\alpha_*$.
The original parity sectors therefore mix.
Equation~\eqref{8-151}, by contrast, has even coefficients.
I solve it in an even--odd basis and reconstruct the asymmetric physical
variables only at the boundaries.

Set
\[
 g=\frac{\ell+2}{2}.
\]
A normalized even and odd pair for each $\eta$ is
\begin{align}
 e_\eta(y)
 &=
 (\cos y)^g
 {}_2F_1
 \left(
 \frac{g+\eta}{2},
 \frac{g-\eta}{2};
 \frac12;
 \sin^2y
 \right),
 \nonumber\\
 o_\eta(y)
 &=
 \sin y\,(\cos y)^g
 {}_2F_1
 \left(
 \frac{g+\eta+1}{2},
 \frac{g-\eta+1}{2};
 \frac32;
 \sin^2y
 \right).
 \label{8-153}
\end{align}
They obey
\[
 e_\eta(0)=1,
 \qquad
 e_\eta'(0)=0,
 \qquad
 o_\eta(0)=0,
 \qquad
 o_\eta'(0)=1,
\]
and their Wronskian is unity.

Place equal physical cuts at $a=R$.
Since
\[
 a^2=\delta\sec y,
\]
the cuts lie at
\[
 Y
 =
 \arccos\frac{\delta}{R^2}.
\]
For each transformed channel define
\begin{equation}
 d_\eta
 =
 \delta
 \left(
 \frac{e_\eta'}{e_\eta}
 +
 \frac{o_\eta'}{o_\eta}
 \right)_{y=Y},
 \quad
 x_\eta
 =
 -\frac{\delta}{
 e_\eta(Y)o_\eta(Y)
 },
 \quad
 D_\partial
 =
 \diag(d_k,-d_0),
 \quad
 X_\partial
 =
 \diag(x_k,-x_0).
 \label{8-154}
\end{equation}
The relative minus signs in the $\eta=0$ channel follow from the
opposite kinetic sign in Eq.~\eqref{8-150}.

At the left and right cuts,
\[
 w_L=\alpha_*+kY,
 \qquad
 w_R=\alpha_*-kY,
\]
and I write
\[
 S_{L,R}=S(w_{L,R}),
 \qquad
 T_{L,R}=T(w_{L,R}).
\]
The physical outer Dirichlet Hessian, with the common normalization
$\cA$ removed, has blocks
\begin{align}
 K^Q_{LL}
 &=
 S_L^{-T}D_\partial S_L^{-1}
 -
 2\delta T_L,
 \nonumber\\
 K^Q_{RR}
 &=
 S_R^{-T}D_\partial S_R^{-1}
 +
 2\delta T_R,
 \nonumber\\
 K^Q_{LR}
 &=
 S_L^{-T}X_\partial S_R^{-1},
 \qquad
 K^Q_{RL}
 =
 (K^Q_{LR})^T .
 \label{8-155}
\end{align}
Thus
\[
 \cK^\partial_{\bm\nu,B}
 =
 \cA K^Q
\]
in this angular block.
The terms proportional to $T_{L,R}$ come from the total derivative in
Eq.~\eqref{8-150}.
They are part of the physical Dirichlet Hessian.
The matrices $S_L$, $S_R$, $T_L$, and $T_R$ carry the asymmetry of the
charged matter profile, while the bulk propagation is controlled by the
even equation~\eqref{8-151}.

The outer matrix $K^Q$ is distinct from the artificial interface
Hessian $\cK^\Sigma$.
Changing the physical boundary ensemble requires the corresponding
Legendre transform.
Likewise, matching to asymptotic multipole sources requires the source
normalization and local subtraction specified by the observable.

The exact transformed equation also gives a closed recurrence for the
overlap expansion.
In the inner radial coordinate $\rho$,
Eq.~\eqref{8-151} becomes
\begin{equation}
 \left[
 P_\ell(D_\rho)
 +
 \rho^{-4}
 \left(
 4\eta^2-D_\rho^2
 \right)
 \right]z_\eta
 =
 0,
 \qquad
 P_\ell(s)
 =
 s(s+2)-\ell(\ell+2),
 \qquad
 D_\rho
 =
 \rho\partial_\rho .
 \label{8-156}
\end{equation}
For the branch beginning with $\rho^\ell$, write
\[
 z_\eta
 =
 \rho^\ell
 \sum_{n\geq0}
 a_n\rho^{-4n},
 \qquad
 a_0=1 .
\]
Matching equal powers gives
\begin{equation}
 a_n
 =
 \frac{
 (\ell-4n+4)^2-4\eta^2
 }{
 P_\ell(\ell-4n)
 }
 a_{n-1}.
 \label{8-157}
\end{equation}
The second indicial root of this scalar equation is
$-\ell-2$.
The descending sequence
$\ell-4n$
reaches this root only for odd $\ell$.
For
\[
 n_*=\frac{\ell+1}{2},
\]
the resonant denominator in Eq.~\eqref{8-157} vanishes.
The coefficient of
$\rho^{-\ell-2}\log\rho$
is
\begin{equation}
 \gamma_\ell(\eta)
 =
 \frac{
 \displaystyle
 \prod_{j=1}^{n_*}
 \left[
 4\eta^2-(2j-1)^2
 \right]
 }{
 4n_*\,16^{\,n_*-1}
 [(n_*-1)!]^2
 } .
 \label{8-158}
\end{equation}
For even $\ell$, the power sequence never reaches the second root and
this resonance is absent.
For odd $\ell$, a zero of any factor in the numerator cancels the
corresponding logarithm.

For $\ell=1$,
\[
 \gamma_1(\eta)
 =
 \eta^2-\frac14 .
\]
After the inverse transformation~\eqref{8-149}, the associated
$\rho^{-3}\log\rho$ term maps to the repeated $a^{-1}$ BPS branch
identified in the preceding subsection.
It is therefore the transformed representation of the $\ell=1$ BPS
logarithmic companion, rather than a new first-order finite-radius
resonance of the physical $(q,c)$ system.
This agrees with the direct recurrence, where the first
finite-radius correction in the $\ell=1$ block closes with an ordinary
power.

For $\ell=3$, the transformed recurrence reaches the second root at
$n_*=2$.
After transforming back to $(q,c)$, the resulting logarithm reproduces
the $\rho^{-3}\log\rho$ structure of
Eq.~\eqref{8-144}.
At fixed outer radius it therefore enters the physical solution at
\[
 \delta^2
 \log\frac{\delta}{\delta_0},
\]
up to the matched reference scale $\delta_0$.
Its first derivative with respect to $\delta$ vanishes at the BPS
endpoint.
The exact transformation thus confirms the resonance found from the
coupled indicial recurrence and supplies its complete two-boundary map.

I next separate genuine fluctuation modes from collective directions.
Spatial rotations leave every radial background field invariant and
generate no radial fluctuation.
Their stabilizer volume belongs to the gauge treatment.
When a common translation is an admissible collective coordinate, its
scalar variation is proportional to
\[
 \delta\phi_i
 \propto
 n_i\sigma'\,
 \epsilon\cdot\hat x .
\]
Its transverse projection vanishes because
\[
 E^Tn=0,
\]
and the corresponding transverse axion projection vanishes as well.
The transverse angular block studied above therefore contains no
translation collective mode.

At a finite Dirichlet cut,
$Y<\pi/2$.
The lowest Dirichlet eigenvalue of
$-\partial_y^2$ on $[-Y,Y]$ is
\[
 \left(\frac{\pi}{2Y}\right)^2>1 .
\]
Since
\[
 0\leq\eta^2\leq k^2<1
\]
on the branch studied here and
$\lambda_\ell/(4\cos^2y)>0$ for $\ell\geq1$,
the operators
\[
 -\partial_y^2
 +
 \frac{\lambda_\ell}{4\cos^2y}
 -
 \eta^2
\]
are strictly positive on the finite-cut Dirichlet domain.
Each transformed scalar channel therefore has trivial zero-frequency
kernel.
The original $(q,c)$ action still carries opposite kinetic signs before
the axion contour rotation.
The positivity statement concerns the transformed Dirichlet
Sturm--Liouville operators, rather than the unreduced Euclidean
gravity--matter action.

I now turn to the homogeneous scalar--gravity channel.
The one-scalar reduction in Eq.~\eqref{8-23} gives a regular
constraint-reduced description.
In the BPS interior scaling region with $\Lambda=0$,
\begin{equation}
 F
 =
 \frac{2a^2}{\beta},
 \qquad
 p_a
 =
 -12a,
 \qquad
 \omega
 =
 \frac{2}{\beta a}.
 \label{8-159}
\end{equation}
Equations~\eqref{8-25}--\eqref{8-27} then give
\begin{equation}
 F_{p_ap_a}
 =
 \frac{\beta}{24}
 (1-3\beta^2),
 \qquad
 F_{p_aa}
 =
 -\beta-\frac32\beta^3,
 \qquad
 F_{aa}
 =
 -30\beta-18\beta^3 .
 \label{8-160}
\end{equation}
Substitution into Eq.~\eqref{8-23} gives, apart from the common factor
$\mathcal V_3/(4\kappa_4^2)$,
\begin{equation}
 -\frac{12}{1-3\beta^2}\,
 a
 \left(
 f'
 -
 \frac{2+3\beta^2}{a}f
 \right)^2
 -
 \frac{60+36\beta^2}{a}f^2 .
 \label{8-161}
\end{equation}
Its Euler equation simplifies exactly to
\begin{equation}
 a^2f''
 +
 af'
 -
 9f
 =
 0 .
 \label{8-162}
\end{equation}
Hence
\[
 f_+\propto a^3,
 \qquad
 f_-\propto a^{-3}.
\]
With fixed normalization of these local branches, their relative size
at the minimal radius is
\begin{equation}
 a_*^6
 =
 \delta^3 .
 \label{8-163}
\end{equation}
This local power ratio identifies the natural mixing order of the
scalar--gravity channel.
The physical outer Hessian still requires propagation through the
finite-radius boundary-value problem.

That propagation can be obtained directly by varying the exact
background at fixed dilaton/saxion.
On every interval with $\sigma'\neq0$, two independent radius
variations and their canonical momenta are
\begin{align}
 f_\alpha
 &=
 \frac{a}{2k}\tan y,
 &
 f_\delta
 &=
 \frac{a}{2\delta}
 \left(
 1-\frac{\tan y\tan w}{k}
 \right),
 \nonumber\\
 p_\alpha
 &=
 -12\sin y\,f_\alpha
 -
 \frac{12a\cos y}{k},
 &
 p_\delta
 &=
 -12\sin y\,f_\delta
 +
 \frac{12a\cos y\tan w}{k\delta}.
 \label{8-164}
\end{align}
Their symplectic Wronskian is constant,
\[
 f_\delta p_\alpha
 -
 f_\alpha p_\delta
 =
 -\frac6k .
\]
The matrix of outward canonical momenta multiplied by the inverse matrix
of boundary values gives the exact Dirichlet-to-Neumann coefficient
whenever the boundary-value map is nonsingular.

At equal cuts $a=R$, set
\[
 T_Y=\tan Y,
 \qquad
 w_L=\alpha_*+kY,
 \qquad
 w_R=\alpha_*-kY .
\]
The off-diagonal scalar--gravity Hessian entry, with $\cA$ removed, is
\begin{equation}
 K^{\rm grav}_{LR}
 =
 \frac{
 24\delta
 }{
 R^2T_Y
 \left[
 2+(T_Y/k)(\tan w_L-\tan w_R)
 \right]
 }
 =
 \frac{
 24k\cot(\pi k)
 }{
 R^6
 }
 \delta^3
 +
 O(\delta^4).
 \label{8-165}
\end{equation}
The displayed expansion applies when the boundary-value map is
nondegenerate and $\cos(\pi k)\neq0$.
It shows that the generic scalar--gravity off-diagonal Hessian begins at
$O(\delta^3)$.
Special coupling values can cancel this coefficient and move the first
nonzero term to higher order.
The corresponding covariance also depends on the diagonal entries and
on the gravitational contour.

At a turning point of the scalar used as the clock, I return to the
first-order constrained system of
Eq.~\eqref{8-13}, choose another regular canonical coordinate, and
continue the same physical boundary map.

The preceding calculations now determine the finite-radius deformation
that enters the spectral analysis.
At fixed physical outer coordinates, let
\[
 \cK^\partial_{\bm\nu,B}(\delta)
\]
denote the two-sided outer-boundary Hessian.
Define
\begin{equation}
 \mathcal D^\partial_{\bm\nu}
 =
 \left.
 \frac{\dd}{\dd\delta}
 \cK^\partial_{\bm\nu,B}(\delta)
 \right|_{0^+}.
 \label{8-166}
\end{equation}
This derivative follows the fixed-charge saddle family of
Appendix~\ref{sec:near-bps-problem}, with the physical outer coordinates
and probe convention held fixed.

If the calculation uses an artificial radial cut, first eliminate the
two Dirichlet interiors and form the joint Hessian on the physical
outer values and the trace at $\Sigma$:
\begin{equation}
 \cK^\partial_{\bm\nu,B}
 =
 \mathbb H_{\partial\partial}
 -
 \mathbb H_{\partial\Sigma}
 \mathbb H_{\Sigma\Sigma}^{-1}
 \mathbb H_{\Sigma\partial},
 \qquad
 \mathbb H_{\Sigma\Sigma}
 =
 \cK^\Sigma_{\bm\nu,B}.
 \label{8-167}
\end{equation}
Every block in Eq.~\eqref{8-167} uses the same boundary domain,
coordinates, regulator, and local subtraction.
The individual interface quantities depend on the position of
$\Sigma$.
Their Schur complement does not.
This is the outer Hessian whose derivative enters the physical
finite-radius response.

For a differentiable outer block,
\begin{equation}
 \cK^\partial_{\bm\nu,B}(\delta)
 =
 \cK^\partial_{\bm\nu,B}(0)
 +
 \delta\mathcal D^\partial_{\bm\nu}
 +
 o(\delta).
 \label{8-168}
\end{equation}
If the observable retains a further outer subspace, let
$\mathfrak J_0$ denote its endpoint stationary extension.
The first deformation of the final reduced matrix is
\begin{equation}
 D^{\rm phys}_{1,\bm\nu}
 =
 \mathfrak J_0^\dagger
 \mathcal D^\partial_{\bm\nu}
 \mathfrak J_0 .
 \label{8-169}
\end{equation}
This is the stationary-extension identity of
Eq.~\eqref{7-33}.
On an exact endpoint kernel with a bounded complementary inverse, it
reduces to direct projection.

The calculations above also separate the different powers of
$\delta$ that occur in the paper:
\begin{center}
\begin{tabular}{ll}
\toprule
Quantity & Leading finite-radius change \\
\midrule
Fixed-radius outer background & $O(\delta^2)$ \\
Inner transverse quadratic action & $O(\delta)$ \\
Homogeneous transverse off-diagonal Hessian & $O(\delta^2)$ \\
Homogeneous transverse cross covariance & $O(\delta)$ \\
Scalar--gravity off-diagonal Hessian at generic stated cuts
& $O(\delta^3)$ \\
First nonvanishing angular resonance
& $O\!\left(\delta^2\log(\delta/\delta_0)\right)$ \\
\bottomrule
\end{tabular}
\end{center}
The powers belong to different stages of the boundary reduction.
The local inner action, the physical Hessian, and its inverse therefore
need not begin at the same order.

In particular, the homogeneous transverse result gives
\begin{equation}
 K_{LL}'(0)=a_1,
 \qquad
 K_{LR}=O(\delta^2),
 \qquad
 G_{LR}
 =
 g_{1,\bm\nu}\delta
 +
 O(\delta^2).
\end{equation}
In the product system, the exact homogeneous transverse quadratic form
remains positive at every allowed finite radius. Its local normalized
potential Hessian is $\mathfrak M=4\beta^2 I$. 
The endpoint derivative of the physical quadratic form on the soft
subspace is positive semidefinite.
The subtractive finite-rank deformations introduced later are separate
spectral models and are not inferred from this product calculation.

For an invertible regulated endpoint block $M_{0,\nu}$ and one fixed
scalar source contraction, the relative Gaussian bosonic two-source
coefficient is
\begin{align}
 \frac{
 \Pi_{B,\bm\nu}(\delta)
 }{
 \Pi_{B,\bm\nu}(0)
 }
 ={}&
 \mathcal R_{\bm\nu,\rm comp}(\delta)
 \left[
 \frac{
 \det M_{\bm\nu,B}(\delta)
 }{
 \det M_{0,\bm\nu}
 }
 \right]^{-1/2}
 \nonumber\\
 &\times
 \frac{
 \cV_{\bm\nu}(\delta)
 +
 G_{\bm\nu,Q}(\delta)
 +
 \mathbf b^{\rm eff}_{\bm\nu}(\delta)^\dagger
 M_{\bm\nu,B}(\delta)^{-1}
 \mathbf b^{\rm eff}_{\bm\nu}(\delta)
 }{
 \cV_{\bm\nu}(0)
 +
 G_{\bm\nu,Q}(0)
 +
 \mathbf b^{\rm eff}_{\bm\nu}(0)^\dagger
 M_{0,\bm\nu}^{-1}
 \mathbf b^{\rm eff}_{\bm\nu}(0)
 } .
 \label{8-170}
\end{align}
The complementary factor is normalized by
\[
 \mathcal R_{\bm\nu,\rm comp}(0)=1
\]
and contains the remaining regulated bosonic contributions.
Equation~\eqref{8-170} applies to an invertible endpoint block with a
nonzero endpoint source contraction.
The singular homogeneous pair is instead treated exactly by
Eq.~\eqref{8-85}.

Dividing an unnormalized two-source coefficient by its zero-source
functional removes the determinant and other source-independent
weights and leaves the connected Gaussian response.
For the unnormalized coefficient those weights remain.
In the homogeneous soft channel,
$G_{LR}(0)=0$, so the derivative of the zero-source weight multiplies
zero at first order.
Consequently the linear coefficient in Eq.~\eqref{8-89} is
$Z_{\bm\nu,0}g_{1,\bm\nu}$ independently of the first derivative of a
finite differentiable weight.

The exact radial analysis therefore supplies three distinct inputs for
the later spectral problem.
The homogeneous transverse block gives the linear cross covariance.
The coupled angular block gives the first resonant
$\delta^2\log\delta$ correction.
The scalar--gravity block begins at cubic order in its off-diagonal
Hessian at generic cuts.
The compactification and the complete fixed-charge measure determine
how these mode-dependent responses enter the full sector coefficient.

\section{Charge structure and endpoint spectrum}
\label{sec:charge-phase}
\subsection{Charge dependence}

The finite-radius response is computed charge sector by charge sector.
I now separate three levels of charge dependence: the local dependence
of the fluctuation operator on the charge bilinears, the dependence of
the reduced scalar coefficient on the charge lattice, and the
matrix-valued information carried by the reduced Hessian.
These levels coincide only under additional factorization assumptions.

For axion charges $\bm\nu\in\Z^r$ and dilaton/saxions $\phi^A$, the
fixed-charge potential is
\begin{equation}
 U_{\bm\nu}(\phi)
 =
 \nu_i\mathcal G_\chi^{ij}(\phi)\nu_j .
 \label{9-1}
\end{equation}
Its covariant second variation contains
\begin{equation}
 \Delta^{(2)}U_{\bm\nu}
 =
 \frac12\xi^A\xi^B
 \nu_i
 \nabla_A\nabla_B
 \mathcal G_\chi^{ij}(\phi)
 \nu_j .
 \label{9-2}
\end{equation}
This term supplies only the local charge-dependent potential.
The physical quadratic form also contains the kinetic terms, target
curvature, gravitational constraint, form-field mixing, and boundary
form.

Let $e_a$ denote a coordinate vector in the final reduced boundary
space, and let
\[
 \Psi_a=\mathfrak J e_a
\]
be its stationary bulk extension after all complementary bosonic
variables have been eliminated.
The exact reduced bosonic matrix is then
\begin{align}
 [M_{\bm\nu,B}]_{ab}
 ={}&
 \mathfrak h_{\bm\nu,B}
 [\Psi_a,\Psi_b]
 \nonumber\\
 ={}&
 \int\dd\tau\,
 \left[
 \dot\Psi_a^\dagger A\dot\Psi_b
 +
 \dot\Psi_a^\dagger B\Psi_b
 +
 \Psi_a^\dagger B^\dagger\dot\Psi_b
 +
 \Psi_a^\dagger C\Psi_b
 \right]
 +
 \mathfrak h_{\bm\nu,\partial}
 [\Psi_a,\Psi_b].
 \label{9-3}
\end{align}
Here
$\mathfrak h_{\bm\nu,B}[\cdot,\cdot]$
is the complete normalized bosonic quadratic form on the physical
domain.
The stationary extension $\mathfrak J$ already contains the response
of the eliminated fields.
Adding a separate relaxation term to Eq.~\eqref{9-3} would therefore
double count their Schur-complement contribution.

The charge dependence is most cleanly differentiated before imposing
the rank-one relation between charge bilinears.
Introduce independent symmetric variables $t_{ij}=t_{ji}$ as in
Appendix~\ref{sec:finite-radius-boundary}.
Around a fixed reference point $t^0$,
\begin{equation}
 M_B(t;\delta)
 =
 M_B(t^0;\delta)
 +
 \sum_{i\leq j}
 (t_{ij}-t^0_{ij})
 X_{ij}(\delta)
 +
 O(\|t-t^0\|^2),
 \qquad
 X_{ij}
 =
 \left.
 \partial_{t_{ij}}M_B
 \right|_{t^0}.
 \label{9-4}
\end{equation}
Every derivative in Eq.~\eqref{9-4} includes both the explicit charge
dependence of the fluctuation operator and the induced displacement of
the background at fixed physical outer data.
It also includes transport of the chosen reduced basis.
Only after differentiation do I restrict to the physical locus
\[
 t_{ij}=\nu_i\nu_j .
\]
Thus Eq.~\eqref{9-4} is a local matrix expansion in the independent
bilinears, rather than a statement that the physical charge dependence
is affine.
Matrix inversion, Schur reduction, and motion of the saddle generally
generate nonpolynomial dependence on $\bm\nu$.
The same charge variation acts on the fermionic block, effective source
vectors, complementary response, exact-mode factors, and
normalization.

The homogeneous product model gives one part of this dependence in
closed form.
With the $C^1$ endpoint matching used in
Eq.~\eqref{8-89}, Eq.~\eqref{8-86} gives
\begin{equation}
 g_{1,\bm\nu}
 =
 \frac{
 6h^2
 }{
 \cA k V_{\bm\nu}v_1
 },
 \qquad
 B_{\bm\nu}
 =
 Z_{\bm\nu,0}g_{1,\bm\nu},
 \qquad
 C_{\bm\nu}^{LR}
 =
 \delta B_{\bm\nu}
 +
 O(\delta^2).
 \label{9-5}
\end{equation}
The factor $g_{1,\bm\nu}$ is the geometric response multiplier.
The factor $Z_{\bm\nu,0}$ is the complete BPS sector weight.
Their product gives the leading unnormalized finite-radius
coefficient.

For the stationary product trajectory, define
\[
 \omega_i
 =
 \frac{\beta^2}{\beta_i^2},
 \qquad
 \sum_i\omega_i=1 .
\]
At the stationary point,
Eq.~\eqref{8-67} gives
$X_i=c/\beta_i^2$.
Taking logarithms and using
$n^TEs=0$ yields
\begin{equation}
 V_{\bm\nu}
 =
 \prod_i
 \left(
 \frac{\nu_i^2}{\omega_i}
 \right)^{\omega_i},
 \qquad
 V_{t\bm\nu}
 =
 t^2V_{\bm\nu},
 \qquad
 g_{1,t\bm\nu}
 =
 t^{-2}g_{1,\bm\nu}.
 \label{9-6}
\end{equation}
These identities hold at fixed active support, fixed couplings, and
fixed right scaling datum $h$.
The geometric multiplier is real and positive.
It changes the magnitude of a complex BPS coefficient while leaving
its phase unchanged.
Sectors with the same $V_{\bm\nu}$ therefore have the same homogeneous
geometric multiplier under the same boundary prescription.

The closed form in Eq.~\eqref{9-6} applies to the stationary product
submanifold.
At a fixed right asymptotic dilaton/saxion vector, a charge sector lies
on this submanifold precisely when
\begin{equation}
 \beta_i^2\nu_i^2
 \e^{2\beta_i\phi_{i,R}}
 =
 \beta_j^2\nu_j^2
 \e^{2\beta_j\phi_{j,R}}
 \qquad
 \hbox{for all active }i,j .
 \label{9-7}
\end{equation}
Sign-related sectors satisfy the same condition because the background
depends on $\nu_i^2$.
They therefore share the same stationary trajectory and the same
geometric response.
Changing relative charge magnitudes generally violates
Eq.~\eqref{9-7} at fixed $\phi_R$ and moves the stationary trajectory.
Those charge variations are obtained from the sensitivity equations
of Appendix~\ref{sec:finite-radius-boundary}, rather than by applying
Eq.~\eqref{9-6} at a fixed background.

The scalar coefficient that enters the charge-sector construction must
be formed only after the complete fixed-charge fluctuation problem has
been reduced.
At quadratic order,
\begin{equation}
 \Pi_{\bm\nu}
 =
 R_b\!\left[
 \cA_{\bm\nu,Q}
 \frac{
 \Pf M_{\bm\nu,F}
 }{
 \sqrt{\abs{\det M_{\bm\nu,B}}}
 }
 \left(
 G_{\bm\nu,Q}
 +
 (\mathbf b^{\rm eff}_{\bm\nu})^\dagger
 M_{\bm\nu,B}^{-1}
 \mathbf b^{\rm eff}_{\bm\nu}
 +
 \cV_{\bm\nu,\rm eff}
 \right)
 \right]
 +
 \Pi_{\bm\nu,\rm higher}.
 \label{9-9}
\end{equation}
Here $\Pi_{\bm\nu,\rm higher}$ denotes the already reduced
non-Gaussian and source-dependent corrections.
The absolute-value notation in Eq.~\eqref{9-9} refers to the real
self-adjoint patch; a complex saddle uses the determinant branch fixed
by its contour.

The same reduced matrix appears in both the bosonic determinant and
the source response.
An eigenvalue $\lambda\to0^+$ contributes
$\lambda^{-1/2}$ to the bosonic determinant factor and
$\lambda^{-1}$ to a source contraction with nonzero overlap.
Exact-mode measures, saturation insertions, fermions, ghosts, contour
phases, and the remaining fluctuation factors must therefore be
included before a charge-sector coefficient is assigned.

For one isolated saddle, remove the classical exponential and define
\begin{equation}
 \Pi_{\bm\nu}[b;\delta]
 =
 \e^{S^{\rm cl}_{\bm\nu}(\delta)}
 W_{\bm\nu}[b;\delta].
 \label{9-10}
\end{equation}
This is the reduced coefficient after the operation $R_b$.
For several saddles I construct
$\Pi_{\bm\nu,\mathfrak s}$ separately on each branch and sum the
branch contributions only after their individual measures, phases, and
zero-mode prescriptions have been fixed.

The scalar charge dependence can now be tested directly on the charge
lattice.
Let $\bm e_a$ denote the unit vector in charge direction $a$.
For two distinct directions define the elementary mixed plaquette ratio
\begin{equation}
 \mathfrak K_{ab}^{\rm pre}(\bm\nu)
 =
 \frac{
 \Pi_{\bm\nu+\bm e_a+\bm e_b}
 \Pi_{\bm\nu}
 }{
 \Pi_{\bm\nu+\bm e_a}
 \Pi_{\bm\nu+\bm e_b}
 } .
 \label{9-11}
\end{equation}
I use Eq.~\eqref{9-11} only where all four coefficients are nonzero and
are defined with the same boundary and reduction data.
Unlike a logarithm, this multiplicative ratio is insensitive to the
choice of a local logarithm branch.

\paragraph{Charge-factorization criterion.}
Suppose the reduced prefactor factorizes as
\begin{equation}
 \Pi_{\bm\nu}
 =
 \prod_c f_c(\nu_c)
\end{equation}
throughout the four sectors entering Eq.~\eqref{9-11}.
Then
\begin{equation}
 \mathfrak K_{ab}^{\rm pre}(\bm\nu)
 =
 1,
 \qquad
 a\neq b .
 \label{9-12}
\end{equation}
Therefore
$\mathfrak K_{ab}^{\rm pre}\neq1$
proves nonfactorized dependence on the two charge directions.

To see this, note that every factor with $c\neq a,b$ cancels between numerator and denominator.
The $a$- and $b$-dependent factors cancel separately.

The converse is local rather than global:
$\mathfrak K_{ab}^{\rm pre}=1$ at one charge plaquette does not prove
factorization of the full charge family.

The geometric multiplier in Eq.~\eqref{9-6} itself factorizes,
\[
 g_{1,\bm\nu}
 \propto
 \prod_i
 |\nu_i|^{-2\omega_i},
\]
as long as all four sectors remain on the same active support and use
the same stationary boundary prescription.
It therefore cancels from
Eq.~\eqref{9-11}.
Under those conditions, the mixed plaquette ratio of
\[
 B_{\bm\nu}
 =
 Z_{\bm\nu,0}g_{1,\bm\nu}
\]
equals the corresponding ratio formed from $Z_{\bm\nu,0}$.
For generic independent charge increments at fixed physical outer
data, however, the saddle moves and the four geometric multipliers
must be recomputed from the charge-sensitivity problem.
This distinction separates a special product-model factorization from
the general charge dependence of the physical coefficient.

The reduced scalar ratio detects charge factorization, but it does not
by itself resolve the matrix geometry that produced the coefficient.
To expose that information, consider a positive Hermitian matrix
$M(t_a,t_b)$ that is affine in two charge parameters near a reference
point:
\[
 M
 =
 M_0+t_aM_a+t_bM_b .
\]
Whiten the two tangent matrices with $M_0$ and define
\begin{equation}
 A
 =
 M_0^{-1/2}M_aM_0^{-1/2},
 \qquad
 B
 =
 M_0^{-1/2}M_bM_0^{-1/2},
 \qquad
 \mathcal F
 =
 -\frac12\log\det M .
 \label{9-13}
\end{equation}
The fourth-order term of
$-\frac12\Tr\log(\id+t_aA+t_bB)$
contains the six words with two $A$'s and two $B$'s.
Cyclicity reduces them to four copies of
$\Tr(A^2B^2)$ and two copies of $\Tr(ABAB)$.
Hence
\begin{equation}
 \partial_a^2\partial_b^2\mathcal F
 =
 2\Tr(A^2B^2)
 +
 \Tr(ABAB)
 =
 3\Tr(A^2B^2)
 -
 \frac12
 \|[A,B]\|_F^2 .
 \label{9-14}
\end{equation}
The second equality uses
\[
 \|[A,B]\|_F^2
 =
 2\Tr(A^2B^2-ABAB)
\]
for Hermitian $A$ and $B$.
Thus the mixed fourth derivative contains both an ordinary positive
contraction and a noncommutative correction.
The derivative alone does not determine the commutator norm.
Once $\Tr(A^2B^2)$ is obtained independently,
Eq.~\eqref{9-14} isolates
$\|[A,B]\|_F^2$.

The affine assumption is essential to this simple formula.
For a general one-parameter family $M(t)$, evaluate all derivatives at
the reference point and define
\[
 X_j
 =
 M^{-1/2}M^{(j)}M^{-1/2}.
\]
Four differentiations give
\begin{equation}
 \mathcal F^{(4)}
 =
 -\frac12
 \Tr\left[
 X_4
 -
 4X_1X_3
 -
 3X_2^2
 +
 12X_1^2X_2
 -
 6X_1^4
 \right].
 \label{9-15}
\end{equation}
Cyclicity of the trace has already combined the distinct orderings in
Eq.~\eqref{9-15}.
Multivariable polarization produces the corresponding mixed
derivatives.
In the gravitational problem, the charge-sensitivity equations supply
the derivatives of both the operator and the moving background.
I therefore differentiate with respect to independent $t_{ij}$ first
and impose
$t_{ij}=\nu_i\nu_j$
only after the matrix derivatives have been formed.

The scalar coefficient admits a parallel hierarchy directly on the
charge lattice.
Define
\[
 \Delta_a f(\bm\nu)
 =
 f(\bm\nu+\bm e_a)-f(\bm\nu).
\]
On a nonvanishing charge domain choose one continuous branch of
$\log\Pi_{\bm\nu}$ and form
\begin{equation}
 \Delta_{a_1}\cdots
 \Delta_{a_k}
 \log\Pi_{\bm\nu}.
 \label{9-16}
\end{equation}
For distinct charge directions, these differences are discrete mixed
cumulants of the reduced coefficient.
If
$\Pi_{\bm\nu}=\prod_a f_a(\nu_a)$,
every mixed difference involving two or more distinct directions
vanishes.
If $\log\Pi_{\bm\nu}$ is a polynomial of total degree $d$, every finite
difference of total order greater than $d$ vanishes.

Neither statement follows from low matrix rank or from simultaneous
diagonalizability of the charge-response matrices.
A scalar example already shows the distinction:
\[
 \log(1+t_a+t_b)
\]
has mixed derivatives of arbitrarily high order even though there is
no matrix noncommutativity at all.
Thus nonzero mixed charge differences diagnose multi-charge dependence
of the reduced scalar coefficient, whereas
$\|[A,B]\|_F$
diagnoses noncommuting tangent directions of the matrix-valued
response.
These are independent structures.

This separation fixes the order of the subsequent analysis.
The charge lattice carries scalar mixed ratios and higher finite
differences only after the fixed-sector coefficient has been reduced.
The endpoint spectrum carries eigenvalue and eigenvector information of
the physical reduced matrix.
Their joint use therefore requires the charge dependence of the matrix,
its source vectors, and its sector weight to remain correlated rather
than replacing them by independent marginal data.

\subsection{BPS matrix ensemble and finite-radius correction}
\label{sec:endpoint-spectrum}
\label{sec:endpoint-positive-matrix}

I now combine the finite-radius deformation with the BPS endpoint
ensemble.
I first identify the reduced BPS matrix and its hard-edge inverse
moments in one fixed charge sector. I then insert the finite-radius
deformation determined by the gravitational boundary problem.

The BPS factorization makes the physical endpoint Hessian positive.
It does not determine a probability measure on positive matrices.
For the microscopic endpoint model I use the central chiral Gaussian
ensemble of Ref.~\cite{ReyChiral2026}.
The charge label $\bm\nu$, the rectangularity $\kappa$, and the Dyson
index $\beta_{\rm D}$ remain distinct data.

Represent the endpoint square root by
\begin{equation}
 \cQ\in
 \mathbb F_{\beta_{\rm D}}^{(N+\kappa)\times N},
 \qquad
 \dd\mu(\cQ)
 \propto
 \exp\left[
 -\frac{\beta_{\rm D}}4
 \Tr(\cQ^\dagger\cQ)
 \right]\dd\cQ .
 \label{9-22}
\end{equation}
Here
$\mathbb F_{\beta_{\rm D}}=\mathbb R,\mathbb C,\mathbb H$
for
$\beta_{\rm D}=1,2,4$,
and $\kappa\geq0$ is the rectangular defect.
The Dyson index $\beta_{\rm D}$ is unrelated to the scalar coupling
$\beta$.

Choose the retained $m$ columns independently of the realization of
$\cQ$ and write
\begin{equation}
 \cQ=(X\;Y),
 \qquad
 X\in
 \mathbb F_{\beta_{\rm D}}^{(N+\kappa)\times m},
 \qquad
 Y\in
 \mathbb F_{\beta_{\rm D}}^{(N+\kappa)\times n},
 \qquad
 n=N-m .
 \label{9-23}
\end{equation}
Then
\begin{equation}
 \cH_0
 =
 \cQ^\dagger\cQ
 =
 \begin{pmatrix}
 X^\dagger X & X^\dagger Y\\
 Y^\dagger X & Y^\dagger Y
 \end{pmatrix}.
 \label{9-24}
\end{equation}
The Gaussian matrix $Y$ has full column rank almost surely.
Eliminating its columns gives the Schur complement
\begin{equation}
 M_0
 =
 X^\dagger X
 -
 X^\dagger Y
 (Y^\dagger Y)^{-1}
 Y^\dagger X
 =
 X^\dagger P_Y^\perp X ,
 \label{9-25}
\end{equation}
with
\begin{equation}
 P_Y^\perp
 =
 \id-
 Y(Y^\dagger Y)^{-1}Y^\dagger,
 \qquad
 \rank P_Y^\perp
 =
 m+\kappa .
 \label{9-26}
\end{equation}

\paragraph{Endpoint law after column elimination.}
Under the central measure~\eqref{9-22}, $M_0$ has the same law as
$Z^\dagger Z$, where
\[
 Z\in
 \mathbb F_{\beta_{\rm D}}^{(m+\kappa)\times m}
\]
is central Gaussian with the same variance convention.
Its eigenvalue density is
\begin{equation}
 \dd\mu_{\rm eig}(\lambda)
 =
 \frac1{Z_{m,\kappa}^{(\beta_{\rm D})}}
 \prod_{a<b}
 |\lambda_a-\lambda_b|^{\beta_{\rm D}}
 \prod_{a=1}^{m}
 \lambda_a^{\frac{\beta_{\rm D}}2(\kappa+1)-1}
 \e^{-\frac{\beta_{\rm D}}4\lambda_a}
 \dd\lambda_a .
 \label{9-27}
\end{equation}
The rectangularity $\kappa$ is unchanged by the elimination.
Moreover, $M_0$ is independent of $Y^\dagger Y$.

To see this, condition on $Y$ and rotate the row space so that
\begin{equation}
 P_Y^\perp
 =
 \diag(\id_{m+\kappa},0).
\end{equation}
The distribution of $X$ is invariant under this rotation and is
independent of $Y$.
Hence the conditional law of
$X^\dagger P_Y^\perp X$
is independent of $Y$ and equals the law of $Z^\dagger Z$.
The singular-value Jacobian gives Eq.~\eqref{9-27}.
See Ref.~\cite{Muirhead}.

For $\beta_{\rm D}=4$, trace and determinant count each quaternion
eigenvalue once.
If the physical operator is represented by a real or complex matrix
with duplicated quaternion eigenvalues, its Gaussian determinant power
must be matched to that representation.
It cannot be inferred from Eq.~\eqref{9-27} alone.

I next determine which inverse observables are integrable at the hard
edge.
Set
\[
 a_\kappa
 =
 \frac{\beta_{\rm D}}2(\kappa+1),
 \qquad
 b_{\rm D}
 =
 \frac{\beta_{\rm D}}4 .
\]
Multiplying the density by $(\det M_0)^u$ shifts the one-eigenvalue
power by $u$.
The Laguerre normalization therefore gives
\begin{equation}
 \mathbb E[(\det M_0)^u]
 =
 b_{\rm D}^{-mu}
 \prod_{j=0}^{m-1}
 \frac{
 \Gamma(a_\kappa+u+j\beta_{\rm D}/2)
 }{
 \Gamma(a_\kappa+j\beta_{\rm D}/2)
 },
 \qquad
 a_\kappa+u>0 .
 \label{D-det-moments}
\end{equation}
For real $u$, this condition is exact.

Under the determinant-weighted measure, integration by parts gives
\[
 \mathbb E_u\Tr M_0^{-1}
 =
 \frac{
 mb_{\rm D}
 }{
 a_\kappa+u-1
 }.
\]
The pairwise Vandermonde derivatives cancel after summation over the
eigenvalues.
Rotational invariance then gives
\begin{equation}
 \mathbb E[
 (\det M_0)^uM_0^{-1}
 ]
 =
 \mathbb E[(\det M_0)^u]\,
 \frac{
 b_{\rm D}
 }{
 a_\kappa+u-1
 }\id_m,
 \qquad
 a_\kappa+u>1 .
 \label{D-weighted-inverse}
\end{equation}
Several thresholds follow immediately:
\[
 \mathbb E M_0^{-1}<\infty
 \quad\Longleftrightarrow\quad
 a_\kappa>1,
\]
\[
 \mathbb E(\det M_0)^{-1/2}<\infty
 \quad\Longleftrightarrow\quad
 a_\kappa>\frac12,
\]
and
\[
 \mathbb E\!\left[
 (\det M_0)^{-1/2}
 \mathbf b^\dagger M_0^{-1}\mathbf b
 \right]
 <\infty
 \quad\Longleftrightarrow\quad
 a_\kappa>\frac32
\]
for a fixed nonzero source vector $\mathbf b$ independent of the
eigenvectors.
A physical source correlated with $M_0$, a fermionic factor, or a
vanishing source overlap changes the joint integrand and must be treated
before these criteria are applied.

The same threshold is visible directly at the hard edge.
Holding the other eigenvalues away from zero,
\[
 \dd\mu_{\rm eig}
 \propto
 \lambda_{\min}^{a_\kappa-1}\dd\lambda_{\min}.
\]
An unweighted inverse moment of order $q$ therefore requires
\[
 a_\kappa>q .
\]
These conditions concern the random positive eigenvalues of the BPS
endpoint matrix.
They are distinct from the deterministic soft eigenvalue
$\lambda_{\rm soft}\propto\delta$
found in Eq.~\eqref{8-83}.

I now turn on the finite-radius deformation.
For a finite regulated block that is analytic at the endpoint, write
\begin{equation}
 M(\delta)
 =
 M_0+\delta D_1+\delta^2D_2+\cdots .
 \label{9-28}
\end{equation}
For fixed invertible $M_0$,
\begin{align}
 \log\frac{\det M(\delta)}{\det M_0}
 ={}&
 \delta\Tr(M_0^{-1}D_1)
 \nonumber\\
 &+
 \delta^2
 \left[
 \Tr(M_0^{-1}D_2)
 -
 \frac12
 \Tr\!\left(
 M_0^{-1}D_1M_0^{-1}D_1
 \right)
 \right]
 +
 O(\delta^3),
 \label{9-29}
\end{align}
and
\begin{align}
 M(\delta)^{-1}
 ={}&
 M_0^{-1}
 -
 \delta
 M_0^{-1}D_1M_0^{-1}
 \nonumber\\
 &+
 \delta^2
 \left[
 M_0^{-1}D_1M_0^{-1}D_1M_0^{-1}
 -
 M_0^{-1}D_2M_0^{-1}
 \right]
 +
 O(\delta^3).
 \label{9-30}
\end{align}
Termwise ensemble averaging of these formulas requires every displayed
inverse moment to exist and the remainder to admit a common integrable
bound.
The hard-edge conditions above determine when this step is legitimate.

The matched deformation need not always be analytic in $\delta$.
For a mode with
\[
 M(\delta)
 =
 M_0+\eta(\delta)D+\cdots,
\]
the radial calculation determines $\eta(\delta)$.
The homogeneous mode gives ordinary powers.
The $\ell=3$ angular resonance gives
\[
 \eta(\delta)
 =
 \delta^2\log\frac{\delta_0}{\delta}.
\]
Its matrix coefficient is fixed by the physical boundary map
Eq.~\eqref{8-155}.
Thus the radial problem determines the strength and orientation of the
deformation before the endpoint ensemble is averaged.

A deterministic finite-rank correction takes the form
\begin{equation}
 M(\delta)
 =
 M_0+UC(\delta)U^\dagger,
 \qquad
 U^\dagger U=\id_s .
 \label{9-31}
\end{equation}
Here $U$ specifies the affected $s$-dimensional subspace.
Define
\[
 S
 =
 U^\dagger M_0^{-1}U .
\]
The matrix determinant lemma gives
\begin{equation}
 \frac{\det M(\delta)}{\det M_0}
 =
 \det\!\left(
 \id_s+C(\delta)S
 \right),
 \label{9-32}
\end{equation}
while the Woodbury identity gives
\begin{align}
 M(\delta)^{-1}
 ={}&
 M_0^{-1}
 -
 M_0^{-1}U
 [\id_s+C(\delta)S]^{-1}
 C(\delta)
 U^\dagger M_0^{-1}.
 \label{9-33}
\end{align}
Neither identity requires $C(\delta)$ to be invertible.
The correction is controlled by three separate data:
its rank, its strength, and the orientation of $U$ relative to the
endpoint eigenvectors.

The finite-rank family is a spectral diagnostic.
It is not an identification of the positive transverse deformation
derived in Appendix~C.
In particular, consider the subtractive model
\[
 C(\delta)
 =
 -\eta(\delta)G,
 \qquad
 G>0,
 \qquad
 \eta(0)=0 .
\]
A zero eigenvalue appears when
\begin{equation}
 \det\!\left(
 \id_s-\eta G S
 \right)
 =
 0.
 \label{9-34}
\end{equation}
%

To turn this exact condition into a probability law, I now specify the
orientation of the perturbation. Assume that $U$ is fixed independently
of $M_0$ and define
\begin{equation}
 \mathcal Y
 =
 \left(U^\dagger M_0^{-1}U\right)^{-1}.
 \label{9-35}
\end{equation}

\paragraph{Fixed-rank endpoint reduction.}
Let $M_0$ have the endpoint law~\eqref{9-27}, and let $U$ be an
$m\times s$ isometry fixed independently of $M_0$. Then $\mathcal Y$
has density
\begin{equation}
 \dd\mu(\mathcal Y)\propto
 (\det\mathcal Y)^{\frac{\beta_{\rm D}}2(\kappa+1)-1}
 \exp\left[-\frac{\beta_{\rm D}}4\Tr\mathcal Y\right]
 \dd\mathcal Y.
 \label{9-36}
\end{equation}
Equivalently, $\mathcal Y$ is an $s\times s$ Wishart matrix generated
by $s+\kappa$ Gaussian rows. The rectangularity remains $\kappa$.

To see this, use orthogonal, unitary, or symplectic invariance to rotate
$U$ to the first $s$ coordinate vectors and partition
\begin{equation}
 M_0=
 \begin{pmatrix}
 A&B\\
 B^\dagger&D
 \end{pmatrix}.
\end{equation}
The upper-left block of $M_0^{-1}$ is
\begin{equation}
 (A-BD^{-1}B^\dagger)^{-1}.
\end{equation}
Hence Eq.~\eqref{9-35} gives
\begin{equation}
 \mathcal Y
 =
 A-BD^{-1}B^\dagger.
\end{equation}
The partitioned Wishart theorem identifies this Schur complement with
an $s\times s$ Wishart matrix having $s+\kappa$ rows.
Equation~\eqref{9-36} follows. See Ref.~\cite{Muirhead}.

At fixed $s$ and $\kappa$, the law in Eq.~\eqref{9-36} contains no
large-$m$ scale. This fact controls the crossing threshold of a generic
fixed finite-rank direction.

Write the subtractive model explicitly as
\begin{equation}
 M(\delta)
 =
 M_0-\eta(\delta)UGU^\dagger,
 \qquad
 G>0,
 \qquad
 \eta(0)=0.
 \label{9-38}
\end{equation}
Since $S=\mathcal Y^{-1}$, the first zero as a function of the positive
strength $\eta$ occurs at
\begin{align}
 \eta_c
 &=
 \frac1{\lambda_{\max}(G^{1/2}SG^{1/2})},
 \nonumber\\
 &=
 \lambda_{\min}
 \left(G^{-1/2}\mathcal Y G^{-1/2}\right).
 \label{9-40}
\end{align}
A physical crossing occurs at a positive solution of
\begin{equation}
 \eta(\delta_c)=\eta_c .
\end{equation}
On a branch where $\eta(\delta)$ increases from zero, the first such
solution is the first loss of positivity.

For $G=g\id_s$, $\eta_c$ is the smallest eigenvalue of $\mathcal Y$
divided by $g$.
The complete fixed-rank crossing law is therefore a smallest-eigenvalue
law of an $s\times s$ Wishart matrix with the same rectangularity
$\kappa$.

The rank-one case is elementary.
Set
\begin{equation}
 U=u,
 \qquad
 u^\dagger u=1,
\end{equation}
and define
\begin{equation}
 y
 =
 \left(
 u^\dagger M_0^{-1}u
 \right)^{-1},
 \qquad
 a_\kappa
 =
 \frac{\beta_{\rm D}}2(\kappa+1).
 \label{9-41}
\end{equation}
Equation~\eqref{9-36} with $s=1$ gives the gamma density
\begin{equation}
 p_y(y)
 =
 \frac{
 b_{\rm D}^{a_\kappa}
 }{
 \Gamma(a_\kappa)
 }
 y^{a_\kappa-1}
 \e^{-b_{\rm D}y},
 \qquad
 y>0.
\end{equation}
For $G=g$, one has
\begin{equation}
 \eta_c=\frac{y}{g}.
\end{equation}
Therefore
\begin{equation}
 p(\eta_c)
 =
 \frac{
 (\beta_{\rm D}g/4)^{a_\kappa}
 }{
 \Gamma(a_\kappa)
 }
 \eta_c^{a_\kappa-1}
 \exp\left[
 -\frac{\beta_{\rm D}g}{4}\eta_c
 \right],
 \qquad
 \eta_c>0 .
 \label{9-42}
\end{equation}
Its mean and variance are
\begin{equation}
 \mathbb E\,\eta_c
 =
 \frac{2(\kappa+1)}{g},
 \qquad
 \operatorname{Var}(\eta_c)
 =
 \frac{
 8(\kappa+1)
 }{
 \beta_{\rm D}g^2
 }.
 \label{9-43}
\end{equation}

For fixed $t>0$,
\begin{equation}
 \Pr\left(
 \eta_c<\frac{t}{m}
 \right)
 =
 \frac{
 \gamma\left(
 a_\kappa,
 \frac{\beta_{\rm D}gt}{4m}
 \right)
 }{
 \Gamma(a_\kappa)
 }.
 \label{9-44}
\end{equation}
As $m\to\infty$,
\[
 \Pr\left(
 \eta_c<\frac{t}{m}
 \right)
 \sim
 \frac1{\Gamma(a_\kappa+1)}
 \left(
 \frac{\beta_{\rm D}gt}{4m}
 \right)^{a_\kappa}.
\]
Thus a rank-one direction fixed independently of $M_0$ has an
$O(1)$ crossing strength in the normalization of Eq.~\eqref{9-27}.
Its probability of crossing already on the microscopic
$m^{-1}$ hard-edge scale is suppressed as
$m^{-a_\kappa}$.

The same rank-one update gives exact source formulas.
The determinant ratio is
\begin{equation}
 \frac{\det M(\delta)}{\det M_0}
 =
 1-
 \frac{\eta(\delta)}{\eta_c}.
 \label{9-45}
\end{equation}
For a fixed source vector $\mathbf b$,
\begin{equation}
 \mathbf b^\dagger
 M(\delta)^{-1}
 \mathbf b
 =
 \mathbf b^\dagger
 M_0^{-1}
 \mathbf b
 +
 \frac{
 \eta(\delta)g
 \abs{
 u^\dagger M_0^{-1}\mathbf b
 }^2
 }{
 1-\eta(\delta)/\eta_c
 }.
 \label{9-46}
\end{equation}
At
$\eta(\delta_c)=\eta_c$,
the determinant vanishes.
The source contraction has a pole precisely when
\[
 u^\dagger M_0^{-1}\mathbf b\neq0.
\]
The quadratic integral then develops a zero direction, and its leading
nonzero higher-order vertex controls the local saddle integral.

The distribution in Eq.~\eqref{9-42} uses the independence of $u$ and
$M_0$.
A physical perturbation whose orientation is determined by the same
microscopic data as $M_0$ requires the joint law of
\[
 (M_0,u,g).
\]
Equations~\eqref{9-45} and \eqref{9-46}, however, remain exact for
every realization.

I finally compare the deformation strength with the microscopic hard
edge.
In the normalization of Eq.~\eqref{9-27}, the bulk eigenvalues are
$O(m)$, whereas the smallest positive eigenvalues occur on the scale
\begin{equation}
 \lambda_{\rm hard}(m)
 \sim
 m^{-1},
 \qquad
 m\to\infty,
 \quad
 \kappa\ {\rm fixed}.
 \label{9-47}
\end{equation}
Equivalently, the normalized matrix $m^{-1}M_0$ has hard-edge scale
$m^{-2}$.

For a general correction define
\begin{equation}
 \chi(\delta)
 =
 \frac{
 \norm{M(\delta)-M_0}_{\rm op}
 }{
 \lambda_{\rm hard}(m)
 }.
 \label{9-48}
\end{equation}
If
$\chi(\delta)\to0$ in probability,
Weyl's eigenvalue bound makes the perturbation negligible on the
microscopic hard-edge scale.
When
$\chi(\delta)=O(1)$,
the matrix elements of the correction in the endpoint eigenbasis
become leading data.
Its norm alone no longer determines the shifted edge.
The rank, orientation, and correlations with $M_0$ must then be kept.

This distinction is particularly sharp for fixed rank.
A fixed subspace independent of $M_0$ has the $O(1)$ threshold derived
above even though the smallest eigenvalues of $M_0$ are $O(m^{-1})$.
A microscopic crossing at strength $O(m^{-1})$ therefore requires an
atypically small compressed eigenvalue or a perturbation correlated
with the hard-edge eigenvectors.

For comparison, the exactly solvable full-rank downward shift
\begin{equation}
 M(\delta)
 =
 M_0-g\eta(\delta)\id,
 \qquad
 g>0,
 \label{9-49}
\end{equation}
moves every eigenvalue by the same amount.
The number of negative eigenvalues is exactly
\begin{equation}
 n_-(\delta)
 =
 \#\left\{
 a:
 \lambda_a(M_0)
 <
 g\eta(\delta)
 \right\}.
 \label{9-50}
\end{equation}
Its zero crossings are therefore governed directly by the hard-edge
counting process.

The full-rank downward shift and the fixed-rank subtractive model are
spectral diagnostics.
The homogeneous transverse deformation derived in Appendix~C instead
has a positive endpoint derivative on its soft physical subspace.
The physical sign, rank, strength, and orientation must therefore be
taken from the radial and boundary calculation before any endpoint
spectral average is performed.

\section{Statistics and bulk spectrum}
\label{sec:statistics}
\subsection{Spectral averages and microscopic measure}

I first fix the charge sector, saddle branch, boundary conditions, and
reduction data.
For each realization of the microscopic data, the reduced matrix has
the matched expansion
\[
 M_{\bm\nu}(\delta)
 =
 M_{0,\bm\nu}
 +
 \sum_\alpha
 \eta_\alpha(\delta)D_{\alpha,\bm\nu}
 +
 \cdots .
\]
Ordinary radial terms have
$\eta_\alpha(\delta)=\delta^k$.
A resonant term has
$\eta_\alpha(\delta)=\delta^k\log(\delta/\delta_0)$.
I reserve $D_{1,\bm\nu}$ for the coefficient of the linear term in
$\delta$,
\[
 D_{1,\bm\nu}
 =
 \left.
 \frac{\dd M_{\bm\nu}}{\dd\delta}
 \right|_{0^+}.
\]
Higher-order and resonant matrices remain separate from $D_1$.

The physical origin of $D_1$ is fixed before any statistical average is
taken.
Let $e_a$ be fixed coordinates in the reduced endpoint space and let
\[
 \Psi_a=\mathfrak J_0e_a
\]
be their stationary endpoint extensions.
Differentiating the complete reduced quadratic form gives
\begin{align}
 (D_{1,\bm\nu})_{ab}
 ={}&
 \int_{\mathcal M}
 \dd\mu(x)\,
 \mathcal I_{1,ab}(x)
 +
 (D_1^\partial)_{ab},
 \label{10-4}\\
 \mathcal I_{1,ab}
 ={}&
 \dot\Psi_a^\dagger A_1\dot\Psi_b
 +
 \dot\Psi_a^\dagger B_1\Psi_b
 \nonumber\\
 &+
 \Psi_a^\dagger B_1^\dagger\dot\Psi_b
 +
 \Psi_a^\dagger C_1\Psi_b .
 \nonumber
\end{align}
For the radial formulation,
$\dd\mu=\dd\tau$ with the radial weights already included in
$A_1,B_1,C_1$.
The integral covers the complete solved region.
The term $D_1^\partial$ contains the derivative of the physical
boundary form and the contributions induced by a varying boundary
basis.

The stationary extensions already include the response of fields
eliminated by their equations of motion.
Their first variations therefore produce no additional relaxation term
in Eq.~\eqref{10-4}.
Any inner--outer decomposition must sum back to the same physical
matrix $D_1$.

I now specify the fixed-charge microscopic ensemble.
Let $\mathbb E_{\bm\nu}$ denote its normalized expectation and define
\begin{equation}
 \overline D_{1,\bm\nu}
 =
 \mathbb E_{\bm\nu}D_{1,\bm\nu},
 \qquad
 \Xi_{\bm\nu}
 =
 D_{1,\bm\nu}
 -
 \overline D_{1,\bm\nu}.
 \label{10-1}
\end{equation}
For a Hermitian reduced matrix, the entry covariance is
\begin{equation}
 \mathcal C_{\bm\nu;ab,cd}
 =
 \mathbb E_{\bm\nu}
 \left[
 (\Xi_{\bm\nu})_{ab}
 \overline{(\Xi_{\bm\nu})_{cd}}
 \right].
 \label{10-2}
\end{equation}
The covariance need not be isotropic.

The endpoint matrix and its finite-radius deformation are generated by
the same microscopic data.
Their joint connected generator is
\begin{equation}
 \mathcal W_{\bm\nu}^{\rm joint}[J_0,J_1]
 =
 \log
 \mathbb E_{\bm\nu}
 \exp\left[
 i\Tr(J_0M_{0,\bm\nu})
 +
 i\Tr(J_1D_{1,\bm\nu})
 \right].
 \label{10-3}
\end{equation}
Mixed derivatives with respect to $J_0$ and $J_1$ measure correlations
between the endpoint matrix and its deformation.
Replacing their joint law by independent samples removes precisely
these mixed cumulants.

Equation~\eqref{10-4} also gives the covariance in terms of the local
response density.
Taking the connected covariance of the complete expression gives
\begin{align}
 \mathcal C_{\bm\nu;ab,cd}
 ={}&
 \int_{\mathcal M^2}
 \dd\mu(x)\dd\mu(y)\,
 \left\langle
 \mathcal I_{1,ab}(x)
 \overline{\mathcal I_{1,cd}(y)}
 \right\rangle_c
 \nonumber\\
 &+
 \mathcal C^\partial_{\bm\nu;ab,cd}.
 \label{10-5}
\end{align}
Here $\mathcal C^\partial$ collects every term containing
$D_1^\partial$: the boundary--boundary covariance and the two
bulk--boundary cross covariances.
If the stationary extensions $\Psi_a$ fluctuate under the microscopic
measure, those fluctuations are part of $\mathcal I_1$.
The required input is therefore the joint law of the bulk coefficients,
boundary data, and stationary extensions.

I next state a sufficient regime in which the deformation matrix itself
approaches a Gaussian matrix law.
Suppose
\begin{equation}
 \Xi_{\bm\nu}
 =
 \frac1{\sqrt L}
 \sum_{\alpha=1}^{L}
 X_{\bm\nu}^{(\alpha)},
 \qquad
 \mathbb E X_{\bm\nu}^{(\alpha)}
 =
 0.
 \label{10-6}
\end{equation}
For an independent triangular array, require convergence of the
covariance for every fixed finite set of real matrix coordinates and
the Lindeberg condition
\begin{equation}
 \frac1L
 \sum_{\alpha=1}^{L}
 \mathbb E
 \left[
 \norm{X_{\bm\nu}^{(\alpha)}}_F^2
 \mathbf 1_{
 \{
 \norm{X_{\bm\nu}^{(\alpha)}}_F
 >
 \epsilon\sqrt L
 \}
 }
 \right]
 \longrightarrow0
 \qquad
 (\epsilon>0).
 \label{10-7}
\end{equation}
Then every fixed finite collection of matrix entries obeys the ordinary
multivariate central limit theorem.

For independent identically distributed contributions with finite
connected moments,
\begin{align}
 &
 \left\langle
 (\Xi_{\bm\nu})_{a_1b_1}
 \cdots
 (\Xi_{\bm\nu})_{a_pb_p}
 \right\rangle_c
 \nonumber\\
 &\qquad=
 L^{1-p/2}
 \left\langle
 (X_{\bm\nu}^{(1)})_{a_1b_1}
 \cdots
 (X_{\bm\nu}^{(1)})_{a_pb_p}
 \right\rangle_c .
 \label{10-8}
\end{align}
Thus all connected cumulants with $p\geq3$ vanish as
$L\to\infty$.

The same scaling extends to sufficiently weakly correlated
contributions.
For every fixed set of matrix coordinates, impose
\begin{equation}
 \sup_{\alpha_1}
 \sum_{\alpha_2,\ldots,\alpha_p}
 \left|
 \kappa_p
 \left(
 (X^{(\alpha_1)})_{a_1b_1},
 \ldots,
 (X^{(\alpha_p)})_{a_pb_p}
 \right)
 \right|
 \leq
 C_p,
 \qquad
 p\geq2,
 \label{10-9}
\end{equation}
with $C_p$ independent of $L$.
The sum over $\alpha_1$ then contributes one factor of $L$, while the
normalization in Eq.~\eqref{10-6} contributes $L^{-p/2}$.
Hence the $p$-th cumulant is
$O(L^{1-p/2})$.
Together with tightness and moment determinacy, these bounds give the
same finite-coordinate Gaussian limit.
A dominant contribution or a non-summable long-range correlation lies
outside this regime.

The fixed-charge ensemble and variation across different charge
sectors are distinct statistical operations.
Different sectors may nevertheless share the same compactification
data.
Their joint microscopic law therefore contains the cross covariance
\begin{equation}
 \mathcal C_{\bm\nu,\bm\mu;ab,cd}
 =
 \mathbb E
 \left[
 (\Xi_{\bm\nu})_{ab}
 \overline{(\Xi_{\bm\mu})_{cd}}
 \right].
 \label{10-10}
\end{equation}
These cross-sector correlations retain information that is lost if
each charge sector is sampled independently.

I also consider variation across a finite shell in the charge lattice.
This is a second ensemble, distinct from
$\mathbb E_{\bm\nu}$.
Let
\[
 G_q=L_q^TL_q,
 \qquad
 \bm q=L_q\bm\nu,
\]
and define
\begin{equation}
 \mathcal S_{R,\Delta}
 =
 \left\{
 \bm\nu\in\mathbb Z^r:
 R^2-\Delta
 \leq
 \bm\nu^TG_q\bm\nu
 \leq
 R^2+\Delta
 \right\}.
 \label{10-11}
\end{equation}
This is an equal-action shell only when the classical action depends on
the same quadratic invariant.
The product model generally does not satisfy that condition.

For a local quadratic approximation in $\bm q$, write
\begin{equation}
 D_1(\bm q)
 =
 D_0
 +
 \sum_{i,j}
 q_iq_jX_{ij},
 \qquad
 X_{ij}=X_{ji}.
 \label{10-12}
\end{equation}
The continuum approximation replaces the lattice shell by the sphere
$S^{r-1}(R)$.
Using
\[
 \mathbb E q_iq_j
 =
 \frac{R^2}{r}\delta_{ij},
\]
one obtains
\begin{equation}
 \mathbb E_{S^{r-1}(R)}D_1
 =
 D_0
 +
 \frac{R^2}{r}
 \sum_iX_{ii}.
 \label{10-13}
\end{equation}
The fourth moment
\[
 \mathbb E q_iq_jq_kq_l
 =
 \frac{R^4}{r(r+2)}
 \left(
 \delta_{ij}\delta_{kl}
 +
 \delta_{ik}\delta_{jl}
 +
 \delta_{il}\delta_{jk}
 \right)
\]
gives
\begin{align}
 \operatorname{Cov}_{S^{r-1}(R)}(D_1)
 ={}&
 \frac{2R^4}{r(r+2)}
 \Bigg[
 \sum_{i,j}
 X_{ij}\otimes\overline{X_{ij}}
 \nonumber\\
 &-
 \frac1r
 \left(
 \sum_iX_{ii}
 \right)
 \otimes
 \overline{
 \left(
 \sum_jX_{jj}
 \right)
 }
 \Bigg].
 \label{10-14}
\end{align}
The finite lattice-shell moments are instead exact finite sums over
$\mathcal S_{R,\Delta}$.
They measure variation across charge sectors.
They are not the fixed-charge microscopic covariance in
Eq.~\eqref{10-2}.

Under the finite-coordinate convergence assumptions above,
$\Xi_{\bm\nu}$ approaches a centered Gaussian matrix in every fixed
finite collection of entries.
For a fixed Hermitian test matrix $J$,
\begin{equation}
 \log
 \mathbb E_{\bm\nu}
 \exp\left(
 i\Tr J\Xi_{\bm\nu}
 \right)
 =
 -\frac12
 \mathfrak C_{\bm\nu}[J,J]
 +
 o(1),
 \label{10-15}
\end{equation}
where
$\mathfrak C_{\bm\nu}[J,J]$
is the variance of
$\Tr(J\Xi_{\bm\nu})$
obtained by contracting
Eq.~\eqref{10-2} with $J$.
The covariance remains anisotropic unless the microscopic ensemble
supplies an additional invariance.

This Gaussian limit concerns finitely many matrix coordinates.
It does not imply independence from $M_{0,\bm\nu}$, unitary invariance,
or a large-matrix spectral law.
The mixed cumulants in
Eq.~\eqref{10-3} remain part of the endpoint problem.
A growing-matrix limit therefore requires a separate joint spectral
analysis of $M_0$ and its finite-radius deformation.

\subsection{Large-matrix limit and spectral correlations}

The finite-coordinate Gaussian limit of the preceding subsection does
not determine a growing-matrix spectral law.
Two distinct large-matrix regimes arise. In the first, the retained
block grows with $m$ and requires a matrix-scale normalization and a
joint local law. In the second, the external block remains finite
while the internal bulk dimension diverges.

The Gaussian matrix in the representation
$M_{0,\bm\nu}=Z_{\bm\nu}^\dagger Z_{\bm\nu}$ has entries of order one.
Its bulk eigenvalues are therefore of order $m$.
Define
\begin{equation}
 \widetilde Z_{\bm\nu}
 =
 m^{-1/2}Z_{\bm\nu},
 \qquad
 \widetilde M_{0,\bm\nu}
 =
 m^{-1}M_{0,\bm\nu}.
 \label{10-16}
\end{equation}
Then
\[
 \widetilde M_{0,\bm\nu}
 =
 \widetilde Z_{\bm\nu}^\dagger
 \widetilde Z_{\bm\nu}.
\]
For the linear finite-radius deformation write
\begin{equation}
 \widetilde M_{\bm\nu}(\delta)
 =
 \widetilde M_{0,\bm\nu}
 +
 \delta\widetilde D_{1,\bm\nu}
 +
 O(\delta^2),
 \qquad
 \widetilde D_{1,\bm\nu}
 =
 m^{-1}D_{1,\bm\nu}.
 \label{10-17}
\end{equation}
Decompose
\[
 \widetilde D_{1,\bm\nu}
 =
 \overline{\widetilde D}_{1,\bm\nu}
 +
 \widetilde\Xi_{\bm\nu},
 \qquad
 \mathbb E_{\bm\nu}\widetilde\Xi_{\bm\nu}=0.
\]
The dense regime considered here assumes centered entries of
$\widetilde\Xi_{\bm\nu}$ of order $m^{-1/2}$, bounded rescaled moments,
and sufficiently weak correlations.
These assumptions are stronger than the finite-coordinate central
limit assumptions of the preceding subsection.

The sample-covariance structure has the exact Hermitian linearization
\begin{equation}
 \widetilde{\mathbb B}_{\bm\nu}(z)
 =
 \begin{pmatrix}
 \delta\overline{\widetilde D}_{1,\bm\nu}
 +\delta\widetilde\Xi_{\bm\nu}
 -z\id_m
 &
 \widetilde Z_{\bm\nu}^\dagger
 \\
 \widetilde Z_{\bm\nu}
 &
 -\id_{m+\kappa}
 \end{pmatrix}.
 \label{10-18}
\end{equation}
Taking the Schur complement of the lower-right block gives
\begin{equation}
 [\widetilde{\mathbb B}_{\bm\nu}(z)^{-1}]_{11}
 =
 \left[
 \widetilde M_{0,\bm\nu}
 +
 \delta\widetilde D_{1,\bm\nu}
 -
 z\id_m
 \right]^{-1}.
 \label{10-19}
\end{equation}
Thus the upper-left block of the linearized resolvent is exactly the
resolvent of the normalized reduced matrix.

The two hard-edge normalizations are now explicit.
For fixed $\kappa$,
\[
 \lambda_{\min}(M_{0,\bm\nu})
 =
 O(m^{-1}),
\]
whereas
\[
 \lambda_{\min}(\widetilde M_{0,\bm\nu})
 =
 O(m^{-2}).
\]
The first scale is the one used in the fixed-rank formulas of
Section~\ref{sec:endpoint-spectrum}.
The second is the microscopic hard-edge scale of the normalized
sample-covariance matrix.

To state a matrix Dyson equation, separate deterministic and centered
parts:
\[
 \widetilde{\mathbb B}_{\bm\nu}(z)
 =
 \widetilde{\mathbb A}_{\bm\nu}(z)
 +
 \widetilde{\mathbb X}_{\bm\nu},
\]
with
\[
 \widetilde{\mathbb A}_{\bm\nu}(z)
 =
 \begin{pmatrix}
 \delta\overline{\widetilde D}_{1,\bm\nu}
 -z\id_m
 &
 0
 \\
 0
 &
 -\id_{m+\kappa}
 \end{pmatrix},
\]
and
\[
 \widetilde{\mathbb X}_{\bm\nu}
 =
 \begin{pmatrix}
 \delta\widetilde\Xi_{\bm\nu}
 &
 \widetilde Z_{\bm\nu}^\dagger
 \\
 \widetilde Z_{\bm\nu}
 &
 0
 \end{pmatrix}.
\]
The joint distribution of
$\widetilde Z_{\bm\nu}$ and
$\widetilde\Xi_{\bm\nu}$ must be retained.
Define the covariance operator
\begin{equation}
 \widetilde{\mathbb S}_{\bm\nu}[R]
 =
 \mathbb E_{\bm\nu}
 \left[
 \widetilde{\mathbb X}_{\bm\nu}
 R
 \widetilde{\mathbb X}_{\bm\nu}
 \right].
 \label{10-20}
\end{equation}
Assume that this correlated block ensemble satisfies a local law and
that the associated self-consistent solution is stable on the spectral
domain under consideration.
The deterministic approximation then obeys
\begin{equation}
 -\widetilde{\mathbb G}_{\bm\nu}(z)^{-1}
 =
 -\widetilde{\mathbb A}_{\bm\nu}(z)
 +
 \widetilde{\mathbb S}_{\bm\nu}
 [\widetilde{\mathbb G}_{\bm\nu}(z)].
 \label{10-21}
\end{equation}
The hypotheses required for such matrix Dyson equations are given in
Refs.~\cite{AjankiErdosKruger,ErdosMDE}.
The present linearization is bipartite: its lower-right block is
deterministic and the covariance operator is degenerate in the same
sense as for ordinary sample-covariance matrices.
Correlations between
$\widetilde Z_{\bm\nu}$ and
$\widetilde\Xi_{\bm\nu}$
add further covariance channels and must enter the same local-law
estimate.

The undeformed central Gaussian ensemble provides a normalization
check.
Set
\[
 c_{\rm asp}
 =
 \frac{m+\kappa}{m}.
\]
In the convention of Eq.~\eqref{9-22},
\[
 \mathbb E
 |\widetilde Z_{ia}|^2
 =
 \frac2m .
\]
For the isotropic solution write the two diagonal blocks of
$\widetilde{\mathbb G}$ as
\[
 m_1\id_m,
 \qquad
 m_2\id_{m+\kappa}.
\]
Equation~\eqref{10-21} becomes
\begin{equation}
 -m_1^{-1}
 =
 z+2c_{\rm asp}m_2,
 \qquad
 -m_2^{-1}
 =
 1+2m_1.
 \label{10-21a}
\end{equation}
Eliminating $m_2$ gives
\begin{equation}
 2zm_1^2
 +
 [z-2(c_{\rm asp}-1)]m_1
 +
 1
 =
 0.
 \label{E-Gaussian-check}
\end{equation}
The physical branch satisfies
\[
 m_1(z)\sim-\frac1z
\]
at large $z$ and
$\operatorname{Im}m_1(z)>0$
for
$\operatorname{Im}z>0$.
For fixed $\kappa$,
$c_{\rm asp}\to1$,
and the limiting bulk support is
\[
 [0,8]
\]
in this variance convention.
This check fixes the normalization of the linearization.

I next extract two observables from the large-matrix law.
Let
$\widetilde\rho_{\bm\nu,\delta}$
denote the limiting empirical spectral density of
$\widetilde M_{\bm\nu}(\delta)$.
Assume uniform integrability of
$\log|\lambda|$
for the two spectral measures.
Then
\begin{align}
 \frac1m
 \log\abs{
 \frac{
 \det M_{\bm\nu}(\delta)
 }{
 \det M_{0,\bm\nu}
 }}
 ={}&
 \frac1m
 \log\abs{
 \frac{
 \det\widetilde M_{\bm\nu}(\delta)
 }{
 \det\widetilde M_{0,\bm\nu}
 }}
 \nonumber\\
 &\longrightarrow
 \int
 \log|\lambda|\,
 \widetilde\rho_{\bm\nu,\delta}(\lambda)
 \dd\lambda
 \nonumber\\
 &\qquad-
 \int
 \log|\lambda|\,
 \widetilde\rho_{\bm\nu,0}(\lambda)
 \dd\lambda .
 \label{10-22}
\end{align}
The logarithmic integrability assumption is essential at the hard edge.

For the source response, define
\[
 \widetilde{\mathbf b}^{\rm eff}_{\bm\nu}
 =
 m^{-1/2}
 \mathbf b^{\rm eff}_{\bm\nu}.
\]
Then the normalization is exact:
\[
 (\mathbf b^{\rm eff}_{\bm\nu})^\dagger
 M_{\bm\nu}^{-1}
 \mathbf b^{\rm eff}_{\bm\nu}
 =
 (\widetilde{\mathbf b}^{\rm eff}_{\bm\nu})^\dagger
 \widetilde M_{\bm\nu}^{-1}
 \widetilde{\mathbf b}^{\rm eff}_{\bm\nu}.
\]
Under the corresponding anisotropic local law,
\begin{align}
 (\widetilde{\mathbf b}^{\rm eff}_{\bm\nu})^\dagger
 \widetilde M_{\bm\nu}(\delta)^{-1}
 \widetilde{\mathbf b}^{\rm eff}_{\bm\nu}
 \longrightarrow
 (\widetilde{\mathbf b}^{\rm eff}_{\bm\nu})^\dagger
 [\widetilde{\mathbb G}_{\bm\nu}(0)]_{11}
 \widetilde{\mathbf b}^{\rm eff}_{\bm\nu},
 \label{10-23}
\end{align}
whenever $z=0$ belongs to a stable domain of the local law.
If zero lies at the hard edge, one must retain a spectral regulator and
control its removal with an edge local law and the inverse-moment bounds
of Section~\ref{sec:endpoint-spectrum}.
A bulk local law alone does not determine the inverse exactly at the
hard edge.

The large-$m$ limit above keeps the retained block itself macroscopic.
A different spectral regime arises when the external dimension is held
fixed while a large internal block is eliminated.
This second limit produces a finite-dimensional reaction matrix rather
than a hard-edge sample-covariance law.

Let the external dimension be $m$ and the internal dimension be $N$.
At a real spectral parameter $z$, write
\begin{equation}
 \cH_{\bm\nu}(z)
 =
 \begin{pmatrix}
 H_{\bm\nu,\rm ext}-z\id_m
 &
 \Gamma_{\bm\nu}^\dagger
 \\
 \Gamma_{\bm\nu}
 &
 H_{\bm\nu,\rm int}-z\id_N
 \end{pmatrix}.
 \label{10-24}
\end{equation}
Eliminating the internal block gives
\begin{equation}
 M_{\bm\nu}(z)
 =
 H_{\bm\nu,\rm ext}
 -
 z\id_m
 -
 \Gamma_{\bm\nu}^\dagger
 (H_{\bm\nu,\rm int}-z\id_N)^{-1}
 \Gamma_{\bm\nu}.
 \label{10-25}
\end{equation}
Define the internal reaction matrix
\begin{equation}
 R_{\bm\nu}(z)
 =
 \Gamma_{\bm\nu}^\dagger
 (z\id_N-H_{\bm\nu,\rm int})^{-1}
 \Gamma_{\bm\nu}.
 \label{10-26}
\end{equation}
Then
\[
 M_{\bm\nu}(z)
 =
 H_{\bm\nu,\rm ext}
 -
 z\id_m
 +
 R_{\bm\nu}(z).
\]
The same Schur complement gives
\begin{equation}
 \det\cH_{\bm\nu}(z)
 =
 \det(H_{\bm\nu,\rm int}-z\id_N)
 \det M_{\bm\nu}(z),
 \label{10-27a}
\end{equation}
and
\begin{equation}
 [\cH_{\bm\nu}(z)^{-1}]_{\rm ext,ext}
 =
 M_{\bm\nu}(z)^{-1}.
 \label{10-27}
\end{equation}

The matrix-Cauchy law belongs to this second, bulk regime.
Consider the unitary equivalent-channel ensemble of
Ref.~\cite{FyodorovKhoruzhenkoNock}.
Take
$N\to\infty$
at fixed external dimension $m$ and evaluate $z$ at a point where the
internal limiting density is positive.
After centering and scaling the reaction matrix, its limiting density is
\begin{equation}
 \mathcal P(R)
 \propto
 \det\left[
 \lambda_{\rm C}^2\id_m
 +
 (R-\epsilon_{\rm C}\id_m)^2
 \right]^{-m},
 \qquad
 \lambda_{\rm C}>0.
 \label{10-28}
\end{equation}
The center $\epsilon_{\rm C}$ and width $\lambda_{\rm C}$ are fixed by
the internal spectral density and the channel coupling.

This distribution is a bulk reaction-matrix law.
It is distinct from the BPS hard-edge ensemble in
Eq.~\eqref{9-27}.
The latter describes small positive eigenvalues of an endpoint square.
Equation~\eqref{10-28} instead describes the response of finitely many
channels to a large internal spectrum at a point of positive bulk
density.

Define the centered and rescaled reaction matrix
\begin{equation}
 \widetilde R
 =
 \frac{
 R-\epsilon_{\rm C}\id_m
 }{
 \lambda_{\rm C}
 }.
 \label{10-29}
\end{equation}
Its Cayley transform is
\begin{equation}
 \mathcal U
 =
 (\id_m-i\widetilde R)
 (\id_m+i\widetilde R)^{-1}.
 \label{10-30}
\end{equation}
Equivalently,
\[
 \mathcal U
 =
 [\lambda_{\rm C}\id_m
 -i(R-\epsilon_{\rm C}\id_m)]
 [\lambda_{\rm C}\id_m
 +i(R-\epsilon_{\rm C}\id_m)]^{-1}.
\]
Since $\widetilde R$ is Hermitian,
$\mathcal U$ is unitary.
If $x_a$ are the eigenvalues of $\widetilde R$ and
$\e^{i\theta_a}$ are the corresponding eigenvalues of $\mathcal U$,
then
\[
 x_a
 =
 -\tan\frac{\theta_a}{2}.
\]
Under this change of variables, the eigenvalue density generated by
\[
 \det(\id_m+\widetilde R^2)^{-m}
\]
becomes the circular-unitary eigenvalue density.
The full unitarily invariant matrix-Cauchy measure therefore maps to
Haar measure on $U(m)$.

The Cayley map makes the angular structure of the bulk response
explicit, but it does not supply correlations between different charge
sectors.
To apply the bulk reaction-matrix law across the charge lattice one
needs the full family
\begin{equation}
 \bm\nu
 \longmapsto
 \left(
 H_{\bm\nu,\rm int},
 \Gamma_{\bm\nu},
 H_{\bm\nu,\rm ext}
 \right).
 \label{10-31}
\end{equation}
For each sector,
\[
 R_{\bm\nu}(z)
 =
 \Gamma_{\bm\nu}^\dagger
 (z\id_N-H_{\bm\nu,\rm int})^{-1}
 \Gamma_{\bm\nu}.
\]
The joint law must retain correlations of the internal operators,
channel couplings, external subspaces, and charge dependence.
Equation~\eqref{10-28} supplies a one-sector marginal under the stated
bulk assumptions.
It does not determine the joint charge-sector law.

The two large-matrix limits therefore probe different structures.
The normalized Wishart problem resolves the BPS hard edge and its
finite-radius deformation.
The fixed-channel reaction-matrix problem resolves a bulk internal
spectrum and maps its Hermitian response to a compact unitary variable.
Combining charge dependence with either spectral regime requires their
joint microscopic law rather than independent sampling of the charge
weights, endpoint matrices, and response matrices.

\section{Finite matrix identities}
\label{app:block-identities}

The identities in this section are finite-dimensional algebraic statements.
Their continuum use requires the common operator domains, measures,
regulators, and determinant prescription of
Appendix~\ref{sec:finite-radius-boundary}.

I begin with bosonic block elimination.
Let
\begin{equation}
 H=
 \begin{pmatrix}
 A&B\\
 C&D
 \end{pmatrix},
 \qquad
 D\ \text{invertible},
 \qquad
 M=A-BD^{-1}C.
 \label{12-1}
\end{equation}
No Hermiticity assumption is needed for the following factorization:
\begin{equation}
 H=
 \begin{pmatrix}
 I&BD^{-1}\\
 0&I
 \end{pmatrix}
 \begin{pmatrix}
 M&0\\
 0&D
 \end{pmatrix}
 \begin{pmatrix}
 I&0\\
 D^{-1}C&I
 \end{pmatrix}.
 \label{F-triangular}
\end{equation}
Both triangular matrices have determinant one. Hence
\begin{equation}
 \det H=\det D\,\det M.
 \label{12-2}
\end{equation}
If $M$ is also invertible, inversion of
Eq.~\eqref{F-triangular} gives
\begin{equation}
 H^{-1}
 =
 \begin{pmatrix}
 M^{-1}
 &
 -M^{-1}BD^{-1}
 \\
 -D^{-1}CM^{-1}
 &
 D^{-1}+D^{-1}CM^{-1}BD^{-1}
 \end{pmatrix}.
 \label{12-3}
\end{equation}

Split a row source and a column source according to the same block
decomposition,
\[
 x^\dagger=(x_P^\dagger,x_Q^\dagger),
 \qquad
 y=\binom{y_P}{y_Q}.
\]
Equation~\eqref{12-3} then gives
\begin{align}
 x^\dagger H^{-1}y
 ={}&
 x_Q^\dagger D^{-1}y_Q
 \nonumber\\
 &+
 \left(
 x_P^\dagger-x_Q^\dagger D^{-1}C
 \right)
 M^{-1}
 \left(
 y_P-BD^{-1}y_Q
 \right).
 \label{12-4}
\end{align}
For Hermitian $H$, the left effective source is the Hermitian conjugate
of the right effective source.
For a general complex bilinear problem, the ordered row and column in
Eq.~\eqref{12-4} remain distinct.

The same structure follows directly from a bosonic Gaussian integral.
Take $H$ real symmetric and positive and write the variables as
$(x,u)$.
With linear source
\[
 \sum_iJ_i
 \left(
 b_{i,P}^Tx+b_{i,Q}^Tu
 \right),
\]
define
\begin{equation}
 \widetilde u
 =
 u+D^{-1}Cx
 -D^{-1}\sum_iJ_i b_{i,Q}.
\end{equation}
Completing the square separates the complementary Gaussian.
Its integration produces the factor
\[
 (\det D)^{-1/2},
\]
and leaves the Schur complement
\[
 M=A-BD^{-1}C.
\]
The source coupled to the retained variable becomes
\begin{equation}
 b_{i,P}^{\rm eff}
 =
 b_{i,P}-BD^{-1}b_{i,Q},
\end{equation}
while the eliminated block produces the direct source response
\begin{equation}
 G_Q^{ij}
 =
 b_{i,Q}^TD^{-1}b_{j,Q}.
\end{equation}
The remaining Gaussian contributes
\[
 (\det M)^{-1/2}
\]
and the retained response
\begin{equation}
 (b_{i,P}^{\rm eff})^T
 M^{-1}
 b_{j,P}^{\rm eff}.
\end{equation}
Thus the complementary determinant, effective source, and contact
response arise from one Gaussian elimination and must be retained
together.

With normalized Lebesgue measure
\[
 \frac{\dd^nx}{(2\pi)^{n/2}},
\]
the standard Gaussian contains no additional powers of $2\pi$.
A different measure carries its normalization explicitly.
Under contour continuation, the quadratic form and its source are
continued together.

I next record the fermionic analogue.
Let
\begin{equation}
 \mathcal D
 =
 \begin{pmatrix}
 A&B\\
 -B^T&D
 \end{pmatrix},
 \qquad
 A^T=-A,
 \qquad
 D^T=-D,
 \qquad
 M_F=A+BD^{-1}B^T,
 \label{F-pfaffian-block}
\end{equation}
where both antisymmetric diagonal blocks have even dimension and $D$ is
invertible.
Define
\begin{equation}
 \mathcal T
 =
 \begin{pmatrix}
 I&0\\
 D^{-1}B^T&I
 \end{pmatrix}.
\end{equation}
Direct multiplication gives
\begin{equation}
 \mathcal T^T\mathcal D\,\mathcal T
 =
 \begin{pmatrix}
 M_F&0\\
 0&D
 \end{pmatrix},
 \qquad
 \det\mathcal T=1.
\end{equation}
Using
\[
 \Pf(\mathcal T^T\mathcal D\,\mathcal T)
 =
 \det\mathcal T\,\Pf\mathcal D
\]
therefore gives
\begin{equation}
 \Pf\mathcal D
 =
 \Pf M_F\,\Pf D.
 \label{F-pfaffian-factor}
\end{equation}
The Berezin orientation is fixed by this ordered block decomposition.
Extracted exact modes retain their own insertion and measure factors.
Equations~\eqref{6-25}, \eqref{6-26}, and \eqref{6-30}
are the determinant, Pfaffian, and source identities on the compatible
reduced spaces.

Finally, let $K(\delta)$ be an invertible differentiable finite matrix.
Differentiating
$K K^{-1}=I$
gives
\begin{equation}
 \partial_\delta K^{-1}
 =
 -K^{-1}
 (\partial_\delta K)
 K^{-1}.
\end{equation}
Differentiating the determinant gives
\begin{equation}
 \partial_\delta\log\det K
 =
 \Tr\!\left(
 K^{-1}\partial_\delta K
 \right).
\end{equation}
For an invertible antisymmetric matrix $\mathcal D(\delta)$ on a fixed
Pfaffian branch,
\begin{equation}
 \partial_\delta\log\Pf\mathcal D
 =
 \frac12
 \Tr\!\left(
 \mathcal D^{-1}
 \partial_\delta\mathcal D
 \right).
\end{equation}
These identities give the matrix derivatives used in
Appendices~\ref{sec:finite-radius-boundary}
and~\ref{sec:charge-phase}.


\end{document}